\documentclass[aps,pra,twocolumn,floats,showpacs,letterpaper,tighten,eqsecnum]{revtex4}
\usepackage{amssymb}
\usepackage{amsbsy}
\usepackage{amsmath}
\usepackage{epsfig}
\usepackage{graphicx}
\usepackage{array}
\usepackage[utf8]{inputenc}
\usepackage{textcomp}
\usepackage{color}
\usepackage{bm}

\newcommand{\bsub}{\begin{subequations}}
\newcommand{\esub}{\end{subequations}}

\newcommand{\ket}[1]{\left| {#1} \right\rangle}
\newcommand{\bra}[1]{\left\langle {#1} \right|}

\DeclareMathOperator{\sgn}{sgn}

\newcommand{\intl}[1]{\int\limits_{#1}}
\newcommand{\vex}[1]{\bm{\mathrm{#1}}}

\newcommand{\sigh}{\hat{\sigma}}

\let\a=\alpha \let\b=\beta \let\g=\gamma \let\d=\partial 
   
\let\l=\lambda     
\let\s=\sigma   

\let\w=\omega   \let\Q=\Theta 
 \let\P=\Phi  
 \let\W=\Omega    
\let\la=\label  
  
\def\bfR{\mathbf{R}}
\def\bfdel{\boldsymbol{\delta}}

\def\nn{\nonumber}

\let\fr=\frac

\def\bpm{\begin{pmatrix}}
\def\epm{\end{pmatrix}}
\def\be{\begin{equation}}
\def\ee{\end{equation}}
\def\bea{\begin{eqnarray}}
\def\eea{\end{eqnarray}}
\def\ba{\begin{array}}
\def\ea{\end{array}}
\def\del{\delta}

\def\td{\tilde}
\def\wtd{\widetilde}

\def\ep{{\epsilon}}

\def\im{\text{Im}}

\def\Tt{\mathbf{T_\tau}}
\def\hG{\hat{G}}
\def\hs{\hat{\sigma}}

\def\bfr{\mathbf{r}}

\def\hJ{\hat{\sigma}}
\def\hts{\hat{s}}
\def\Tr{\text{Tr}}
\def\hrho{\hat{\rho}}
\def\gms{\gamma_\mathrm{s}}
\def\gmc{\gamma_\mathrm{c}}
\def\la{\lambda_\mathrm{A}}
\def\rmd{\mathrm{d}}

\def\Tt{\mathbf{T_\tau}}

\def\mcpt{\!\!\right.\right. & \left.\left.\!\!}

\def\nvsp{\!\!\!\!\!\!\!\!\!\!\!\!\!\!\!\!\!\!\!\!\!\!\!\!\!\!\!\!\!\!\!\!}
\def\mti{\mathcal{I}}

\newcommand{\bk}{{\bf k}}

\newcommand{\trasp}{\mathsf{T}}
\newcommand{\hmu}{\hat{\mu}}
\newcommand{\htau}{\hat{\tau}}

\newcommand{\bfK}{\mathbf{K}}

\newcommand{\ots}{\otimes}

\newcommand{\calH}{\mathcal{H}}

\newcommand{\uaw}{\uparrow}
\newcommand{\daw}{\downarrow}
\newcommand{\htbs}{\hat{\boldsymbol{\sigma}}}
\newcommand{\hsig}{\hat{\sigma}}

\newcommand{\tmfr}{\hat{\mathfrak{t}}}
\newcommand{\htQ}{\hat{Q}}
\newcommand{\htdQ}{\hat{Q}^\dagger}
\newcommand{\sfF}{\mathsf{F}}
\newcommand{\sfS}{\mathsf{S}}
\newcommand{\rmj}{\mathrm{j}}
\newcommand{\sfP}{\mathsf{P}}
\newcommand{\htW}{\hat{W}}
\newcommand{\rms}{\mathrm{s}}
\newcommand{\rmc}{\mathrm{c}}

\newcommand{\htY}{\hat{Y}}
\newcommand{\rmA}{\mathrm{A}}
\newcommand{\rmI}{\mathrm{I}}
\newcommand{\rmO}{\mathrm{O}}
\newcommand{\rmS}{\mathrm{S}}
\newcommand{\mfrD}{\mathfrak{D}}
\newcommand{\rmB}{\mathrm{B}}

\begin{document}

\title{Surface transport coefficients for three-dimensional topological superconductors}
\author{Hong-Yi~Xie} \email{hongyi.xie@rice.edu}
\affiliation{Department of Physics and Astronomy, Rice University, Houston, Texas 77005, USA}
\author{Yang-Zhi~Chou}
\affiliation{Department of Physics and Astronomy, Rice University, Houston, Texas 77005, USA}
\author{Matthew~S.~Foster}
\affiliation{Department of Physics and Astronomy, Rice University, Houston, Texas 77005, USA}

\date{\today\\}

\begin{abstract}
We argue that surface spin and thermal conductivities of three-dimensional 
topological superconductors are universal and topologically quantized at low 
temperature. For a bulk winding number $\nu$, there are $|\nu|$ ``colors'' of 
surface Majorana fermions. Localization corrections to surface transport coefficients 
vanish due to time-reversal symmetry (TRS). We argue that Altshuler-Aronov interaction 
corrections vanish because TRS forbids color or spin Friedel oscillations. We confirm 
this 
within a perturbative expansion in the interactions, 
and to lowest order in a large-$|\nu|$ expansion.
In both cases, we employ an asymptotically exact treatment of quenched disorder
effects that exploits the chiral character unique to two-dimensional, 
time-reversal-invariant Majorana surface states.
\end{abstract}

\maketitle


\section{Introduction}

In the quantum Hall effect, transport measurements unambiguously reveal 
the chiral edge states. The precisely quantized Hall conductance is 
a topological quantum number that is insensitive to the sample geometry 
and protected from the effects of disorder or interactions.  
Transport has played a lesser role in the characterization of three-dimensional 
(3D) topological insulators, in part because it has proven difficult to separate bulk and 
surface contributions due to unintended doping \cite{QZ2011}. A more fundamental limitation is 
that transport coefficients do not directly reflect the $\mathbb{Z}_2$ topological invariant when time-reversal 
symmetry (TRS) is preserved. Instead, the Dirac surface states of topological insulators 
are distinguished by the absence of a two-dimensional (2D) metal-insulator transition, 
with a disorder-dependent electrical conductivity that flows to ever larger values on 
longer scales due to weak antilocalization \cite{Bardarson07,Nomura07,QGM2010,footnote--TIwithInts}.

In this paper, we show that 3D topological superconductors (TSCs) 
\cite{HK2010, QZ2011, SRFL2008, Kit, RSFL2010, Foster2012, FXCup}
may provide a closer
analog of the 2D quantum Hall effect. A bulk TSC is characterized by an integer-valued
winding number $\nu$, and belongs to one of three classes CI, AIII, or DIII \cite{SRFL2008}.  
At the surface, there are $|\nu|$ degenerate species (or ``colors'') 
of surface Majorana fermion bands \cite{SRFL2008,FXCup}. These are protected by TRS in all three classes. We will argue that surface
transport coefficients (spin and thermal conductivities) are universal, being determined
only by the bulk winding number. An important consequence is that low-temperature 
surface spin and heat transport can provide a ``smoking gun'' for Majorana surface states.

For a TSC with conserved spin and no interactions, it is known that the zero-temperature 
($T = 0$) spin conductivity is unmodified by nonmagnetic disorder \cite{ludwig1994,Tsv1995,mirlin2006}. 
Without interactions, the ratio of the thermal conductivity to temperature is also universal in the limit $T \rightarrow 0$ \cite{SRFL2008,FXCup}.  
These results are insufficient to establish universal transport, however, because interactions usually 
induce Altshuler-Aronov (AA) conductance corrections in the presence of disorder \cite{AA1985}. These occur due to 
carrier scattering off of self-consistent potential fluctuations \cite{AAG99,ZNA2001}, and can even cause Anderson 
localization \cite{AA1985,Finkel1983,BK1994}. 

Here we argue that all interaction corrections to TSC surface transport coefficients vanish.
The physical picture is simple: disorder cannot induce static modulations in the color, spin, or mass 
densities of the surface Majorana fluid unless time-reversal symmetry is broken (externally or spontaneously).
Then there is no mechanism for short-ranged interactions to relax momentum at zero temperature. 
To support our claim, 
we show that perturbative interaction
corrections to the surface spin conductivity 
vanish in every disorder realization. 
We also show that AA corrections to the spin and 
thermal surface conductances vanish to leading order in a large winding number expansion. 
The quantization of surface transport coefficients hints at a deeper topological origin, 
which we will contemplate in the conclusion. 

The results in the absence of interactions are as follows.
In a system in which spin is at least partially conserved (classes CI and AIII \cite{SRFL2008,Kit,RSFL2010,FXCup}), 
spin transport is well-defined. Both spin and heat can be conducted by the Majorana surface bands. 
Neglecting interactions, the $T=0$ surface spin conductivity assumes the universal value
\cite{ludwig1994,Tsv1995,mirlin2006}
\be \label{LET-spin-cond}
	\s_{xx}^\rms = \frac{|\nu|}{\pi h} \left( \frac{\hbar}{2} \right)^2, \;\; \text{classes CI and AIII},
\ee       
where $\nu \in \mathbb{Z}$ ($2 \mathbb{Z}$) denotes the bulk winding number for class AIII (CI) TSCs.
If spin is not conserved due to spin-orbit coupling (class DIII), the Majorana surface states still conduct
energy. The low temperature thermal conductivity is
\be \label{LET-heat-cond}
	\kappa_{xx} =  \frac{|\nu|}{\pi h}\frac{\pi^2 k_\rmB^2 T}{3 \gamma}, 
	\;\;
	\gamma = \left\{
	\begin{array}{ll}
	1, & \text{classes CI and AIII,} \\
	2, & \text{class DIII.}
	\end{array}
	\right.
\ee
Equation~(\ref{LET-heat-cond}) follows from the Wiedemann-Franz law \cite{SF2000,SRFL2008,NRFN2012,Nakai2014,FXCup}.

Interactions play a dual role in quantum transport \cite{AAG99}. On one hand,
real inelastic scattering cuts off quantum interference at finite temperature, suppressing weak localization 
on scales larger than the dephasing length. Interference corrections are absent in a TSC, but interactions 
could modify transport coefficients in another way.

\begin{figure}[b]
\centering
\includegraphics[width=0.46\textwidth]{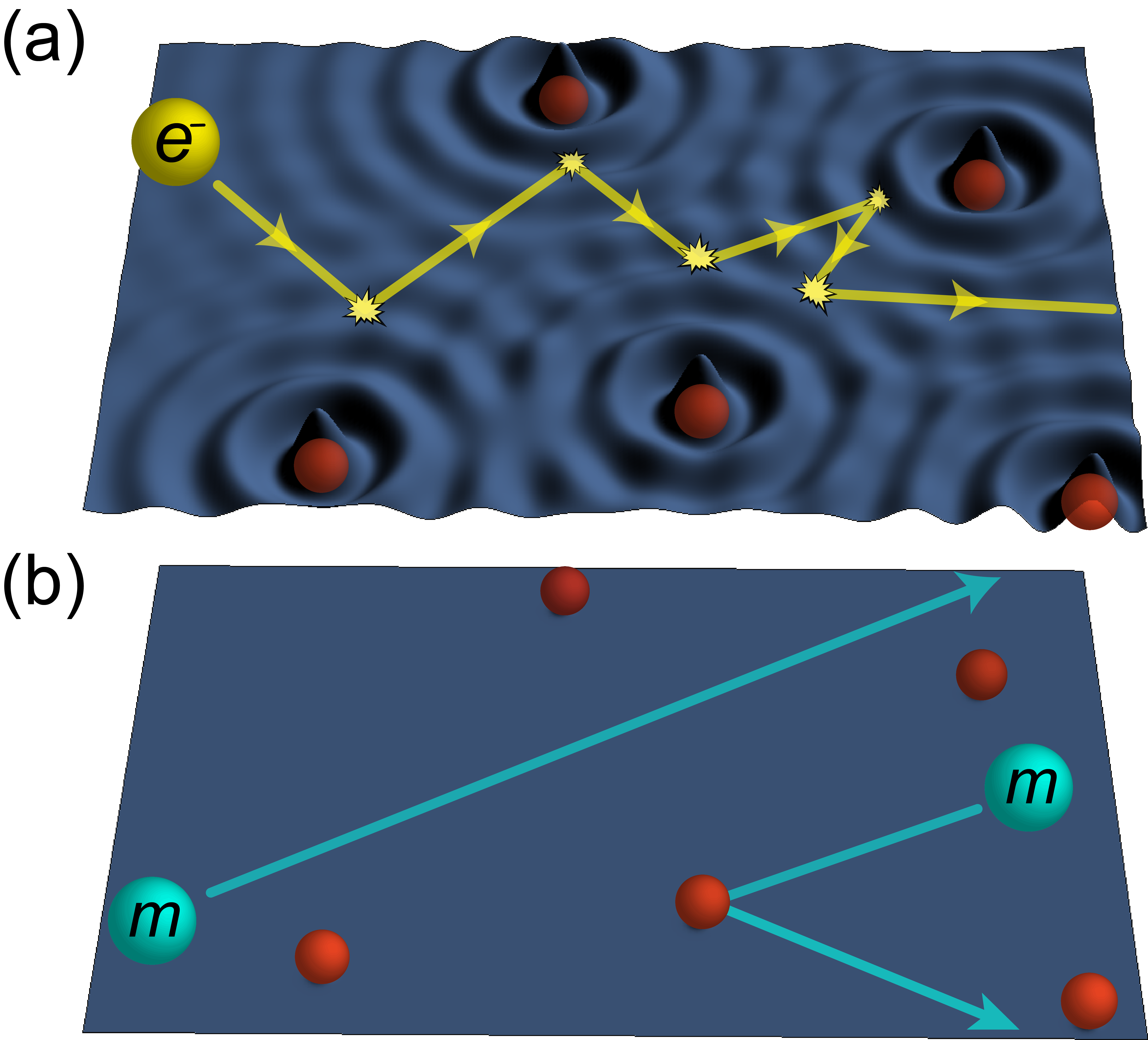}
\caption{In a disordered metal (a), electrons scatter off of both impurities and density Friedel oscillations;
the latter produce Altshuler-Aronov (AA) quantum conductance corrections \cite{AA1985,AAG99,ZNA2001}. 
The Majorana fluid (b) at the surface of a TSC remains featureless in any realization of nonmagnetic disorder, as
no relevant (spin, mass, or color) density can become nonzero without breaking time-reversal symmetry.
We therefore expect that surface transport coefficients for bulk TSCs are free of AA corrections. 
By contrast, AA corrections are ubiquitous in other 2D Dirac systems, including the surface states of 3D topological insulators 
\cite{QGM2010}, and nodal quasiparticles in time-reversal invariant, nontopological superconductors \cite{MJeng2001,DellAnna2002,dellanna2006,foster0608}.}
\label{RippleFig}
\end{figure}

Disorder induces inhomogeneous fluctuations in single-particle wave functions, and these 
can
produce density oscillations near impurities. 
AA conductance corrections \cite{AA1985} 
arise due to the coherent scattering of electrons off of the self-consistent potential due to these oscillations \cite{AAG99,ZNA2001}. 
These corrections are ubiquitous in 
both the standard Wigner-Dyson \cite{AA1985,Finkel1983,BK1994} and exceptional Altland-Zirnbauer \cite{wm2008} classes,
including nontopological superconductors \cite{MJeng2001,DellAnna2002,dellanna2006,foster0608}. 
AA 
corrections can induce Anderson localization even when the noninteracting system would remain metallic. 
For example, in a 2D electron gas with strong spin-orbit scattering, the 
correction
to the electric conductance due to 
Coulomb interactions overwhelms weak antilocalization \cite{BK1994} 
in the diffusive regime. This is precisely what happens for a single surface state band  
enveloping a 3D topological insulator with a properly insulating bulk. In that case delocalization
may survive at a strongly coupled fixed point \cite{QGM2010,footnote--TIwithInts}. 

In a 3D TSC, the physics is uniquely different, owing to the anomalous form
that TRS assumes at the surface \cite{SRFL2008,Foster2012,FXCup}. 
Nonmagnetic disorder couples only to the color or spin 
currents of the $|\nu|$-fold degenerate Majorana quasiparticle bands \cite{SRFL2008,FXCup}. 
Disorder cannot induce static oscillations in the color or spin densities so long as 
TRS is preserved. Mass terms for the Majorana bands are also forbidden.
Interactions can renormalize the existing disorder profile, but this does not modify 
transport coefficients in a TSC. 

In this paper, we argue that interaction corrections vanish to all orders, due 
to the featureless character of the surface Majorana fluid; see Fig.~\ref{RippleFig}.  
To support this argument, we explicitly verify that Eqs.~(\ref{LET-spin-cond}) and 
(\ref{LET-heat-cond}) are unmodified by interactions in two limits. First we 
show that the 
interaction contributions to the $T=0$ surface spin conductivity vanish in every disorder realization for classes CI and AIII. 
We demonstrate this to first (second) order for class CI (AIII), and sketch an all-orders proof for AIII.
Then we consider a large winding number ($|\nu| \gg 1$) expansion, using the Wess-Zumino-Novikov-Witten Finkel'stein 
nonlinear sigma models (WZNW-FNLsMs) introduced in Ref.~\cite{FXCup}. We find that the AA corrections to 
the spin (CI, AIII) or thermal (DIII) conductances are suppressed in the conformal limit for all three TSC classes,
to the lowest nontrivial order in $1/|\nu|$.
An important caveat that we do not address here is whether nonsingular 
interaction corrections arise to the thermal conductivity that violate the Wiedemann-Franz 
relation in classes CI and AIII \cite{CA2005}. 

This paper is organized as follows.
In Sec.~\ref{results-sec} we define models and present the results of our
calculations. 
We qualitatively sketch the key elements responsible for the cancellation of AA
corrections in both schemes, without getting into details.
We also discuss implications and open questions. 
The rest of the paper consists of three technical sections that
are mutually independent.
In Sec.~\ref{micro-sec} we construct a lattice model for a class AIII 
topological superconductor, and derive the form of the surface state theory
assumed in Sec.~\ref{results-sec}.
We present the calculation of Altshuler-Aronov corrections to 
the surface spin conductivity using the Kubo formula
in Sec.~\ref{Kubo-sec}. Finally we derive the WZNW-FNLsM results 
in Sec.~\ref{Fnlsm-sec}.

\section{Approach and main results \label{results-sec}}

\subsection{Majorana surface bands}

The key signature of a 3D TSC with a winding number
$\nu$ is the presence of gapless quasiparticle bands at the surface \cite{SRFL2008,Kit,RSFL2010}.
At energies below the bulk superconducting gap,
these can be viewed as $|\nu|$ ``colors''
of surface Majorana fermions. 
The surface states near zero energy measured relative to the bulk chemical potential 
are protected from the opening of a gap and from Anderson localization, so
long as TRS is preserved \cite{SRFL2008,FXCup}.
The three 3D TSC classes differ by the amount of spin
rotational symmetry preserved in the bulk and at the surface.
Classes CI, AIII, and DIII respectively possess spin SU(2), spin U(1), and no spin 
symmetry. 

The low-energy effective field theory \cite{SRFL2008,FXCup} for noninteracting Majorana surface bands is given by 
\be \label{LET-ham-nd}
\begin{split}
	H^{(0)} 
	=&\, 
	\int \! \rmd^2 \bfr \, 
	\eta^\dagger(\bfr)
	\,
	\hat{h}
	\,
	\eta(\bfr),
	\\
	\hat{h}
	=&
	\,
	\htbs
	\cdot
	\left[-i \boldsymbol{\nabla} + \mathbf{A}_j(\bfr) \, \tmfr^j + {\bm{\mathcal{A}}}(\bfr) \right].
\end{split}
\ee 
In Eq.~(\ref{LET-ham-nd}), $\eta \rightarrow \eta_{\sigma,\kappa}$ is a fermion 
field with indices in pseudospin
$\sigma \in \{1,2\}$ and color $\kappa \in \{1,2,\cdots,|\nu|\}$ spaces;
here $\htbs = \{\sigh^1,\sigh^2\}$ denotes the vector of pseudospin Pauli matrices. 
The pseudospin degree of freedom is some admixture of Nambu (particle-hole), orbital, and
in the case of class DIII physical spin-$1/2$ spaces \cite{SRFL2008,FB2010,FXCup,YSTY2011,YYST2012}.
A bulk microscopic model is necessary to fix the interpretation, but not the structure
of the theory. The potentials $\mathbf{A}_j$ and ${\bm{\mathcal{A}}}$ encode quenched disorder,
as defined below.

For classes CI and AIII, $\eta(\bfr)$ is a complex-valued Dirac spinor; the U(1) charge is the 
conserved spin projection along the $z$-spin axis \cite{SRFL2008,FXCup}.
The U(1) current encodes the $z$-spin density and associated spin current
\be \label{LET-ci-current-1}
	\eta^\dagger \eta(\bfr)=2 S^z(\bfr), \quad  \eta^\dagger \, \htbs \, \eta(\bfr)= 2\mathbf{J}^z(\bfr).
\ee
By contrast, in class DIII $\eta(\bfr)$ is a real spinor that can be taken to satisfy \cite{FXCup} 
$\eta^\dagger = - i \eta^\trasp \sigh^1.$	
Only the energy density and energy current 
(components of the energy-momentum tensor) are conserved 
in class DIII.  

As is typical for a topological phase \cite{HK2010,QZ2011}, symmetries are implemented in an anomalous fashion at the 
surface of a 3D TSC. In particular, TRS appears as the \emph{chiral} condition \cite{SRFL2008,FXCup,BerLec2002}
\begin{align}\label{LET-t-rev}
	-  \sigh^3 \, \hat{h} \, \sigh^3 = \hat{h}.
\end{align}
Given our basis choice, 
Eq.~(\ref{LET-t-rev}) is unique \cite{SRFL2008} and 
implies that  
external time-reversal invariant 
perturbations appear in the surface theory as \emph{vector potentials}. 
In particular, for a system with $|\nu| \geq 2$ colors and spin SU(2) symmetry, 
nonmagnetic disorder induces 
intercolor scattering in the form of the non-Abelian potential $A_j^\alpha(\bfr) \, \tmfr^j$ 
$(\alpha \in \{1,2\})$ in Eq.~(\ref{LET-ham-nd})
\cite{footnote--charge density}. 
The color space symmetry generators $\{\tmfr^j\}$ satisfy a particular Lie algebra $G(|\nu|)$ for each TSC class \cite{footnote--LAclasses}.
In addition, class AIII admits the abelian vector disorder potential $\mathcal{A}^\alpha(\bfr)$,
which couples to the U(1) spin current in Eq.~(\ref{LET-ham-nd}). 
This term is forbidden by spin SU(2) symmetry in class CI \cite{FXCup}, and vanishes exactly for DIII. 
TRS also forbids the accumulation of nonzero spin or color densities.
We denote the spin density as $\mathbf{S}(\bfr)$. In classes CI and AIII,  
the $z$ component is defined above in Eq.~(\ref{LET-ci-current-1}). The non-Abelian color density is 
$\eta^\dagger \tmfr^j \eta(\vex{r})$.
The complete set of Hermitian fermion bilinears (without derivatives) also
includes mixed spin-color potentials, as well as Dirac mass operators \cite{FXCup}.
All of these are odd under time reversal.  

We illustrate these key attributes of TSC surface states in Sec.~\ref{micro-sec}.
Starting from a bulk microscopic model, we derive Eqs.~(\ref{LET-ham-nd}) and (\ref{LET-t-rev}) 
for the Majorana surface fluid of a class AIII TSC.

To treat interactions, we enumerate four-fermion terms consistent with bulk 
time-reversal and spin symmetries. We do not consider long-ranged Coulomb interactions since these 
should be screened by the bulk superfluid. We also neglect interactions that break color symmetry, 
but our results are independent of this.

For class CI (AIII), because spin SU(2) [U(1)] symmetry is preserved, we expect a spin exchange 
interaction of the type $\mathbf{S}(\bfr) \cdot \mathbf{S}(\bfr)$ 
[$S^z(\bfr) S^z(\bfr)$] is important.
In all three classes, TRS implies that a BCS interaction could induce 
a pairing instability \emph{at the surface}. 
The Dirac mass operator 
$m(\bfr) = \eta^\dagger \, \hsig^3 \, \eta(\bfr)$
is time-reversal odd, and $\langle m(\bfr)\rangle \neq 0$ means opening a gap. 
The mass term can be interpreted as an \emph{imaginary} surface pairing amplitude. For example,
in class CI this is the spin singlet operator \cite{Foster2012,FXCup} 
$m(\bfr)  \sim -i \, C_\uaw^\dagger(\bfr) \, C_\daw^\dagger(\bfr) + i \, C_\daw(\bfr) \, C_\uaw(\bfr)$, 
where $C_{\mu}(\bfr)$ annihilates an electron. We therefore can write an attractive BCS interaction 
as $\sim -m^2(\bfr)$. 
The interacting Hamiltonian in each class takes the form \cite{FXCup}
\begin{subequations}  \label{LET-action-int}
\begin{align}
	H_\text{CI}^{(I)}   
	&= 
	\int   \! \rmd^2\bfr \, \left[ \Gamma_\rms \, \mathbf{S}(\bfr) \cdot \mathbf{S}(\bfr) + \Gamma_\rmc \, m(\bfr) \, m(\bfr) \right], 
	\\
	H_\text{AIII}^{(I)} 
	& = 
	\int \! \rmd^2\bfr \, \left[ \Gamma_\rms \, S^z(\bfr) \, S^z(\bfr) + \Gamma_\rmc \, m(\bfr) \, m(\bfr) \right], \label{LET-action-int-AIII} \\
	H_\text{DIII}^{(I)} 
	& = 
	\int \! \rmd^2\bfr \,\Gamma_\rmc \, m(\bfr) \, m(\bfr),
\end{align}
\end{subequations}
where $\Gamma_\rms$ and $\Gamma_\rmc$ are repulsive spin exchange and BCS pairing interaction strengths, respectively.

\subsection{Spin conductivity, interaction expansion}

The spin conductivity in Eq.~(\ref{LET-spin-cond}) is simply the 
$T=0$ ballistic 
Landauer 
result 
expected for $|\nu|$ species of 
2D noninteracting,
massless Dirac fermions 
\cite{ludwig1994, TTTRB, RMFL2007, MBT2009}. 
Here the 
electric charge $e$ is replaced with the spin quantum $\hbar/2$. That Eq.~(\ref{LET-spin-cond}) holds
in the presence of disorder \cite{Tsv1995,mirlin2006} is due to the chiral symmetry in Eq.~(\ref{LET-t-rev}), which 
is just TRS for the surface state quasiparticles \cite{SRFL2008,FXCup}. 
The chiral symmetry allows the retarded (R) and advanced (A) single-particle Green's functions
to be interchanged,
\be \label{LET-chiral-gf}
	-\hat{\sigma}^3 \, \hat{G}^{R/A}(\ep;\vex{r},\vex{r'}) \, \hat{\sigma}^3 = \hat{G}^{A/R}(-\ep;\vex{r},\vex{r'}).   
\ee
Using Eq.~(\ref{LET-chiral-gf}), the noninteracting Kubo formula can be written in terms
of a product of retarded Green's functions. The Ward identity [Eq.~(\ref{ward-id})] can then be 
used to reduce this to the short-distance limit of a single function,
\[
	\s_{xx}^\rms 
	= 
	-\frac{1}{4\pi} 
	\lim_{\vex{r} \rightarrow \vex{r'}} 
	\im\left\{
	\textrm{Tr}
	\left[
	\bm{\sigma}\cdot(\vex{r} - \vex{r'}) \, \hat{G}^{R}(0;\vex{r},\vex{r'})
	\right]
	\right\}.
\]
Although this expression must be properly regularized, it is clear that the
dc conductivity is dominated by the ultraviolet, and is independent of the
disorder [which affects $\hat{G}^{R}(\ep;\vex{r},\vex{r'})$ only on scales
larger than the mean-free path]. The correct noninteracting 
result in Eq.~(\ref{LET-spin-cond}) can be 
understood as a consequence of the axial anomaly in 2+0-D \cite{mirlin2006}.

For classes CI and AIII, Eq.~(\ref{LET-heat-cond}) follows from Eq.~(\ref{LET-spin-cond})
via the Wiedemann-Franz relation. Alternatively, one can obtain Eq.~(\ref{LET-heat-cond}) 
using the Landauer formula for the thermal conductance of 2D ballistic 
Dirac fermions, doped to the Dirac point \cite{TTTRB, RMFL2007, MBT2009}.
We argued in Ref.~\cite{FXCup} that Eq.~(\ref{LET-heat-cond}) with $\gamma = 2$ 
holds for class DIII, wherein spin is not conserved. This 
is derived by artificially doubling the theory to obtain a fictitious U(1) charge
and applying Wiedemann-Franz to Eq.~(\ref{LET-spin-cond}), and then halving this 
result. See also Refs.~\cite{SF2000,NRFN2012, Nakai2014}.

\begin{figure}
\centering
\includegraphics[width=0.47\textwidth]{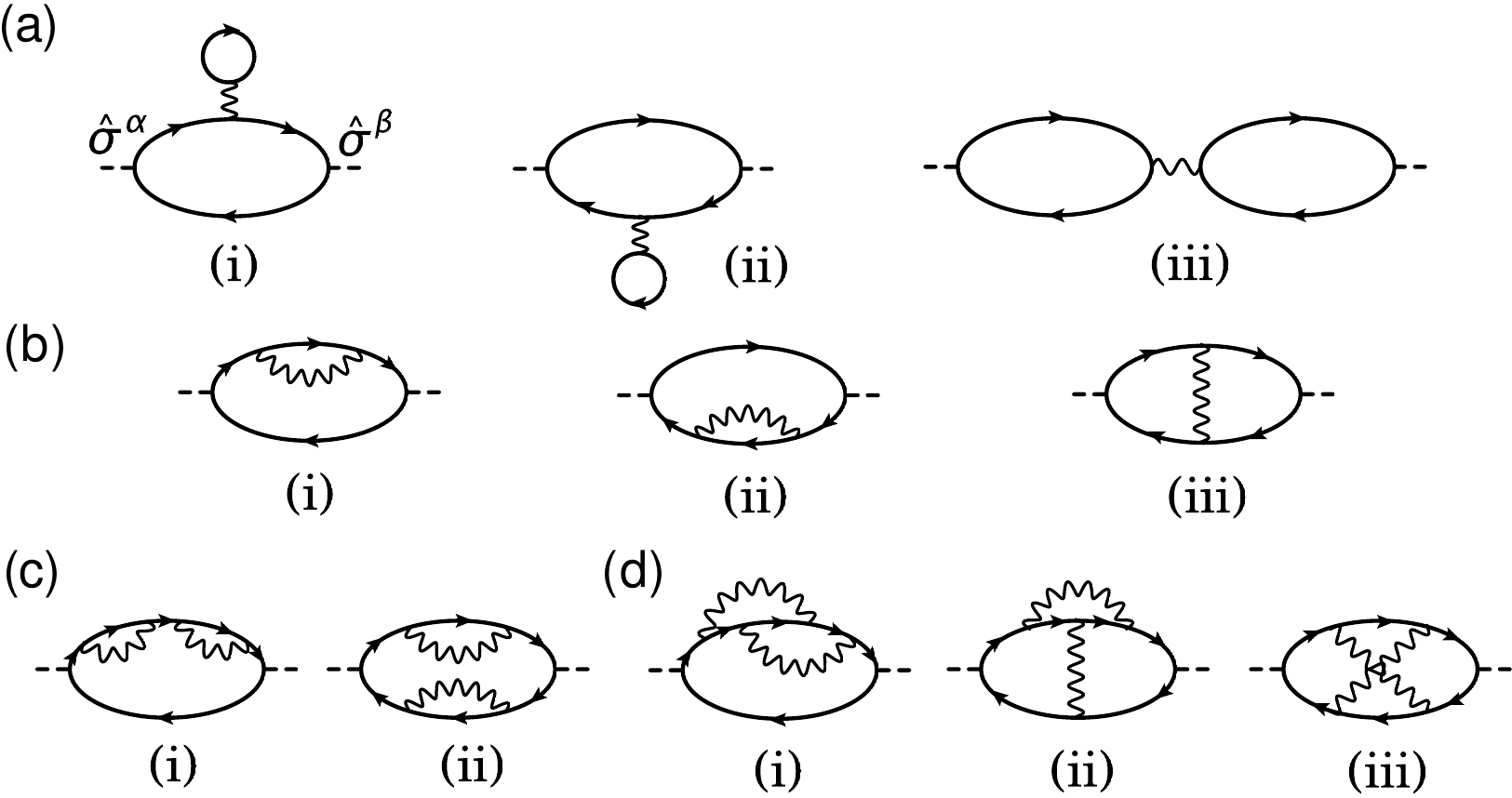}
\caption{
First-order Hartree (a) and Fock (b) interaction corrections to the spin 
conductivity in classes CI and AIII,
and examples of second-order corrections [(c), (d)] for class AIII. 
The dashed lines indicate the spin current operators. 
The solid lines represent the exact noninteracting 
Matsubara Green's functions in an arbitrary, fixed realization of quenched surface disorder. 
The wavy lines correspond to the interaction potentials. Both current operators and interaction potentials are local in space.
Panels (c) and (d) depict categories of second-order corrections that are subleading in the inverse winding number.
Each correction in (a) and (b) vanishes individually, while those in each category (c) and (d) sum to zero.
In Sec.~\ref{2nd-sec}, we show that a similar cancellation occurs for all second-order corrections.
We sketch a proof that the same mechanism works to all orders in Sec.~\ref{high-sec}, implying that
Altshuler-Aronov corrections do not exist for Majorana surface transport coefficients.}
\label{LET-HF-diagrams}
\end{figure}

For classes CI and AIII, the surface spin conductivity $\s_{xx}^\rms$ obtains via the Kubo formula.
For a fixed realization of disorder, the leading order interaction (Hartree-Fock) corrections are represented 
by the Feynman diagrams in 
Figs.~\ref{LET-HF-diagrams}(a)~and~\ref{LET-HF-diagrams}(b).
In Sec.~\ref{Kubo-sec}, we show
that short-ranged interactions do not contribute to 
$\s_{xx}^\rms$ 
in the Hartree-Fock approximation, at zero temperature. 
The Hartree terms a(i) and a(ii) and the Fock terms b(i) and b(ii) vanish individually 
due to the chirality of the Green's functions, Eq.~(\ref{LET-chiral-gf}). 
The terms a(iii) and b(iii) do not contribute to 
$\s_{xx}^\rms$
due to the Ward identity.
The absence of Hartree-Fock corrections 
is distinct from the 
Fermi liquid case \cite{AA1985, AAG99, ZNA2001}. 

We compute all second-order corrections for class AIII, and find that these vanish as well.
These calculations are detailed in Sec.~\ref{2nd-sec}.
The main idea is that $\s_{xx}^\rms$ corrections can be grouped into classes, with each class corresponding 
to a particular ``free energy'' bubble. Examples of two such classes are the second-order groups (c) and (d) shown in Fig.~\ref{LET-HF-diagrams}.
The corrections in each class sum over all possible ways of inserting two current operators into the bubble,
but this sum vanishes. This is because AA corrections at $T = 0$ only involve Green's functions
at zero energy, so that retarded and advanced versions are equivalent [Eq.~(\ref{LET-chiral-gf})]. 
The Ward identity then implies that a sum over diagrams reduces to a sum over the relative positions
of interaction vertices in a free energy bubble, and this is equal to zero.
We sketch a
proof that all higher-order corrections vanish via the same mechanism in 
Sec.~\ref{high-sec}.

\subsection{Spin and thermal conductivities, large winding number expansion}

We also compute interaction corrections in a large winding number expansion,
employing the WZNW-FNLsMs 
\cite{FXCup}
for the interacting surface Dirac fermions described by Eqs.~(\ref{LET-ham-nd}) and (\ref{LET-action-int}).
These low-energy effective field theories 
are derived
directly from the non-Abelian bosonization of the Dirac
fermions, without recourse to the self-consistent Born approximation or a gradient expansion. 
In each class, the model contains a parameter $\lambda$ that is proportional to the dimensionless
spin (thermal) resistance in classes CI and AIII (DIII). The universal transport coefficients in Eqs.~(\ref{LET-spin-cond}) and
(\ref{LET-heat-cond}) obtain for the noninteracting models tuned to a conformal fixed point such that $\lambda = 1/K$,
where $K = |\nu|$ ($K = |\nu|/2$) in classes AIII and DIII (CI).
The models are defined explicitly in Sec.~\ref{Fnlsm-sec}, Eqs.~(\ref{act-sept})--(\ref{o-const}).

Perturbing the sigma models around the noninteracting conformal fixed point, 
we 
derive
the following one-loop RG equations for $\l$ 
in Sec.~\ref{Fnlsm-sec}:
\begin{subequations}  \label{LET-RG-eq-int-l}
\begin{align}
\text{CI:} \quad    d \lambda / d l = & \, \l^2 \left[1-(K\l)^2\right] [1+ \mathcal{J}(\gms,\gmc)],   
	\label{LET-RG-eq-int-l-c1} \\
\text{AIII:} \quad  d \lambda / d l = & \, \l^2 \left[1-(K\l)^2\right] \mti(\gms,\gmc),      
	\label{LET-RG-eq-int-l-a3}\\
\text{DIII:} \quad  d \lambda / d l = & \,-\l^2 \left[1-(K\l)^2\right][2 + \mathcal{K}\left(\gmc\right)].    
	\label{LET-RG-eq-int-d3}
\end{align}
\end{subequations} 
In Eq.~(\ref{LET-RG-eq-int-l}), $\gms$ and $\gmc$ are rescaled versions of the interaction strengths
that appear in Eq.~(\ref{LET-action-int}),
$\gamma_{\rms,\rmc} = 4 \, \Gamma_{\rms,\rmc} / \pi h,$
where $h$ is a sigma model parameter  
that couples to frequency 
[Eqs.~(\ref{action-normal}) and (\ref{fnlsm-aiii-c})].
The functions $\mathcal{J}$, $\mti$, and $\mathcal{K}$ are defined as
\begin{subequations}  \label{LET-I-J-functions}
\begin{align}
	\!\!\mathcal{J}(\gms,\gmc) & = \textstyle{3 \left[ 1 + \frac{1-\gms}{\gms} \ln{(1-\gms)}  \right]- \frac{1}{4} \mathcal{K}(\gmc)}, 
	\\
	\!\!\mti(\gms,\gmc) & =  \textstyle{2 \left[ 1 + \frac{1-\gms}{\gms} \ln{(1-\gms)}  \right]- \frac{1}{2} \mathcal{K}(\gmc)}, 
	\\
	\!\!\mathcal{K}(\gmc) & = \textstyle{2 e^{-1 / \gmc}\left[E_i\left(\frac{1}{\gmc}+ \ln 2\right) - E_i\left(\frac{1}{\gmc}\right) \right]}\!,
\end{align}
\end{subequations}
and represent the AA corrections \cite{Finkel1983,dellanna2006,AA1985,BK1994}.
Here $E_i(z)$ denotes the exponential integral function.
For class AIII, there is an additional equation for the 
parameter $\l_A$ [see Eqs.~(\ref{fnlsm-aiii-c}) and (\ref{ab-dis})],
which governs the strength of the Abelian random potential ${\bm{\mathcal{A}}}(\bfr)$ in Eq.~(\ref{LET-ham-nd}):
\be  \label{LET-eq-la-aiii}
	d \la / d l = \, \textstyle{\l^2 \left[ 1 - (K\l)^2 \right] \left[ 1 + \frac{2 \la}{\l} \mti(\gms,\gmc)\right]}. 
\ee
Equations~(\ref{LET-RG-eq-int-l}) and (\ref{LET-eq-la-aiii}) incorporate interaction effects to all orders
in $\gms$ and $\gmc$, but are valid only to the lowest order in $1/K$. The WZNW-FNLsM is controlled in the limit of 
large winding numbers ($K \gg 1$). 
Simplified versions of Eqs.~(\ref{LET-RG-eq-int-l})--(\ref{LET-eq-la-aiii}) computed to linear order in $\gamma_c$
were stated without proof in \cite{FXCup}.
  
Equations~(\ref{LET-RG-eq-int-l}) and (\ref{LET-eq-la-aiii}) imply that even in the presence of interactions, $\l = 1/K$ is a fixed point 
for TSCs in all classes. 
Although this is valid to the lowest order in $1/K$ or $\l$, it may possibly be exact
(as it is in the noninteracting case \cite{Witten84,FXCup}). 
By comparison, the vanishing of the 
interaction corrections for classes CI and AIII discussed above
is perturbative in the interactions, but exact to all orders in $1/K$.

\subsection{Discussion and directions for future work}

References~\cite{SML,MS2012} 
suggested that the 
surface spin or thermal response 
of a TSC 
induces 
a topological term in the effective field theory, 
though in the context of TRS breaking spin or thermal Hall effects. 
The topological terms
relate
to ``anomalies'' appearing in the theories describing the responses. 
These anomalies are 
believed 
to be insensitive to whether the underlying fermions are interacting or not. 
An important question is whether the $(2+0)$-dimensional 
axial anomaly invoked in the noninteracting case \cite{mirlin2006} can be 
generalized to $2+1$ dimensions to argue for the universality of Eqs.~(\ref{LET-spin-cond}) 
and (\ref{LET-heat-cond}).

Next we address a few caveats and potential complications.
First we note that even without interactions, Eqs.~(\ref{LET-spin-cond}) and 
(\ref{LET-heat-cond}) neglect the influence of 
strongly
irrelevant operators \cite{Nakai2014,FXCup},
but these should be negligible at sufficiently low temperatures. 
Second, in the absence of interactions, the finite-energy states in class DIII (CI) are believed
to be delocalized (localized) by weak disorder \cite{wm2008}. In class AIII,
for a single color the finite-energy states remain delocalized \cite{ludwig1994,OGM2007,NRKMF2008,CFup}.
The fate of such states for $|\nu| \geq 2$ in class AIII remains unanswered to our knowledge. 
Without interactions, transport coefficients vanish at nonzero temperature
if delocalization is confined to a single state, as in the plateau transition
of the quantum Hall effect \cite{IQHP2000}. Thus Eqs.~(\ref{LET-spin-cond}) and (\ref{LET-heat-cond})
would (may) not apply to class CI (AIII) at $T > 0$ without interactions.
In reality, Eqs.~(\ref{LET-spin-cond}) and (\ref{LET-heat-cond}) should hold
to the leading approximation for sufficiently low $T$, with temperature-dependent
corrections determined by inelastic scattering \cite{IQHP2000}.

While transport is unaffected so long as TRS 
and the bulk gap are
preserved, 
disorder afflicts the Majorana surface physics in other ways. In particular, dirty TSCs 
with $|\nu| > 1$ possess 
surface state wave functions that are delocalized, 
yet strongly inhomogeneous. 
These are characterized by
universal multifractal statistics \cite{ludwig1994,MCW96,CKT1996,Foster2012,FXCup,CFup}.
Although the color and spin densities are everywhere equal to zero, the local density of
states will reflect this inhomogeneity and could be measured by STM. 
Wave function multifractality can strongly enhance interaction effects \cite{Foster2012,FXCup}.
In fact, arbitrarily weak interactions always destabilize class CI surface states by inducing 
spontaneous TRS breaking \cite{Foster2012}; for weak interactions or disorder, this will
occur at very low temperatures.  
By contrast, class AIII and DIII surface states can survive to zero temperature \cite{FXCup}. 
The absence of quantum conductance corrections implies that the transition to an insulating state
due to interactions will be of Mott type, i.e., first order at zero temperature. 
A Mott transition to a state with surface topological order is also possible with strong interactions 
\cite{Fidkowski13,Wang13,Wang14,Metlitski14}. 
It would be interesting to investigate the latter scenario within the WZNW-FNLsM, which can be formulated 
even for the clean system (at level one).

An interesting question is whether surface transport remains quantized when the bulk
gap is closed, as occurs at the ``plateau transition'' between different $\nu$. 
A pair of surface states will typically delocalize into the bulk at such a transition 
and annihilate. The transport coefficients characterizing the remaining surface states
will remain quantized if the bulk-surface coupling is neglected, due to the chiral TRS. 
With nonzero coupling, the situation is less clear because the bulk can mediate effective 
long-ranged interactions at the surface. Another open question regards the surface transport
for superconductors with protected nodal lines in the bulk \cite{Sato2006,Beri2010,SchnyderRyu,ZhaoWang13}.

Our main conclusion is that bulk TSCs generalize the key aspect of the quantum Hall effect,
which is topologically quantized transport coefficients due to protected gapless surface states. 
Perhaps the most interesting open question is whether 3D bulk phases with topological order can support
exotic, \emph{gapless} surface states with fractionally quantized 
surface spin or thermal conductivities. That is, is there a topological superconductor analog of the 
fractional quantum Hall effect in 3D?

\section{Lattice model for class AIII \label{micro-sec}}

In this section we present a toy model on the diamond lattice for a bulk class AIII TSC. 
This is a modified version of the class CI model in Ref.~\cite{SUP-SRL2009}. We show
how the Majorana theory in Eq.~(\ref{LET-ham-nd}) emerges at the surface, with time-reversal
symmetry encoded as in Eq.~(\ref{LET-t-rev}).

\begin{figure}
\centering
\includegraphics[width=0.2\textwidth]{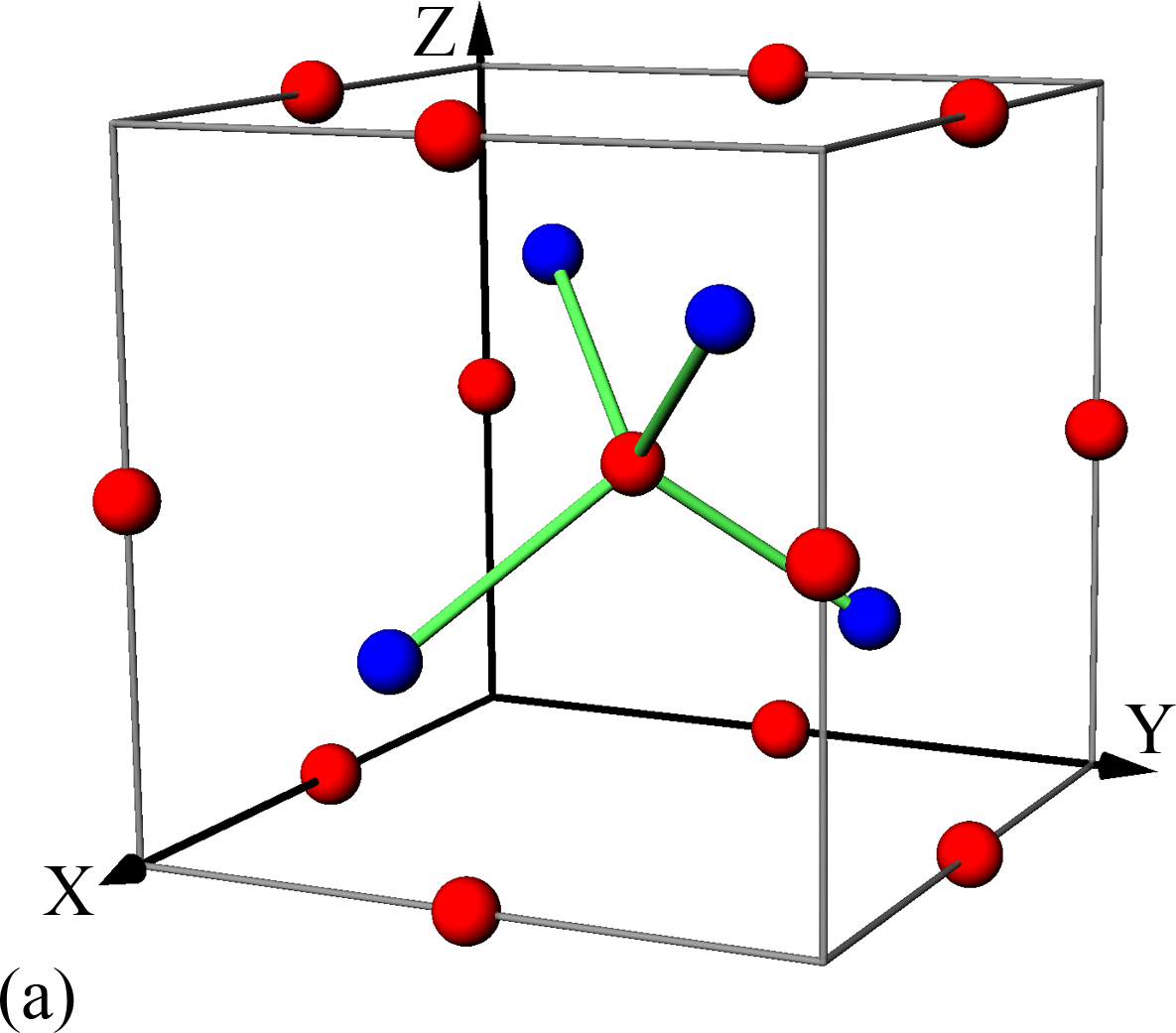}
\includegraphics[width=0.2\textwidth]{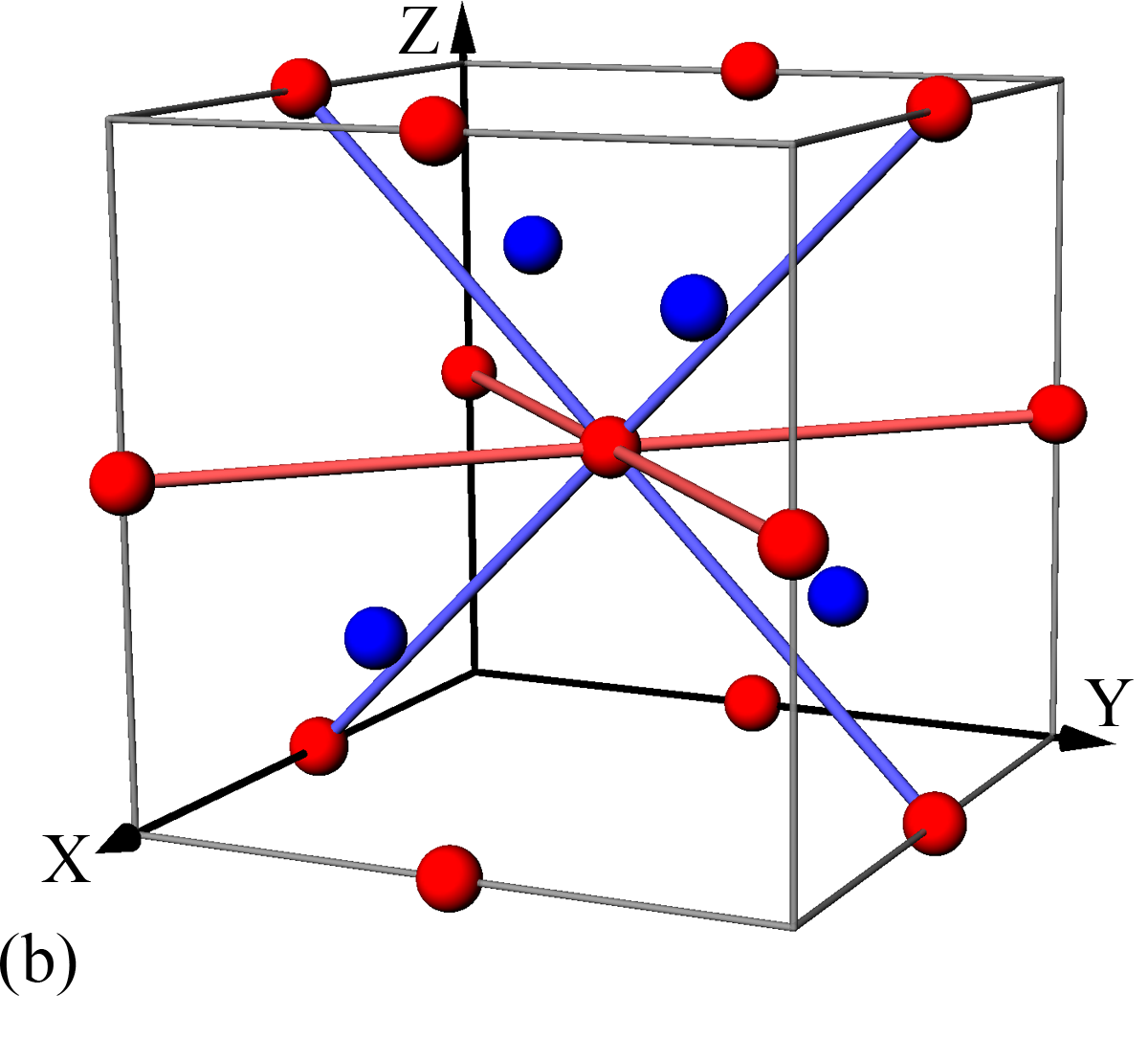}\\
\includegraphics[width=0.2\textwidth]{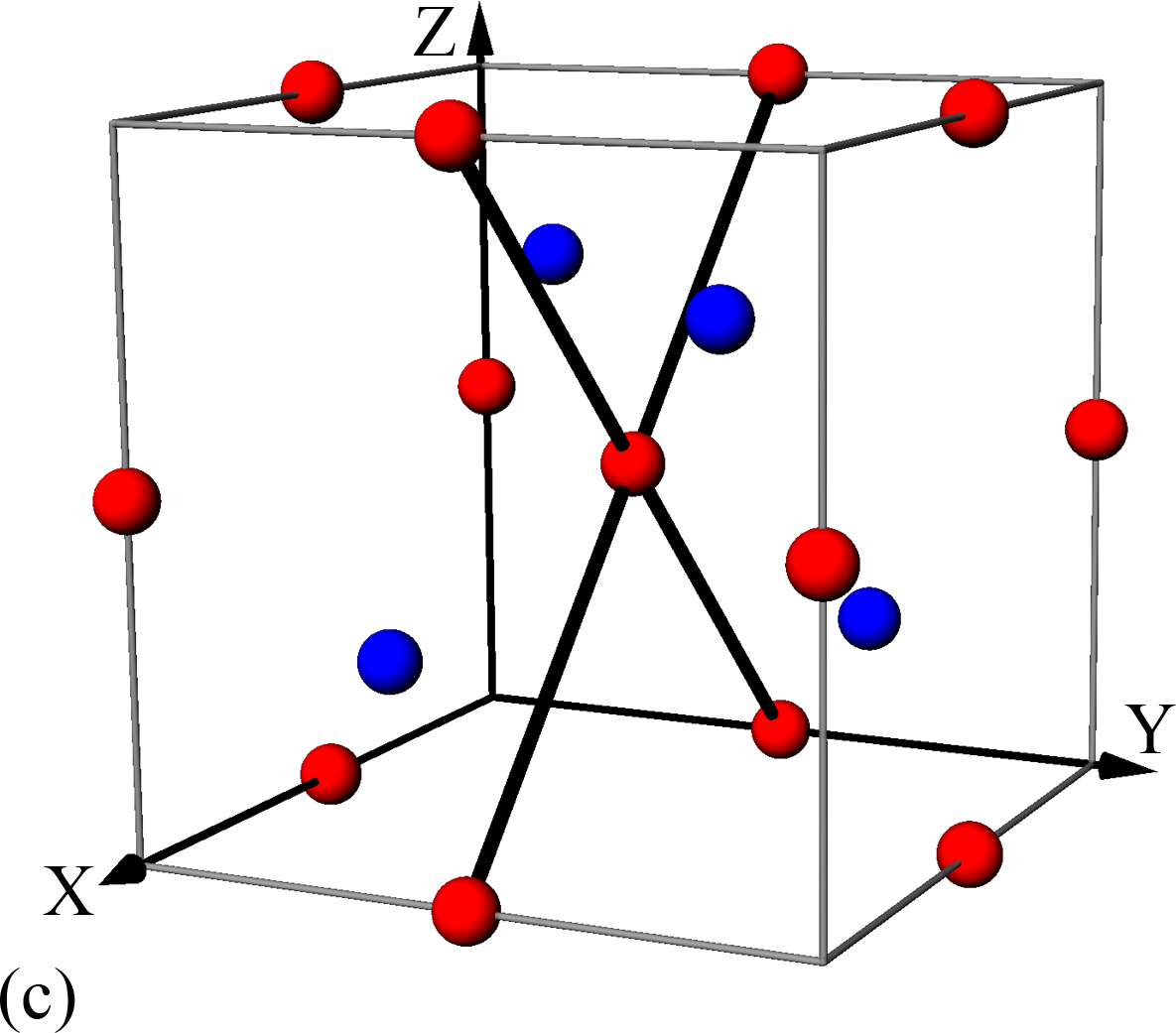}
\includegraphics[width=0.2\textwidth]{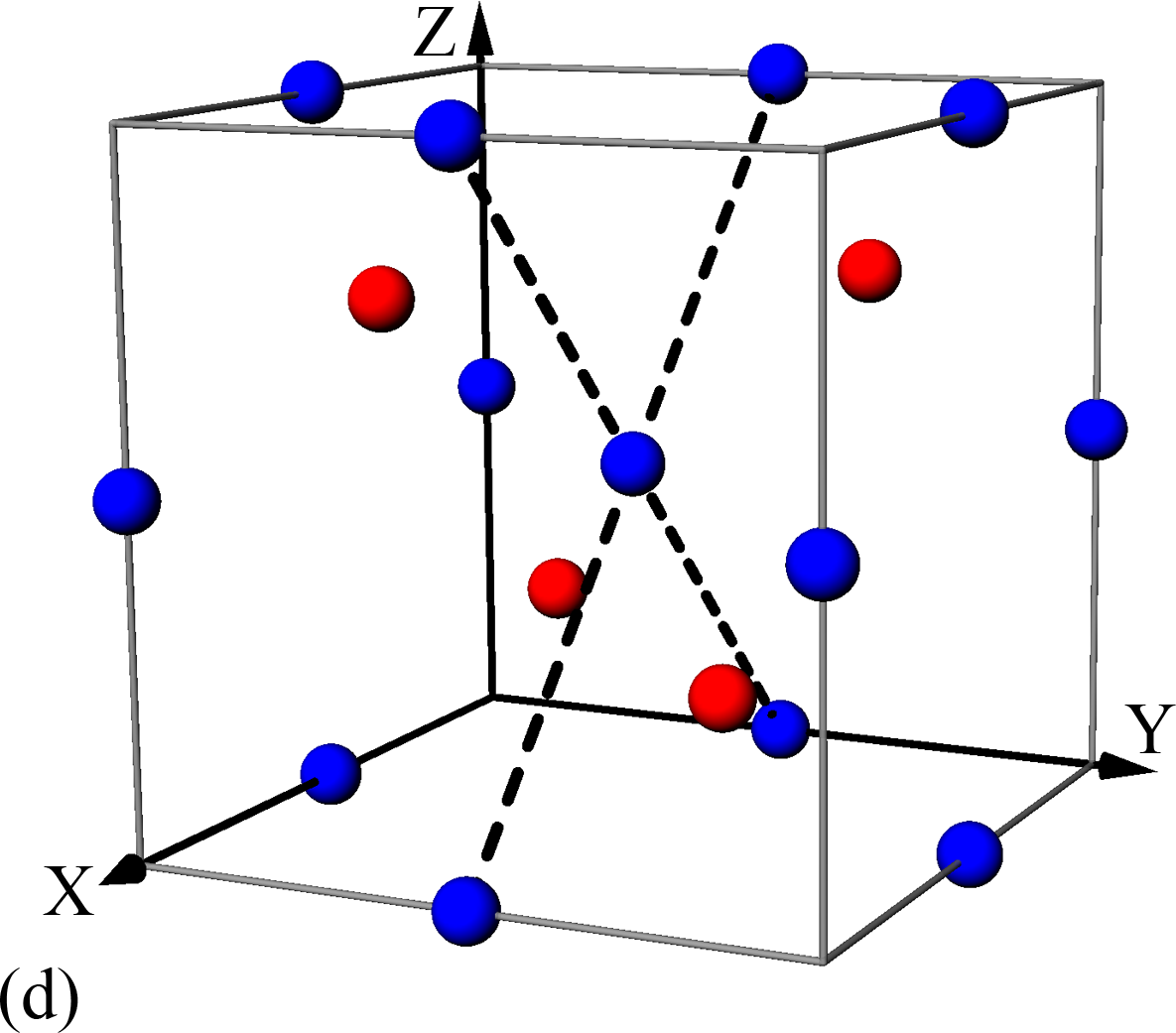}\\
\caption{Class AIII
model on the diamond lattice.
Sublattice A (B) sites are indicated by red (blue) spheres, and 
are subject to the potential $\mu_s$ ($-\mu_s$).
(a) Nearest-neighbor hopping with amplitude $t^\prime$. 
(b) BCS pairing within each of the sublattices. The red lines indicate 
$d$-wave spin-singlet pairing in the $xy$ plane. 
The blue lines indicate
$p$-wave $z$-axial spin-triplet pairing in the 
$yz$ plane. 
(c) and (d) Next-nearest-neighbor hopping 
in the 
$xz$ plane. 
The solid and dashed lines mean that the hopping amplitudes are 
$+t$ and $-t$,
respectively.} \label{diamond-lattice}
\end{figure}

The diamond lattice is composed of two face-centered cubic sublattices, 
which we denote as A and B alternately (see Fig.~\ref{diamond-lattice}). 
Each site is surrounded by four nearest-neighbor sites and twelve next-nearest-neighbor sites.  
We choose a set of the primitive vectors of the Bravais lattice as \cite{SUP-AMbook} 
\be 
	\mathbf{a}_1 = {\textstyle{\frac{1}{2}}}(\hat{\mathbf{y}}+\hat{\mathbf{z}}), 
	\quad 
	\mathbf{a}_2 = {\textstyle{\frac{1}{2}}}(\hat{\mathbf{x}}+\hat{\mathbf{z}}), 
	\quad 
	\mathbf{a}_3 = {\textstyle{\frac{1}{2}}}(\hat{\mathbf{x}}+\hat{\mathbf{y}}), 
\ee 
where we have assumed that the lattice constant is unity. 
The sites of the Bravais lattice are 
\be 
	\bfR = n_1 \mathbf{a}_1 + n_2 \mathbf{a}_2 + n_3 \mathbf{a}_3, 
\ee 
with $n_{1,2,3}$ integers. Moreover, the set of vectors pointing from a site on sublattice A to its nearest neighbors on sublattices B are  
\be  \label{nn-vecs}
\begin{split}
	b_\text{nn} = & \left\{ (\mathbf{a}_1+\mathbf{a}_2 + \mathbf{a}_3)/4, \, (-3\mathbf{a}_1+\mathbf{a}_2 + \mathbf{a}_3)/4, \right. \\ 
		      &  \left (\mathbf{a}_1-3\mathbf{a}_2 + \mathbf{a}_3)/4, \, (\mathbf{a}_1+\mathbf{a}_2 - 3 \mathbf{a}_3)/4   \right\}.
\end{split}
\ee
The set of vectors pointing from one site to its next-nearest neighbors are
\be  \label{nnn-vecs}
\begin{split}
	b_\text{nnn} = & \left\{ \pm\mathbf{a}_1, \, \pm\mathbf{a}_2, \, \pm\mathbf{a}_3, \right. \\
			& \left. \pm(\mathbf{a}_1-\mathbf{a}_2),\, \pm(\mathbf{a}_1-\mathbf{a}_3), \, \pm(\mathbf{a}_2-\mathbf{a}_3) \right\}.
\end{split}
\ee

\subsection{Topological superconductors on the diamond lattice with spin U(1) symmetry}  \label{AIII-model}

The Hamiltonian of the topological superconductor model on the diamond lattice consists of three parts,
\be  \label{ham-diamond}
	H =  H_{\text{nn}} + H_\text{pair} + H_{\text{nnn}},
\ee
which are defined as the follows. 

First, $H_\text{nn}$ describes isotropic nearest-neighbor hopping [see Fig.~\ref{diamond-lattice}(a)],      
\be  \label{diamd-soc}
	H_{\text{nn}} = t^\prime \, \sum_{\bfR} \sum_{\mu = \uparrow, \downarrow } \sum_{\bfdel \in b_{\text{nn}}} \left[ C_{\text{A} \mu}^\dagger(\bfR) \, C_{\text{B} \mu}(\bfR+\bfdel) + \text{H.c.} \right],
\ee
where $C_{\text{i} \mu}^\dagger(\bfR)$ [$C_{\text{i} \mu}(\bfR)$] is the creation (annihilation) operator of a spin $\mu \in \{\uparrow,\downarrow\}$
electron at site $\bfR$ on the sublattice 
$\text{i} \in \{\text{A, B}\}$, and
$t'$ 
is the hopping strength.

Second, $H_\text{pair}$ describes the BCS pairing potentials of electrons within each of the sublattices [see Fig.~\ref{diamond-lattice}(b)],  
\be  \label{diamd-pair}
\begin{split}
	H_\text{pair} = & \frac{1}{2} \sum_{\text{i} = \text{A}, \text{B}} \sum_{\bfR} \sum_{ \bfdel \in b_{\text{nnn}}}  \sum_{\mu,\nu = \uparrow,\downarrow} \\
			& \times \left[ \Delta_{\mu \nu}(\bfdel) \, C_{\text{i} \mu}^\dagger(\bfR) \, C_{\text{i} \nu }^\dagger(\bfR+\bfdel) + \text{H.c.} \right].
\end{split} 
\ee
Here the anisotropic pairing amplitudes are encoded in the $2 \times 2$ matrices
\begin{equation}  \label{real-pairing}
	\hat{\Delta}(\bfdel) 
	= 
	\begin{cases}  
		i \, \sqrt{3} 	  \, \Delta\, \hmu^2 \, \, (\text{d-wave spin-singlet}), & \bfdel \perp \hat{\mathbf{z}}, \\ 
		- i \sgn(\delta_y) \, \Delta\, \hmu^1 \, \, (\text{p-wave spin-triplet}), & \bfdel \perp \hat{\mathbf{x}}, \\
		0, & \bfdel \perp \hat{\mathbf{y}}.
	\end{cases}
\end{equation}
where $\Delta$ is real and $\hmu^{\a=1,2,3}$ are the set of Pauli matrices acting on the physical spin space. 
We 
choose anisotropic 
pairing 
potentials in space for the sake of engineering more topologically nontrivial phases. 

The last term $H_\text{nnn}$ includes a staggered on-site chemical potential and 
next-nearest-neighbor hopping of electrons [see Figs.~\ref{diamond-lattice}(c) and~\ref{diamond-lattice}(d)],  
\be  \label{diamd-hop}
\begin{split}
	H_\text{nnn} = & \sum_{\text{i} = \text{A}, \text{B}} \sum_{\bfR} \sum_{\mu =\uparrow, \downarrow} V_{\text{i},s} \, C_{\text{i}\mu}^\dagger(\bfR) \, C_{\text{i}\mu}(\bfR) \\
	& + \sum_{\text{i} = \text{A}, \text{B}} \sum_{\bfR} \sum_{ \bfdel \in  b_{\text{nnn}}} \sum_{\mu =\uparrow, \downarrow} \\ 
	& \phantom{++} \times \left[ t_{\text{i}}(\bfdel) \, C_{\text{i}  \mu}^\dagger(\bfR) \, C_{\text{i} \mu}( \bfR+\bfdel) + \text{H.c.}\right]
\end{split} 
\ee
where $V_{\text{i},s} = \mu_s$ ($- \mu_s$) for $\text{i} \in \text{A}$ ($\text{i} \in \text{B}$), and 
\be
	t_{\text{i}}(\bfdel) 
	= 
	\begin{cases} 
		t, & \text{i} \in \text{A}, \bfdel \perp \hat{\mathbf{y}} \\ 
		-t, &  \text{i} \in \text{B}, \, \bfdel \perp \hat{\mathbf{y}} \\
		0, &  \mbox{otherwise},
	\end{cases}
\ee 
with $\mu_s$ and $t$ real.

\begin{figure}
\centering
\includegraphics[width=0.27\textwidth]{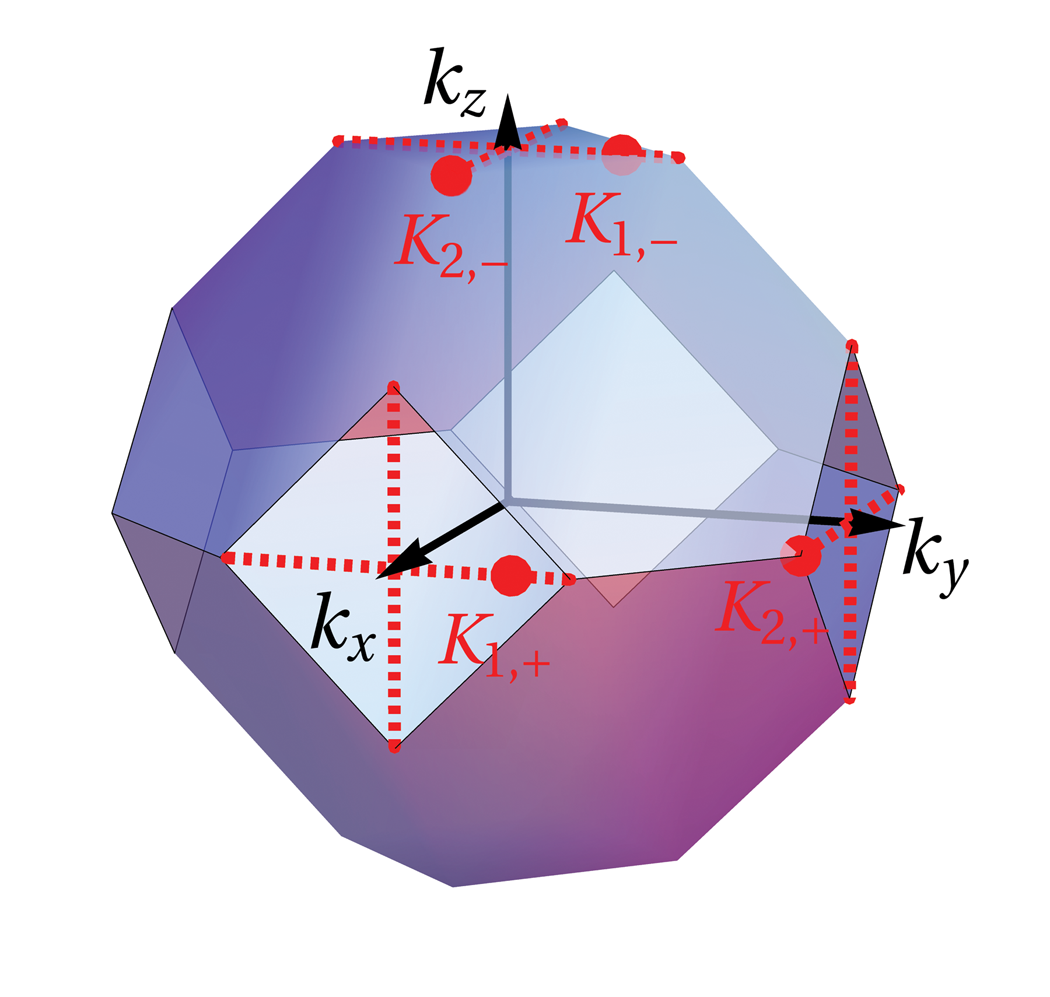}
\caption{
Diagram showing the emergence of the Dirac points of the Hamiltonian (\ref{ham-k-up}) in the first Brillouin 
zone of the diamond lattice, obtained from Eq.~(\ref{bands}). 
The red dashed lines indicate the Fermi surface 
at half filling when $t = \mu_s = \Delta = 0$. 
Nonzero $\Delta$ gaps 
most of the 
the Fermi surface
except for the 
massless 
Dirac nodes indicated by the red spots [Eq.~(\ref{aiii-nodes})]. 
Mass gaps open
at those Dirac nodes 
for nonzero $t$ or $\mu_s$, 
so that topologically trivial or nontrivial superconductors can be 
obtained depending on the values of 
these. 
See Fig.~\ref{fig-w-num}.}   
\label{Bzone}
\end{figure}

In reciprocal space the Hamiltonian takes the form
\be  \label{diamd-ham-k}
	H =
	\intl{\text{B.Z.}} \! \frac{\rmd^3 \bk}{(2 \pi)^3}
	\psi^\dagger(\bk) \, \calH(\bk) \, \psi(\bk),
\ee
where 
\begin{align}
	\psi(\bk) 
	=&\, 
	\begin{bmatrix} 
	C_{\text{A} \uparrow} (\bk) & 
	C_{\text{B} \uparrow} (\bk) & 
	C_{\text{A} \downarrow}^\dagger (-\bk) & 
	C_{\text{B} \downarrow}^\dagger (-\bk) 
	\end{bmatrix}^\trasp\!\!. 
	\label{basis-spinor} 
\end{align}
The model is constructed to preserve time-reversal and spin U(1) symmetries. 
Time-reversal symmetry appears as the antiunitary transformation
\begin{align}
	\begin{gathered}
	C_{\text{i} \uparrow}(\bk) \mapsto - C_{\text{i} \downarrow}(-\bk),
	\quad
	C_{\text{i} \downarrow}(\bk) \mapsto C_{\text{i} \uparrow}(-\bk),
	\\
	\psi(\bk) \mapsto i \,\hsig^2 \, \left[\psi^\dagger(\bk)\right]^\trasp,
	\end{gathered}
\end{align}
where $\hsig^{1,2,3}$ is the set of Pauli matrices acting on the particle-hole space.
This imposes a chiral condition on the Hamiltonian,
\be  \label{SLS}
	- \hsig^2 \, \calH(\bk) \, \hsig^2 = \calH(\bk).
\ee 
The conserved U(1) charge of Eq.~(\ref{diamd-ham-k}) is the spin projection along the $z$ axis. 

Following the analogous construction for class CI \cite{SUP-SRL2009},
the Hamiltonian 
in Eq.~(\ref{diamd-ham-k}) is defined as 
\be
	\begin{split}
	\calH(\bk) = & \, \Theta(\bk) \, \hsig^3 \ots \htau^3 + \Delta(\bk) \,  \hsig^1  \\ 
                  & \, + \hsig^3 \ots \left[ \P_\text{R}(\bk) \, \htau^1 + \P_\text{I}(\bk) \, \htau^2 \right], 
	\label{ham-k-up}
	\end{split}
\ee
where the Pauli matrices $\htau^{1,2,3}$ act on the sublattice space. 
The potential functions in Eq.~(\ref{ham-k-up}) are defined as follows. 
(i) Nearest-neighbor hopping of electrons
[Eq.~(\ref{diamd-soc})] leads to
\begin{align}  
\begin{aligned}\label{PHI-ri}
	\P_\text{R}(\bk) = & 4 t^\prime \cos\left(\fr{k_x}{4}\right) \cos\left(\fr{k_y}{4}\right) \cos\left(\fr{k_z}{4}\right),  \\	
	\P_\text{I}(\bk) = & 4 t^\prime \sin\left(\fr{k_x}{4}\right) \sin\left(\fr{k_y}{4}\right) \sin\left(\fr{k_z}{4}\right). 	
\end{aligned}
\end{align} 
We consider the 
case of half-filling, 
where 
the Fermi surface 
is 
formed by a set of one-dimensional filaments on the diamond faces of the first Brillouin zone.
These are depicted as 
red dashed lines in Fig.~\ref{Bzone}.
(ii) BCS pairing [Eq.~(\ref{diamd-pair})] yields
\begin{align}
	\label{bcs-pot-k}
	&\,
	\Delta(\bk) 
	= 
	\Delta_\text{d}(\bk) + \Delta_\text{p}(\bk),
	\\
&\,
\label{bcs-pot-dp}
\begin{aligned}
	\Delta_\text{d}(\bk) 
	=&\, 
	4  \sqrt{3} \, \Delta \, \cos{\left( \fr{k_x}{2} \right)} \cos{\left( \fr{k_y}{2} \right)}, 
	\\ 
	\Delta_\text{p}(\bk) 
	=&\, 
	4 \, \Delta \, \sin{\left( \fr{k_y}{2} \right)}\cos{\left( \fr{k_z}{2} \right)},	
\end{aligned}
\end{align}
being the Fourier transform of $\hat{\Delta}(\bfdel)$ in Eq.~(\ref{real-pairing}).
The p-wave pairing $\Delta_\text{p}(\bk)$ breaks the spin SU(2) symmetry down to the subgroup of U(1) rotations
about the $z$ axis, but preserves time-reversal symmetry (class AIII). 
These pairing potentials gap out most of 
the Fermi surface, 
leaving	
four 
isolated	
Dirac nodal points 
denoted by the 
red spots in Fig.~\ref{Bzone}. 
(iii) Staggered on-site chemical potential and next-nearest-neighbor hopping [Eq.~(\ref{diamd-hop})] give
\be  \label{nnn-hop-pot}
	\Theta(\bk) = 4 \, t \, \cos{\left( \fr{k_x}{2} \right)} \cos{\left( \fr{k_z}{2} \right)}+ \mu_s.
\ee
Nonzero $\mu_s$ or $t$ 
opens 
mass gaps at the Dirac nodes. The gapped 
phase can be a trivial or topological superconductor.	

The energy eigenvalues of $\calH(\bk)$ are 
\be  \label{bands}
	E_{\pm}(\bk)  = \pm \sqrt{\Theta^2(\bk) + \Delta^2(\bk) + \P_\text{R}^2(\bk)+\P_\text{I}^2(\bk)},
\ee
with 
a twofold degeneracy for each momentum $\bk$. 
For $t = \mu_s = 0$ and nonzero $t^\prime$ and $\Delta$, there are 
four 
massless 
Dirac nodes 
at 
\be  \label{aiii-nodes}
\begin{split}
	\bfK_{1,+} = & \, 2 \pi \,(1, 1/3, 0) , \quad \bfK_{1,-}= 2 \pi \,(0, 1/3, 1), \\
	\bfK_{2,+} = & \, 2 \pi \,(1/2, 1, 0),  \quad \bfK_{2,-}= 2 \pi \,(1/2, 0, 1),
\end{split}
\ee
as shown in 
Fig.~\ref{Bzone}. 
For nonzero $t$ and $\mu_s$, the Dirac masses at these nodes are 
\begin{align}\label{dirac-masses}
	\bfK_{1,\pm}\!:\; M_1 \equiv \mu_s - 4 t, 
	\;\;
	\bfK_{2,\pm}\!:\; M_2 \equiv \mu_s. 
\end{align}	
Thus gapless bulk quasiparticles survive for $\mu_s = 4 t$ and $\mu_s = 0$. 
When crossing one of 
these 
cut lines in the $t$-$\mu_s$ plane, 
the energy gap 
closes and reopens, potentially signaling
a topological phase transition.

\begin{figure}
\centering
\includegraphics[width=0.235\textwidth]{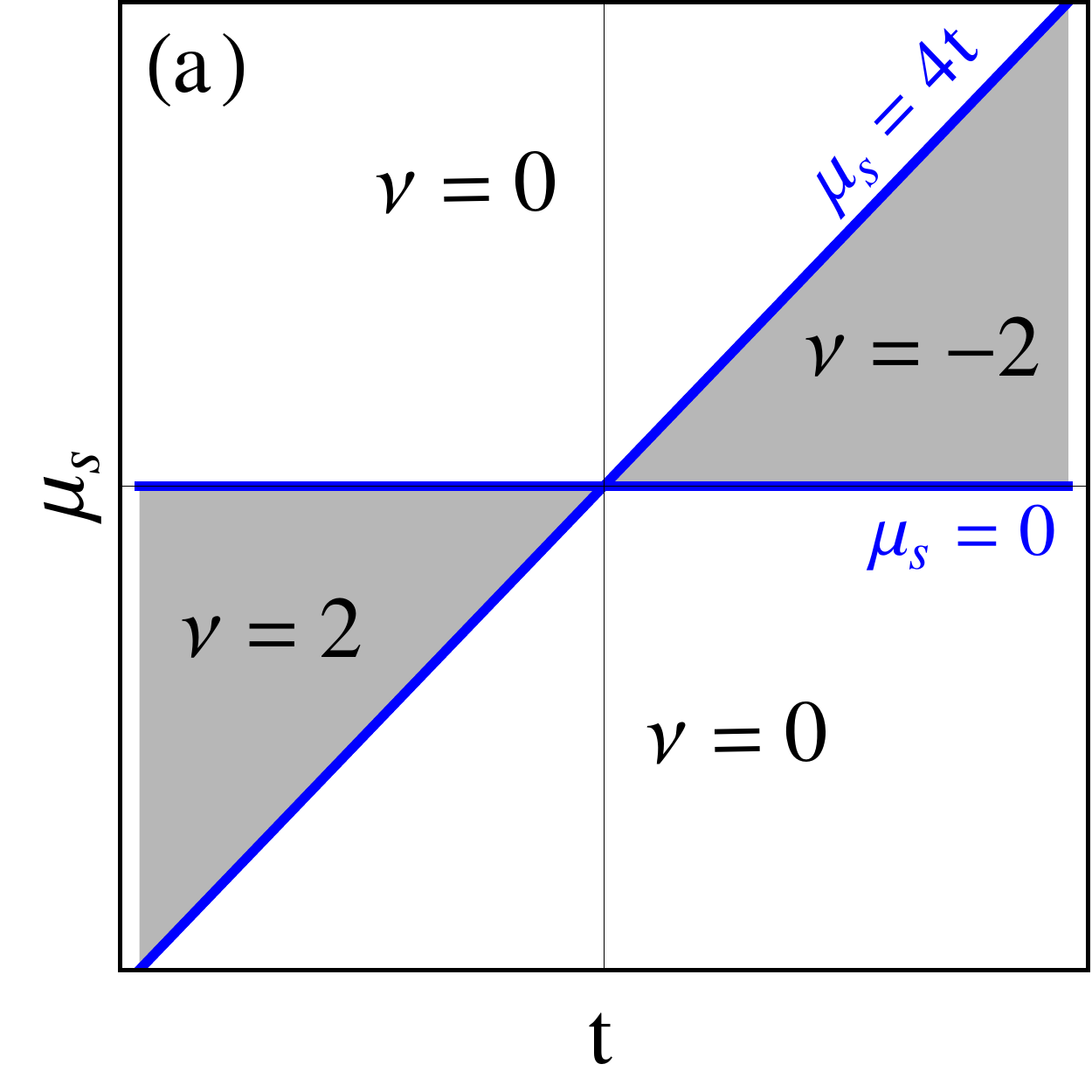}
\includegraphics[width=0.238\textwidth]{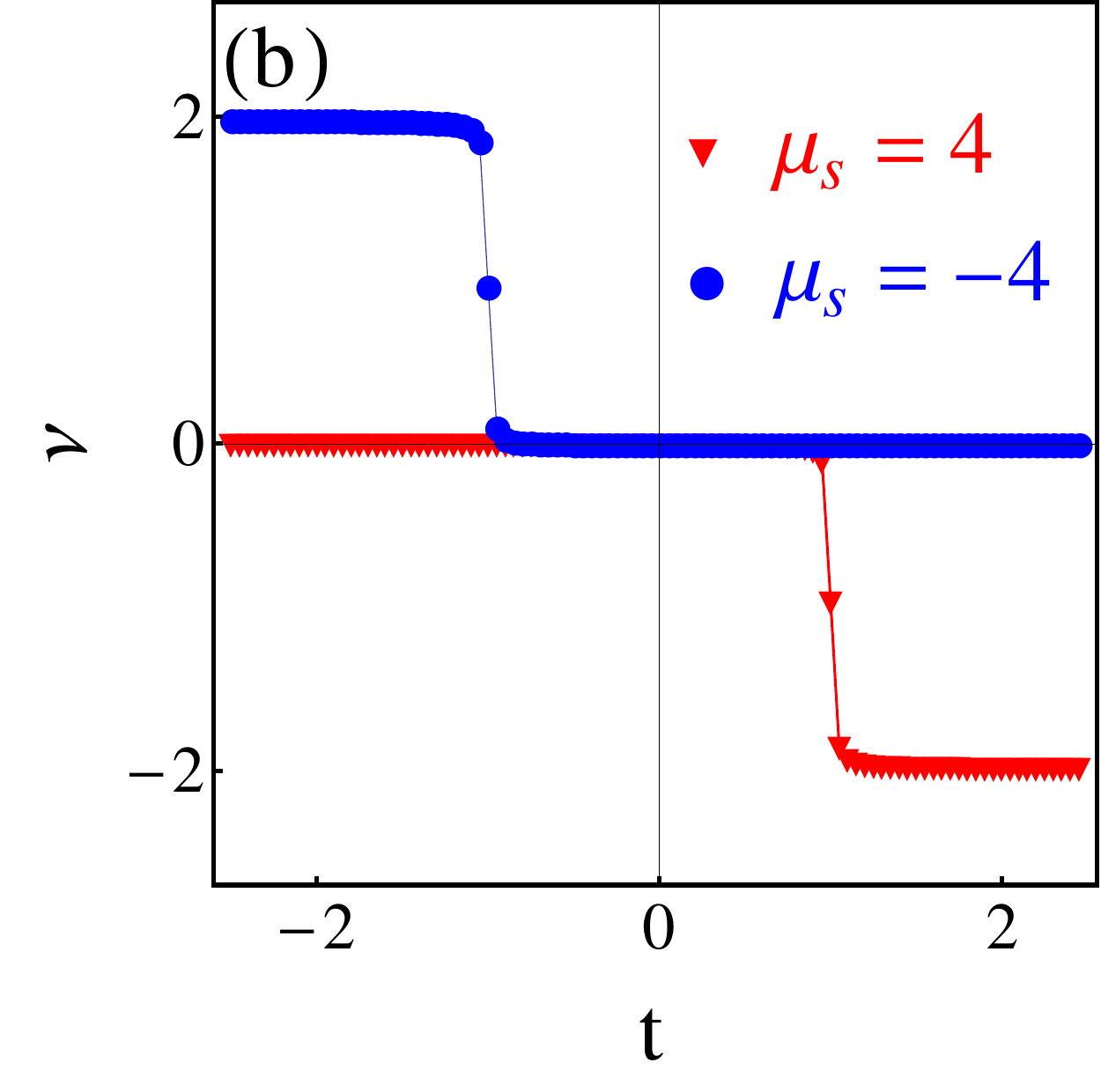}
\caption{Topological superconducting
phases of 
Eq.~(\ref{diamd-ham-k}) as a function of 
the
next-nearest-neighbor hopping strength $t$ and 
the
staggered chemical potential $\mu_s$. (a) 
Phase diagram. The blue lines indicate the phase boundaries where the bulk gap closes. 
The gray area indicates the topologically nontrivial phases with $|\nu|=2$, where the 
superconductor possesses two gapless surface bands. 
(b) Numerical results 
for 
the winding number [Eq.~(\ref{wind-num})], 
with 
$\Delta = 2$ and $t^\prime = 4$. 
} \label{fig-w-num}
\end{figure}

After a particle-hole space rotation 
such that $(\hsig^1, \, \hsig^2, \, \hsig^3) \to (\hsig^1, \, -\hsig^3, \, \hsig^2)$,
$\calH(\bk)$ takes the off-diagonal form 
\[  
	\calH(\bk) \to \begin{pmatrix}  0 & \hat{D}(\bk) \\ \hat{D}^\dagger(\bk) & 0  \end{pmatrix},
\]
\be  \label{Dmatrix}
	\hat{D}(\bk) =  \Delta(\bk) - i \, \P_\text{R}(\bk) \, \htau^1 - i \, \P_\text{I}(\bk) \, \htau^2 -i \, \, \Q(\bk) \, \htau^3.
\ee
One then introduces the unitary matrix 
\be
	\hat{q}(\bk) = - \hat{D}(\bk) /  E_+(\bk),	
\ee 
with $E_+(\bk)$ defined in Eq.~(\ref{bands}). 
The 
integer-valued	
winding number is  \cite{SRFL2008}
\be  \label{wind-num}
	\nu[\hat{q}] = \intl{\text{B.Z.}} \! \frac{\rmd^3 \bk}{24 \pi^2} \varepsilon^{\a \b \g} \Tr \left[ (\hat{q}^{-1} \d_\a \hat{q})(\hat{q}^{-1} \d_\b \hat{q})(\hat{q}^{-1} \d_\g \hat{q}) \right].
\ee
The phase diagram 
is shown in Fig.~\ref{fig-w-num}, where the gray 
regions indicate 
nontrivial phases with 
$\nu = \pm 2$ (two surface colors). 

\begin{figure}
\centering
\includegraphics[width=0.235\textwidth]{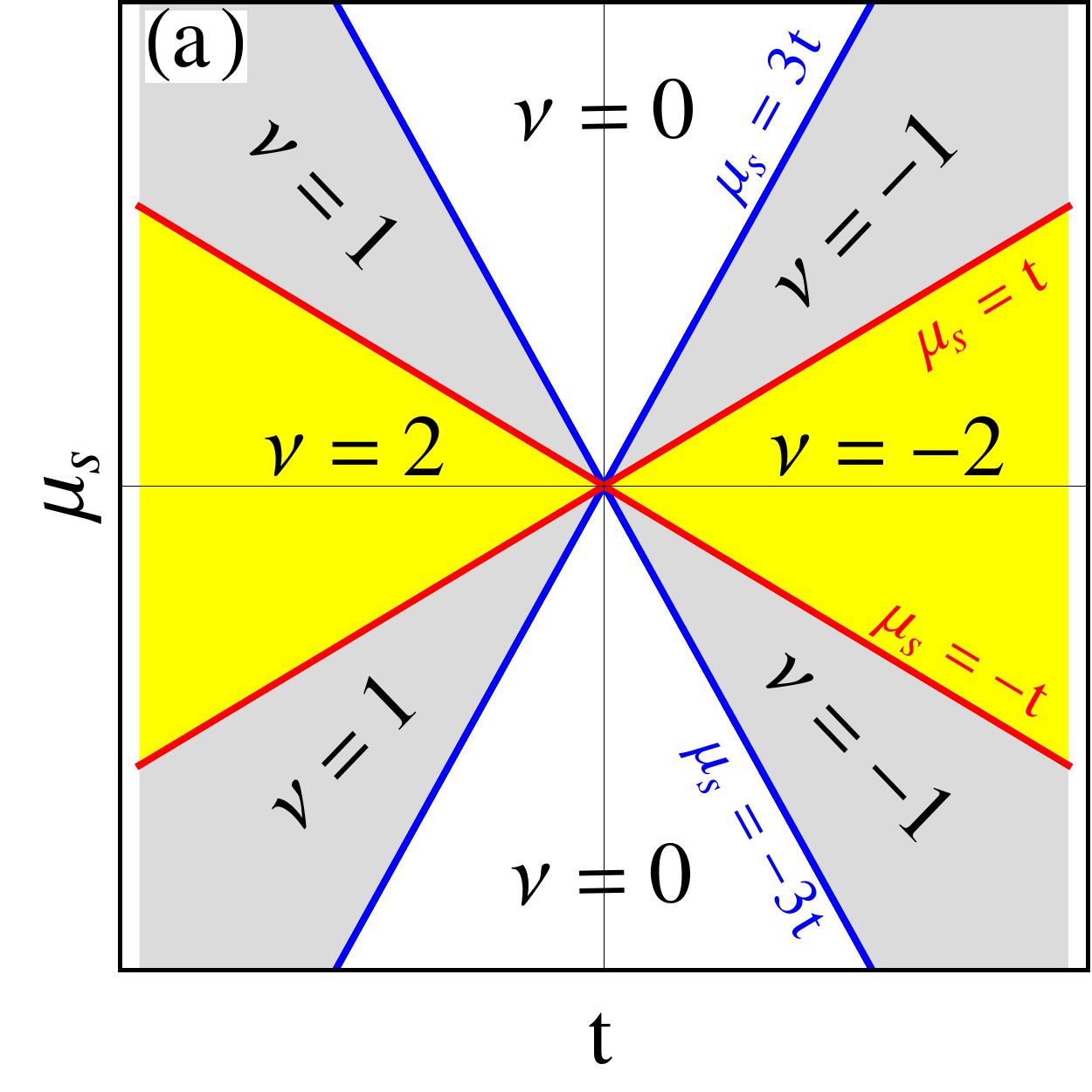}
\includegraphics[width=0.238\textwidth]{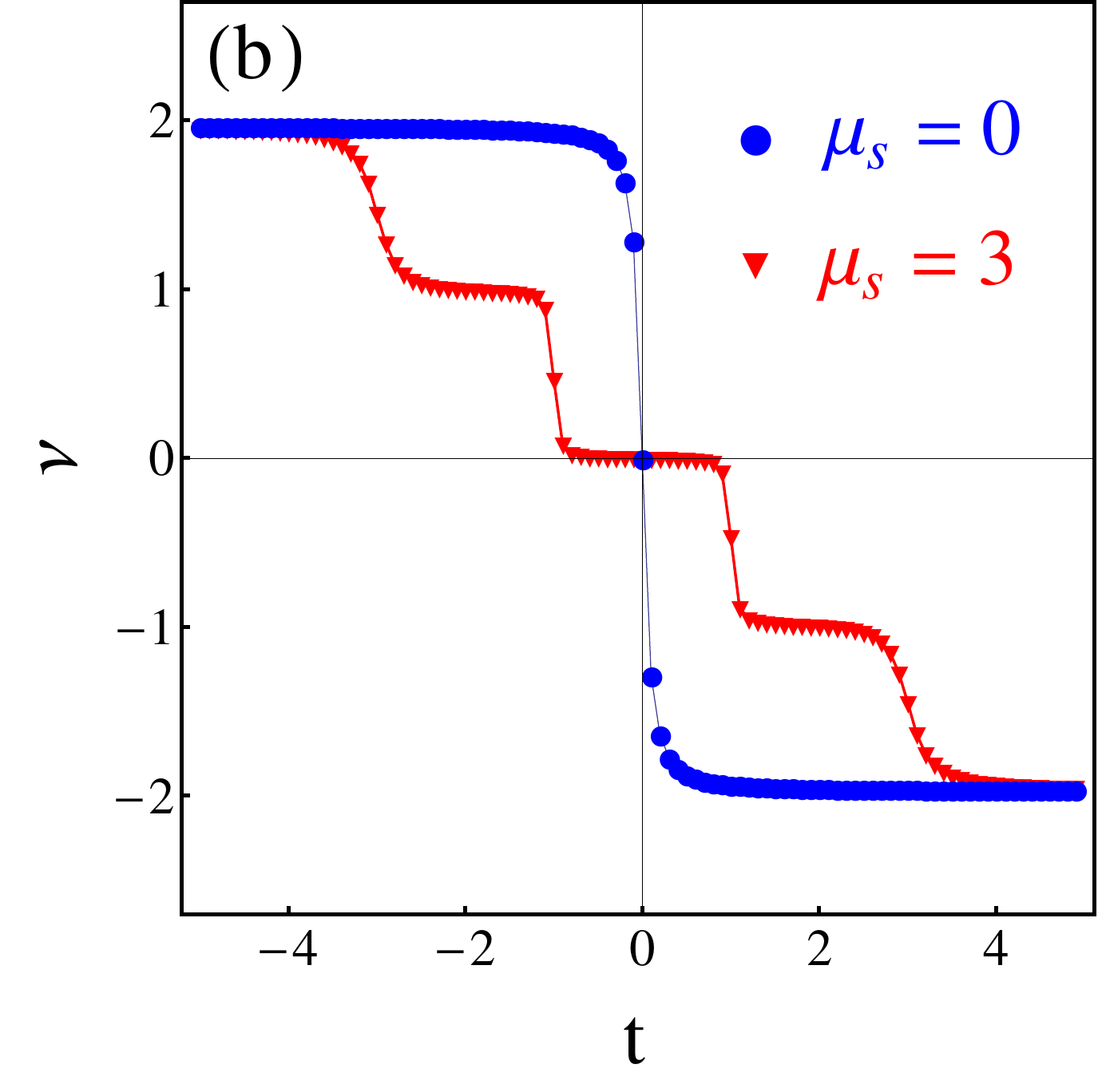}
\caption{Topological phases of the BCS Hamiltonian (\ref{diamd-ham-k}) with pairing potential in 
Eq.~(\ref{pairing-app}) and next-nearest-neighbor hopping in Eq.~(\ref{nnn-hop-app}). 
(a) Phase diagram. The blue and red lines 
indicate the phase boundaries where the bulk gap closes. 
The gray and yellow areas indicate the topological phase with $|\nu|=1$ (one surface color) 
and $|\nu|=2$ (two surface colors), respectively. 
(b) Numerical results of the winding number [Eq.~(\ref{wind-num})] 
for $\Delta = 2$ and $t^\prime = 4$.} 
\label{fig-w-num-app}
\end{figure}

In order to produce richer topological phases, one can choose a different type of BCS pairing 
and next-nearest-neighbor hopping, for example,
\begin{equation}
	\hat{\Delta}(\bfdel) 
	= 
	\begin{cases}  
		i \, \sqrt{3}		\, \Delta\, \hmu^2 \, \, (\text{d-wave spin-singlet}), & \bfdel \perp \hat{\mathbf{z}}, \\ 
                -i \sgn(\delta_y)	\, \Delta\, \hmu^1 \, \, (\text{p-wave spin-triplet}), & \bfdel \perp \hat{\mathbf{x}} \\
                -i \sgn(\delta_x) 	\, \Delta\, \hmu^1 \, \, (\text{p-wave spin-triplet}), & \bfdel \perp \hat{\mathbf{y}},
	\end{cases}
\end{equation}
and 
\be
	t_{\text{i}}(\bfdel) 
	= 
	\begin{cases} 
		t/2, 	& \text{i} \in \text{A}, \bfdel \perp \hat{\mathbf{y}}, \text{or}, \, \text{i} \in B, \, \bfdel \perp \hat{\mathbf{x}} \\ 
		-t/2, 	& \text{i} \in \text{A}, \, \bfdel \perp \hat{\mathbf{x}}, \, \text{or}, \, \text{i} \in \text{B}, \, \bfdel \perp \hat{\mathbf{y}} \\
		0, 	&  \bfdel \perp \hat{\mathbf{z}}.
	\end{cases}
\ee
The corresponding potential functions in momentum space are
\begin{align}\label{pairing-app}
	\begin{aligned}
	\Delta_\text{d}(\bk) = & 4  \sqrt{3} \, \Delta \, \cos{\left( \fr{k_x}{2} \right)} \cos{\left( \fr{k_y}{2} \right)}, \\ 
	\Delta_\text{p}(\bk) = & 4 \, \Delta \, \left[ \sin{\left( \fr{k_x}{2} \right)}+\sin{\left( \fr{k_y}{2} \right)} \right] \cos{\left( \fr{k_z}{2} \right)}.
	\end{aligned}
\end{align}
and 
\be  \label{nnn-hop-app}
	\Theta(\bk) 
	= 
	2 \, t \, \left[
		\cos{\left( \fr{k_x}{2} \right)}
		-
		\cos{\left( \fr{k_y}{2} \right)}
		\right] 
		\cos{\left( \fr{k_z}{2} \right)}
	+ 
	\mu_s.
\ee
Substituting Eqs.~(\ref{pairing-app}) and (\ref{nnn-hop-app}) together with Eq.~(\ref{PHI-ri}) into 
Eqs.~(\ref{ham-k-up}) and (\ref{bands}), and evaluating winding number by Eq.~(\ref{wind-num}), 
we obtain the phase diagram 
shown in Fig.~\ref{fig-w-num-app}. In addition to the topological 
phase with winding number $|\nu|=2$, we obtain the phase with $|\nu|=1$.

\subsection{Low-energy surface state theory}   \label{twisted-mass}

A low-energy field
theory 
can be obtained by expanding $\calH(\bk)$ 
in Eq.~(\ref{ham-k-up}) 
around the four Dirac nodes in Eq.~(\ref{aiii-nodes}), when $\mu_s$ and $t$ are small 
compared to $t^\prime$ and $\Delta$. 
This incorporates two independent 
pairs of 3D
mass-degenerate 
Dirac fermions, with masses given by Eq.~(\ref{dirac-masses}).
To obtain an effective theory for the 
surface states, 
we 
allow the Dirac masses 
to vary with some spatial coordinate, say, the $z$ direction.
We consider 
\be  \label{dirac-mass}
M_1 = M_1(z) = M_0 \, \text{sgn}(z), \quad M_2 = \text{Const.}
\ee 
More generally, $M_2$ is allowed to vary along $z$
as long as its sign 
remains	
fixed. 
The phase diagram shown in Fig.~\ref{fig-w-num}(a) implies that the phase with $z > 0$ 
is a topological trivial superconductor, while $z < 0$ 
is nontrivial. 
Since $|M_2| \neq 0$ everywhere in space, the Dirac fermions 
located at 
$\bfK_{2,\pm}$ are 
gapped 
and can be neglected.

In real space, the bulk low-energy Dirac theory 
arising from 
the nodes $\bfK_{1,\pm}$
can be expressed as
\be  \label{bulk-ham-d}
	H_D = \int_{\bfr,z} \! \Psi^\dagger(\bfr,z) \, \calH_D(\bfr,z) \, \Psi(\bfr, z),
\ee
where $\bfr$ is the 2D coordinate on the interface $z=0$, $\int_{\bfr,z}$ denotes the 
spatial integration $\int \! \rmd^2\bfr \rmd z$, 
and 
$\Psi(\bfr, z)$ is an eight-component spinor, a direct product of 
particle-hole, sublattice, and node (``color'') $\bfK_{1,\pm}$ degrees of freedom. 
After performing a unitary transformation and rescaling to eliminate velocity anisotropies,
the Dirac Hamiltonian is 
\be  \label{Dirac-bulk}
	\calH_D(\bfr,z) = -i \htbs \cdot \boldsymbol{\nabla} - i \, \hsig^3 \ots \htau^2 \, \partial_z + M_1(z) \, \hsig^3 \ots \htau^3.
\ee
The chiral symmetry (\ref{SLS}) is brought to the form
\be  \label{chiral-bulk}
	-(\hsig^3 \ots \htau^1) \calH_D (\hsig^3 \ots \htau^1) = \calH_D.
\ee
The spinor $\Psi(\bfr, z)$ can be decomposed into 
surface and bulk parts	\cite{SUP-JR1976},
\be  \label{decomp-surf}
	\Psi(\bfr, z) = \begin{bmatrix} \eta(\bfr) \\ -\eta(\bfr) \end{bmatrix} \sqrt{\frac{M_0}{2}}  e^{-M_0|z|} + \text{bulk states},
\ee
where $\Psi(\bfr, z)$ is explicitly graded by $\htau^3$. Inserting Eq.~(\ref{decomp-surf}) 
into Eq.~(\ref{bulk-ham-d}), disregarding the contribution of the bulk states, and integrating over $z$, 
we finally obtain the 2D Dirac Hamiltonian describing the surface states:
\be \label{surf-ham}
	H_{\text{surf.}} = \int_\bfr \eta^\dagger(\bfr) \, \hat{h} \, \eta(\bfr), 
	\;\;
	\hat{h}  = - i \, \htbs \cdot \boldsymbol{\nabla}.
\ee

At the surface, 
the physical time-reversal symmetry encoded in the chiral 
condition (\ref{SLS}) assumes the form in Eq.~(\ref{LET-t-rev}), 
which is 
the projection of 
Eq.~(\ref{chiral-bulk}) to the 
subspace $\htau^1 = -1$ [Eq.~(\ref{decomp-surf})]. 
Equation~(\ref{LET-t-rev}) prohibits	
a 
\emph{surface}
Dirac mass term
$m =\eta^\dagger \hsig^3 \eta$.
Weak interactions cannot open a gap unless 
time-reversal symmetry is broken,
either spontaneously or by external means.


\section{Kubo formalism for spin conductivity \label{Kubo-sec}}

In this section we 
compute 
the 
interaction 
corrections to the surface state dc spin conductivity in a disordered class CI or AIII TSC,
at zero temperature.
Class CI preserves spin SU(2) in every disorder realization, while class AIII preserves the   
spin-$z$ component \cite{SRFL2008,FXCup}. We 
evaluate the Kubo formula for the conserved $z$-spin current $\mathbf{J}^z$ defined in Eq.~(\ref{LET-ci-current-1}).	
We show that
quantum conductance corrections due to short-ranged interactions vanish. 
These results hold in every fixed realization of the disorder, 
including the clean limit, and are obtained without detailed knowledge of
the noninteracting Green's functions. Instead, we exploit only general properties
such as the Ward identity and the chiral symmetry in Eq.~(\ref{LET-chiral-gf}) of the main text.
We explicitly evaluate corrections to first (second) order in the interaction strengths for
class CI (AIII). We also sketch a proof for the cancellation of corrections to all
orders for class AIII.
  
The 
Kubo formula for the
dc spin conductivity 
is 
\cite{SUP-mahanbook}
\be  \label{def-conduc}
	\sigma_\rms^{\alpha\beta} 
	= 
	\lim_{\omega \to 0}\,
	{\frac{1}{\omega \mathcal{V}}\, 
	\im \left\{ \Pi_{\alpha\beta}(i \Omega_n)|_{i \Omega_n \to \omega + i \delta}\right\}},
\ee
where $\mathcal{V}$ is the system volume and the current-current correlation function is defined by
\be \label{cc-cf}
	\Pi_{\alpha\beta}(i \Omega_n)=\int_{0}^{1/T} \! d\tau \, e^{i \Omega_n \tau} \, \left\langle \Tt \, j_\alpha(\tau) \, j_\beta(0)    \right\rangle.
\ee
Here $\alpha,\beta = 1, 2$ are directions in 
real
space, $T$ is the temperature,
$j_\alpha(\tau)$ is the current operator at imaginary time $\tau$,  and the bracket $\langle \cdots \rangle$ 
denotes the thermal
average. 
The spin-$z$ current operator 
is [see Eq.~(\ref{LET-ci-current-1})]
\be  \label{current}
	j_\alpha(\tau)  = \int_\mathbf{r} \eta^\dagger(\mathbf{r},\tau) \, \hJ^\alpha \, \eta(\mathbf{r},\tau), \quad \a \in \{1,2\}.
\ee
In the absence of interactions
and the presence of nonmagnetic disorder,
the Hamiltonian 
is given by Eq.~(\ref{LET-ham-nd}). 

We represent the exact
noninteracting 
single-particle retarded and advanced Green's 
functions by matrices whose elements are defined by
\be \label{def-greenf} 
	\left[ \hat{G}_{12}^{R/A}(\ep) \right]_{ab} 
	\equiv  
	\sum_{j}{\frac{\varphi_j(a, \bfr_1) \, \varphi_j^\ast(b, \bfr_2)}{\ep - \ep_j \pm i \delta }},  
\ee
where $a, b$ are shorthand indices for the pseudospin and color degrees of freedom, and $j$ labels 
the exact single-particle wave function at an eigenenergy $\ep_j$. 
Apart from the general relation 
\be \label{sym-gf}
	\left[ \hat{G}_{12}^{R/A} (\ep) \right]^\dagger =  \hat{G}_{21}^{A/R} (\ep),
\ee
time-reversal symmetry [Eq.~(\ref{LET-t-rev})]
allows one to relate the two types of Green's functions via
Eq.~(\ref{LET-chiral-gf}).
The Matsubara Green's function 
satisfies 
\be \label{sym-mgf}
	- \hat{\sigma}^3 \, \hat{G}_{12}(i \w_n) \, \hat{\sigma}^3 = \hat{G}_{12} (-i \w_n).
\ee
In what follows we will exploit the Ward identities \cite{mirlin2006,SUP-mahan2},	
\begin{subequations} \label{ward-id}
	\be  \label{ward-1}
\int_{\bfr_3} \hG_{13}^{R/A}(\ep) \, \hJ^\a \, \hG_{32}^{R/A}(\ep)=  - i \, (\bfr_1 -\bfr_2)^\a \, \hG_{12}^{R/A}(\ep),
\ee
\be   \label{ward-2}
	\int_{\bfr_3} \hG_{13}(i \w_n) \, \hJ^\a \, \hG_{32}(i \w_n)=  - i \, (\bfr_1 \! - \! \bfr_2)^\a \, \hG_{12}(i \w_n),
\ee
\end{subequations}
and
the following 
relations between the 
components of the spin 
U(1) 
current operator in Eq.~(\ref{current}):
\begin{align}  \label{current-sym}
	-\hat{\sigma}^3 \, \hJ^\alpha \, \hat{\sigma}^3 =  \hJ^\alpha,\;\; 	
	\hat{\sigma}^3 \hJ^\a = i \, \varepsilon_{\a\b} \, \hJ^\b,     	 
\end{align} 
where $\varepsilon_{\a\b}$ is the 2D Levi-Civita symbol.

We consider squared $z$-spin $(S^z)^2$ and Dirac mass $m^2$ interactions, as appear
in Eq.~(\ref{LET-action-int}) for class AIII. For class CI, the Hartree and Fock corrections 
will be the same for $(S^{\alpha})^2$ ($\alpha \in \{x,y,z\}$) interactions, by SU(2) symmetry. 
The
$z$-spin density 
and 
Dirac mass (Cooper pair density \cite{Foster2012,FXCup}) operators are 
\be
	S^z (\mathbf{r}) = \eta^\dagger \, \eta(\mathbf{r}), 
	\quad m(\mathbf{r}) = \eta^\dagger \, \hs^3 \, \eta(\mathbf{r}).	
\ee
We define	
\be
	\hat{s} \in \left\{ \hat{1}, \hs^3 \right\}, 
\ee
so that both are encoded as $\eta^\dagger \hat{s} \eta$.

\subsection{Hartree-Fock spin conductivity}  \label{HF-approx}
The lowest-order diagrams for the interaction correction
to Eq.~(\ref{cc-cf}),
which is denoted by $\del \Pi_{\a \b}^{(1)}$, are shown in 
Fig.~\ref{LET-HF-diagrams}, where 
(a) corresponds to the Hartree diagrams and (b) 
to the Fock ones. 
The Hartree diagrams give 
\begin{subequations} \label{Hdiag}
\begin{align}  
	&
	\del \Pi_{\a \b}^{(1a)}  (i \W_n) 
	=   
	-\Gamma_{\rms,\rmc} \, (-1)^2 \, T^2 \sum_{i \w_p, i \w_q } 
	\;
	\int_{\bfr_1, \bfr_2, \bfr_3}	\nonumber\\
	&
	\times \!\! \left\{ \Tr \left[ \hJ^\a \, \hG_{12} (i \w_p)  \,  \hJ^\beta \, \hG_{23} (i \w_p + i \W_n) \, \hts \, \hG_{31}(i \w_p + i \W_n) \right] \right. \nn \\  
	&\phantom{+}\qquad 
	\left. \times \Tr\left[ \hts \, \hG_{33}(i \w_q) \right] \right. \label{H1} \\ 
	&\phantom{+}
	\left.  +  \Tr \left[  \hJ^\b \, \hG_{21} (i \w_p )  \,  \hJ^\a \, \hG_{13} (i \w_p - i \W_n) \, \hts \, \hG_{32}(i \w_p - i \W_n) \right] \right. \nn \\  
	&\phantom{+}\qquad 
	\left. \times \Tr\left[ \hts \, \hG_{33}(i \w_q) \right] \right. \label{H2} \\
	&\phantom{+}
	\left. + \Tr \left[ \hts \, \hG_{31} (i \w_p ) \, \hJ^\a \, \hG_{13}( i \w_p - i \W_n) \right] \right. \nn \\  
	&\phantom{+}\qquad 
	\left. \times \Tr \left[ \hts \, \hG_{32}(i \w_q ) \, \hJ^\b  \, \hG_{23}(i \w_q + i \W_n)  \right]  \right\}, 
	\!\!\!
	\label{H3}
\end{align}
\end{subequations}
where the terms (\ref{H1})-(\ref{H3}) come from the diagrams a(i)-a(iii), respectively. The Fock diagrams 
give
\begin{subequations} \label{Fdiag}
\begin{align}  
	\del \Pi_{\a \b}^{(1b)} (i \W_n) 
	=  
	-\Gamma_{\rms,\rmc} & \, (-1) \, T^2 \sum_{i \w_p, i \w_q } 
	\; 
	\int_{\bfr_1, \bfr_2, \bfr_3}	\nonumber\\
	\, \times \left\{ \Tr \left[ \hJ^\a  \, \hG_{12}(i \w_p \mcpt )\, \hJ^\beta \, \hG_{23}(i \w_p + i \W_n) \, \hts \, \hG_{33}(i \w_q) \, \right.\right. \nn \\
	& \nvsp \left. \left. \times \hts \, \hG_{31}(i \w_p + i \W_n)  \, \right] \right. \label{F1}   \\ 
	\left. +  \Tr \left[ \hJ^\b  \, \hG_{21}(i \w_p \mcpt ) \, \hJ^\a \, \hG_{13}(i \w_p - i \W_n) \, \hts \, \hG_{33}(i \w_q) \, \right.\right. \nn \\
	& \nvsp \left. \left. \times \hts \, \hG_{32}(i \w_p - i \W_n) \,   \right] \right. \label{F2}  \\
	\left. + \Tr \left[ \hts \, \hG_{31}(i \w_p)  \, \mcpt \hJ^\a \, \hG_{13}(i \w_p - i \W_n) \, \hts \, \hG_{32}(i \w_q) \,\right.\right. \nn \\
	& \nvsp \left. \left. \times \hJ^\b \, \hG_{23}(i \w_q + i \W_n) \right]  \right\}, \label{F3}
\end{align}
\end{subequations} 
where (\ref{F1})-(\ref{F3}) come from the diagrams b(i)-b(iii), respectively. 

The Hartree terms~(\ref{H1}) and (\ref{H2}) vanish individually, 
due to Eq.~(\ref{sym-mgf}) and the fact that $[\hat{s},\sigh^3] = 0$. 
The Fock terms~(\ref{F1}) and (\ref{F2}) give the same contribution. 
Equations~(\ref{H3}) and (\ref{F3}) do not contribute to 
the 
dc conductivity, because they are of the order of $\W^2$ when $\W \to 0$.
Therefore, we only have to further analyze the contributions of Eqs.~(\ref{F1}) and (\ref{F2}):        
\begin{align}  \label{Fdiag-2}
	\del \Pi_{\a \a}^{(1b)} & (i \W_n) 
	= 
	2 \, \Gamma_{\rms,\rmc} \, T \sum_{i \w_p} 
	\; 
	\int_{\bfr_1, \bfr_2, \bfr_3}
	\Tr \left[ 
	\hts \, \hrho_3 \, \hts \, 
	\right.
	\nonumber\\
	\times
	& 
	\left.	
	\hG_{31} (i \w_p - i \W_n)
	\, \hJ^\a 
	\, \hG_{12}(i \w_p) \, \hJ^\a \, \hG_{23}(i \w_p - i \W_n) 
	\right].
\end{align}
Here we have defined the local 
$z$-spin
density matrix 
\be  \label{spin-z-dens}
\begin{split}
	\left( \hrho_3 \right)_{ab} 
	&\equiv T \sum_{i \w_q} \left[ \hG_{33}(i \w_q) \right]_{ab} \\
        & = \sum_{j}{\varphi_j(a, \bfr_3) \, \varphi_j^\ast(b, \bfr_3)} f(\ep_j),
\end{split}
\ee
where $f(\ep_j)$ is the Fermi-Dirac distribution function. One can easily prove that $-\hat{\s}^3 \, \hat{\rho}_3 \, \hat{\s}^3 = \hat{\rho}_3 $.
Applying the standard analytic continuation technique \cite{SUP-mahanbook} we obtain 
\be  \label{dc-cond-cont}
\begin{split} 
	\del & \Pi_{\a \a}^{(1b)} (\w) =  - 2 \Gamma_{\rms,\rmc}  \int_{-\infty}^{+\infty} \frac{d\ep}{2 \pi i} 
	\; 
	\int_{\bfr_1, \bfr_2, \bfr_3}
	\Tr 
	\bigg[
	\hts \, \hrho_3 \, \hts \\
	& \times \left.\left\{
	\left[f(\ep + \w) - f(\ep)\right] \, \hG_{31}^A(\ep) \, \hJ^\a \, \hG_{12}^R(\ep+\w) \, \hJ^\a \, \hG_{23}^A(\ep)\right.\right.  \\
	& \left.\left.\qquad +f(\ep) \, \hG_{31}^R(\ep) \, \hJ^\a \, \hG_{12}^R(\ep+\w) \, \hJ^\a \, \hG_{23}^R(\ep) \right. \right. \\ 
	& \left.      \qquad -f(\ep + \w) \, \hG_{31}^A(\ep) \, \hJ^\a \, \hG_{12}^A(\ep+\w) \, \hJ^\a \, \hG_{23}^A(\ep) \right\} \bigg].
\end{split}
\ee
Therefore, the 
correction to 
Eq.~(\ref{LET-spin-cond})
reads
\begin{align} \label{fconds}
	\delta
	\s_\rms^{\a\a}  
	= & \frac{1}{\mathcal{V}}\lim_{\w \to 0} \frac{1}{ 2 i \w} \left[ \del \Pi_{\a \a}^{(1b)} (\w) -\del {\Pi_{\a \a}^{(1b)}}^\ast (\w) \right] \nn \\
        = & 
	\frac{\Gamma_{\rms,\rmc}}{2 \pi \mathcal{V}}	
	\int_{-\infty}^{+\infty} \! d\ep \, \frac{d f(\ep)}{d \ep} \, 
	\; 
	\int_{\bfr_1, \bfr_2, \bfr_3}	
	\Tr \left\{ \hts \, \hrho_3 \, \hts \right. \nn \\
       & \times \left.\left[ \hG_{31}^A(\ep) \, \hJ^\a \, \hG_{12}^R(\ep) \, \hJ^\a \, \hG_{23}^A(\ep) \right.\right. \nn \\
       & \left.\left.\quad\, + \hG_{31}^R(\ep) \, \hJ^\a \, \hG_{12}^A(\ep) \, \hJ^\a \, \hG_{23}^R(\ep) \right. \right. \nn \\   
       & \left.\left.\quad\, - \hG_{31}^R(\ep) \, \hJ^\a \, \hG_{12}^R(\ep) \, \hJ^\a \, \hG_{23}^R(\ep) \right.\right. \nn \\
       & \left.\left.\quad\, - \hG_{31}^A(\ep) \, \hJ^\a \, \hG_{12}^A(\ep) \, \hJ^\a \, \hG_{23}^A(\ep)\right] \right\},  	
\end{align}
where we have used 
Eq.~(\ref{sym-gf}).   

At zero temperature, Eq.~(\ref{fconds}) can be expressed entirely in terms of retarded Green's functions,
\begin{align} \label{fconds-II}
	\delta\s_\rms^{\a\a}  
        =& 
	-
	\frac{\Gamma_{\rms,\rmc}}{2 \pi \mathcal{V}}	
	\sum_{\beta = 1}^2
	\int_{\bfr_1, \bfr_2, \bfr_3}
	\Tr \Big[
	\left(\hJ^3 \hrho_3 \hJ^3 - \hrho_3 \right)
	\nn\\
	&\qquad\quad
	\times
	\hts \,
	\hG_{31}^R(0) \, \hJ^\b \, \hG_{12}^R(0) \, \hJ^\b \, \hG_{23}^R(0)
	\hts
	\Big].
\end{align}
To derive this, we replace all advanced Green's functions with retarded ones 
using Eq.~(\ref{LET-chiral-gf}), and employ Eq.~(\ref{current-sym}).
Finally, we use the Ward identity (\ref{ward-1}) to show that this expression is
zero. Integrating over $\bfr_1$ yields
\begin{align} \label{fconds-III-1}
	\delta\s_\rms^{\a\a}  
        =&\, 
	i
	\frac{\Gamma_{\rms,\rmc}}{2 \pi \mathcal{V}}	
	\sum_{\beta = 1}^2
	\int_{\bfr_2, \bfr_3}
	(\bfr_3 - \bfr_2)^\beta
	\nn\\
	\times&
	\Tr \Big[
	\left(\hJ^3 \hrho_3 \hJ^3 - \hrho_3 \right)
	\hts
	\hG_{32}^R(0) \, \hJ^\b \, \hG_{23}^R(0)
	\hts
	\Big].
\end{align}
Integrating over $\bfr_2$ instead gives
\begin{align} \label{fconds-III-2}
	\delta\s_\rms^{\a\a}  
        =& 
	i
	\frac{\Gamma_{\rms,\rmc}}{2 \pi \mathcal{V}}	
	\sum_{\beta = 1}^2
	\int_{\bfr_1, \bfr_3}
	(\bfr_1 - \bfr_3)^\beta
	\nn\\
	\times&
	\Tr \Big[
	\left(\hJ^3 \hrho_3 \hJ^3 - \hrho_3 \right)
	\hts
	\hG_{31}^R(0) \, \hJ^\b \, \hG_{13}^R(0)
	\hts
	\Big]
	\nonumber\\
	=&
	-
	\delta\s_\rms^{\a\a}. 
\end{align}
The Hartree and Fock corrections vanish for both spin- and 
mass-squared interactions.

\subsection{Second-order corrections \label{2nd-sec}}
\begin{figure}[b!]
\centering
\includegraphics[width=0.5\textwidth]{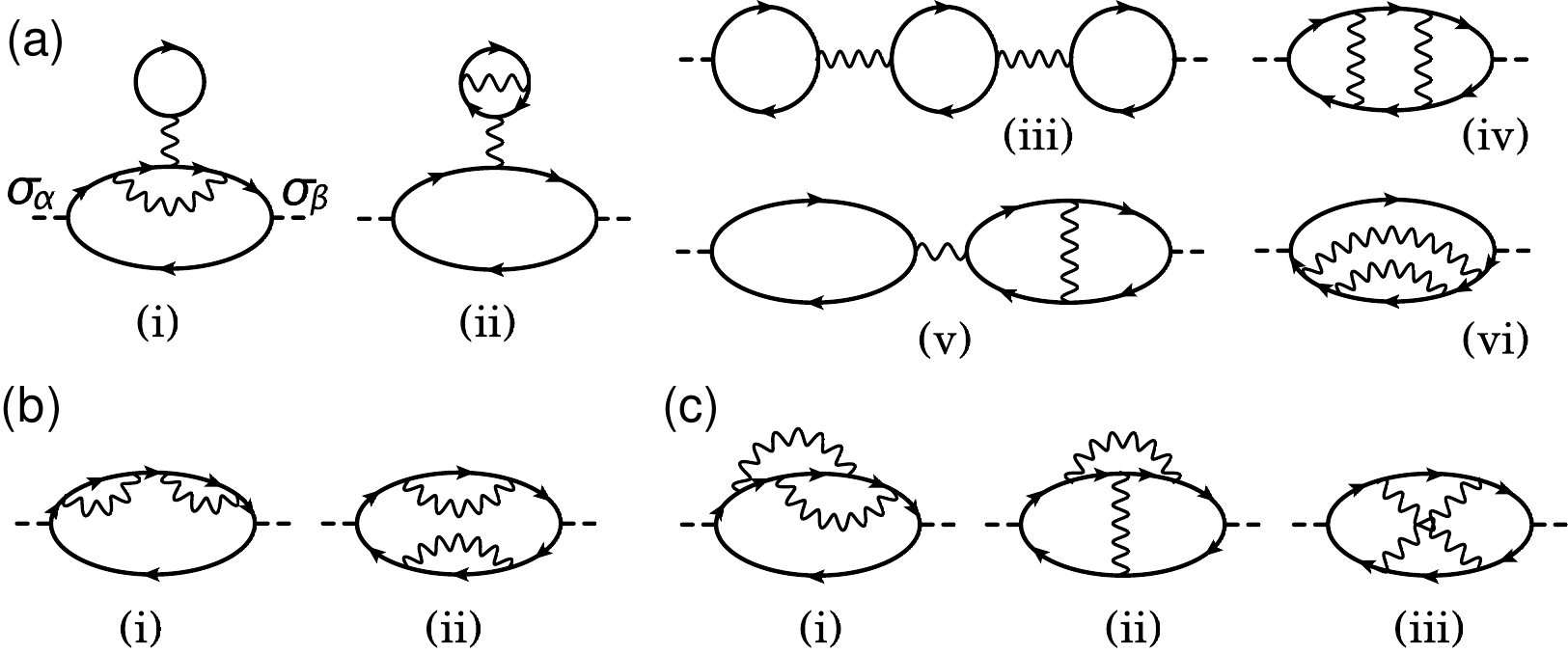} \\
\vspace{0.2cm}
\includegraphics[width=0.5\textwidth]{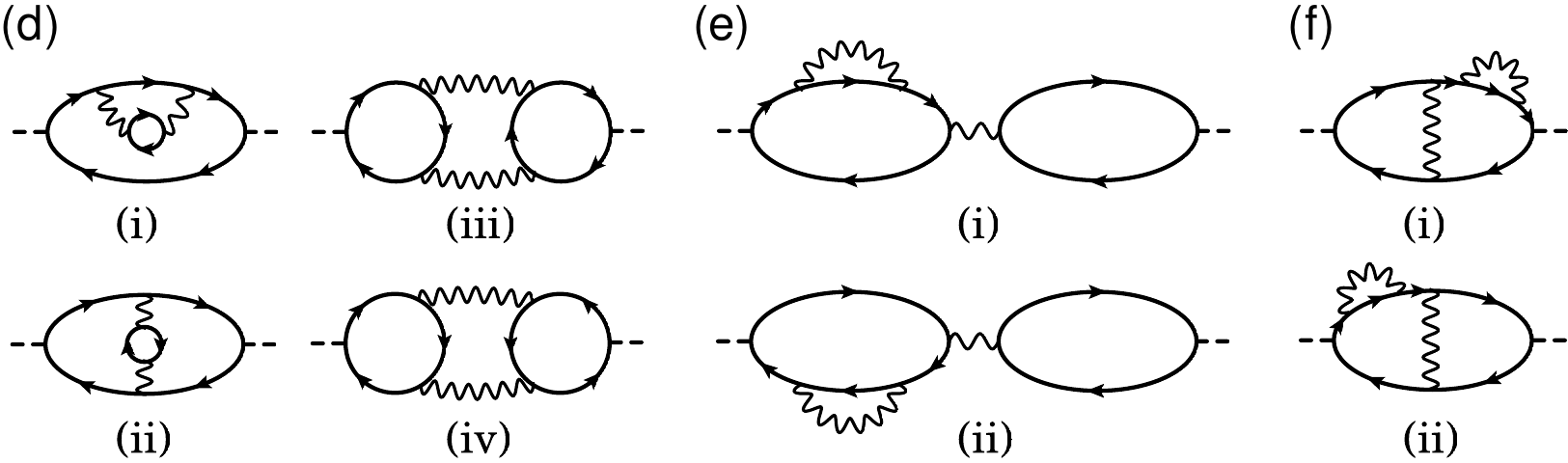} \\
\caption{The second-order interaction corrections to the spin current-current correlation function in 
class
AIII. The Feynman rules are interpreted in 
the caption for
Fig.~\ref{LET-HF-diagrams}
of the main text.}  \label{second-order}
\end{figure}

The second-order interaction corrections to 
the
spin conductivity 
in class AIII
are represented by the Feynman diagrams in Fig.~\ref{second-order}. 
Clearly the diagrams including at least one Hartree bubble, for example, a(i) and a(ii), are individually zero 
due to the chiral condition (\ref{sym-mgf}). The diagrams a(iii)--a(v) do not contribute to the dc conductivity 
because they are of the order of $\W^2$ when $\W \to 0$. Moreover, comparing a(vi) to Fig.~\ref{LET-HF-diagrams}(b)(i) one can 
readily prove that the dc conductivity correction arising from a(vi) vanishes at zero temperature. 
Here we show that the diagrams in each category (b)--(f) 
depicted in Fig.~\ref{second-order}
altogether give null contribution.

\subsubsection{Category b}
The current-current correlation functions represented by b(i) and b(ii) read 
\begin{subequations}
\begin{align}
& \delta\Pi_{\a\a}^{[\mathrm{b(i)}]}(i\W_n) = - \Gamma_{\rms,\rmc}^2 \, T  \sum_{i\w_p}\int_{\bfr_1,\bfr_2,\bfr_3,\bfr_4} \nn \\ 
& \times \Tr \left[\hat{G}_{41}(i \w_p + i \W_n) \, \hat{\s}^\a \, \hat{G}_{12}(i \w_p) \, \hat{\s}^\a \,  \hat{G}_{23}(i \w_p + i \W_n) \right. \nn \\ 
& \left. \hspace{0.8cm} \times \, \hat{s} \, \hat{\rho}_3 \, \hat{s} \,\hat{G}_{34}(i \w_p + i \W_n)  \hat{s} \, \hat{\rho}_4 \, \hat{s} \, \right], \\
& \delta\Pi_{\a\a}^{[\mathrm{b(ii)}]}(i\W_n) = -\Gamma_{\rms,\rmc}^2 \, T  \sum_{i\w_p}\int_{\bfr_1,\bfr_2,\bfr_3,\bfr_4} \nn \\
& \times \Tr \left[ \hat{G}_{41}(i \w_p + i \W_n) \, \hat{\s}^\a \hat{G}_{13}(i \w_p) \, \hat{s} \, \hat{\rho}_3 \, \hat{s} \,  \hat{G}_{32}(i \w_p) \right. \nn \\ 
& \left. \hspace{0.8cm} \times \,\hat{\s}^\a \,\hat{G}_{24}(i \w_p + i \W_n) \hat{s} \, \hat{\rho}_4 \, \hat{s}  \right],
\end{align}
\end{subequations}
where the local $z$-spin density matrix $\hat{\rho}_{3}$ is defined in Eq.~(\ref{spin-z-dens}). After analytic continuation we obtain
\begin{subequations}
\begin{align}
& \delta\Pi_{\a\a}^{[\mathrm{b(i)}]}(\w) =  \Gamma_\mathrm{s,c}^2 \, \int_{-\infty}^{+\infty}\frac{\rmd\ep}{2 \pi i}\int_{\bfr_1,\bfr_2,\bfr_3,\bfr_4} \nn \\
& \times \Tr \left\{ [f(\ep)-f(\ep-\w)] \, \hat{\s}^\a \hat{G}_{12}^A(\ep-\w) \, \hat{\s}^\a \,  \hat{G}_{23}^R(\ep) \hat{s} \, \hat{\rho}_3 \, \hat{s} \right. \nn \\ 
& \hspace{0.8cm} \times \left. \hat{G}_{34}^R(\ep) \hat{s} \hat{\rho}_4 \, \hat{s} \, \hat{G}_{41}^R(\ep) + f(\ep-\w) \, \hat{\s}^\a \hat{G}_{12}^R(\ep-\w) \,  \right. \nn \\ 
& \hspace{0.8cm} \times \hat{\s}^\a \,  \hat{G}_{23}^R(\ep)  \hat{s} \, \hat{\rho}_3 \, \hat{s} \,\hat{G}_{34}^R(\ep) \hat{s} \hat{\rho}_4 \, \hat{s} \, \hat{G}_{41}^R(\ep) - f(\ep) \, \hat{\s}^\a \nn \\
& \hspace{0.8cm} \times \left. \hat{G}_{12}^A(\ep-\w)  \hat{\s}^\a \,  \hat{G}_{23}^A(\ep)  \hat{s} \, \hat{\rho}_3 \, \hat{s} \,\hat{G}_{34}^A(\ep) \hat{s} \hat{\rho}_4 \, \hat{s} \, \hat{G}_{41}^A(\ep) \right\}, \\
& \delta\Pi_{\a\a}^{[\mathrm{b(ii)}]}(\w) = \Gamma_\mathrm{s,c}^2 \, \int_{-\infty}^{+\infty}\frac{\rmd\ep}{2 \pi i}\int_{\bfr_1,\bfr_2,\bfr_3,\bfr_4}  \nn \\
& \times \Tr \left\{ [f(\ep)-f(\ep-\w)] \, \hat{\s}^\a \hat{G}_{13}^A(\ep-\w) \, \hat{s} \, \hat{\rho}_3 \, \hat{s}  \,  \hat{G}_{32}^A(\ep-\w) \right. \nn \\ 
& \hspace{0.3cm} \times \hat{\s}^\a \,\hat{G}_{24}^R(\ep)  \hat{s} \hat{\rho}_4 \, \hat{s} \, \hat{G}_{41}^R(\ep) + f(\ep-\w) \, \hat{\s}^\a \hat{G}_{13}^R(\ep-\w) \,  \nn \\
& \hspace{0.3cm} \times \hat{s} \, \hat{\rho}_3 \, \hat{s}  \,  \hat{G}_{32}^R(\ep-\w) \, \hat{\s}^\a \,\hat{G}_{24}^R(\ep)  \hat{s} \hat{\rho}_4 \, \hat{s} \, \hat{G}_{41}^R(\ep)- f(\ep) \, \hat{\s}^\a  \nn \\
& \hspace{0.3cm} \times \left. \hat{G}_{13}^A(\ep-\w) \, \hat{s} \, \hat{\rho}_3 \, \hat{s}  \,  \hat{G}_{32}^A(\ep-\w) \, \hat{\s}^\a \,\hat{G}_{24}^A(\ep)  \hat{s} \hat{\rho}_4 \, \hat{s} \, \hat{G}_{41}^A(\ep) \right\}.
\end{align}
\end{subequations}
Therefore, via Eq.~(\ref{def-conduc}) the conductivity corrections read
\begin{subequations}
\begin{align} 
& \delta\s_\mathrm{s}^{\a\a}[\mathrm{b(i)}] 
 = \frac{\Gamma_{\rms,\rmc}^2}{4\pi\mathcal{V}} \int_{-\infty}^{+\infty}\!\rmd \ep \frac{\rmd f(\ep)}{\rmd \ep} \int_{\bfr_1,\bfr_2,\bfr_3,\bfr_4} \nn \\ 
& \times \Tr \left[ \hat{\s}^\a \hat{G}_{12}^A(\ep) \, \hat{\s}^\a \,  \hat{G}_{23}^R(\ep)  \hat{s} \, \hat{\rho}_3 \, \hat{s} \,\hat{G}_{34}^R(\ep) \hat{s} \hat{\rho}_4 \, \hat{s} \, \hat{G}_{41}^R(\ep) \right. \nn \\
&\hspace{0.8cm} + \hat{\s}^\a \hat{G}_{12}^R(\ep) \, \hat{\s}^\a \,  \hat{G}_{23}^A(\ep)  \hat{s} \, \hat{\rho}_3 \, \hat{s} \,\hat{G}_{34}^A(\ep) \hat{s} \hat{\rho}_4 \, \hat{s} \, \hat{G}_{41}^A(\ep) \nn \\
&\hspace{0.8cm} - \hat{\s}^\a \hat{G}_{12}^R(\ep) \, \hat{\s}^\a \,  \hat{G}_{23}^R(\ep)  \hat{s} \, \hat{\rho}_3 \, \hat{s} \,\hat{G}_{34}^R(\ep) \hat{s} \hat{\rho}_4 \, \hat{s} \, \hat{G}_{41}^R(\ep) \nn \\
&\hspace{0.8cm} - \left. \hat{\s}^\a \hat{G}_{12}^A(\ep) \, \hat{\s}^\a \,  \hat{G}_{23}^A(\ep)  \hat{s} \, \hat{\rho}_3 \, \hat{s} \,\hat{G}_{34}^A(\ep) \hat{s} \hat{\rho}_4 \, \hat{s} \, \hat{G}_{41}^A(\ep) \right], \label{b-i} \\
& \delta\s_\mathrm{s}^{\a\a}[\mathrm{b(ii)}] = \frac{\Gamma_{\rms,\rmc}^2}{4\pi\mathcal{V}} \int_{-\infty}^{+\infty}\!\rmd \ep \frac{\rmd f(\ep)}{\rmd \ep} \int_{\bfr_1,\bfr_2,\bfr_3,\bfr_4} \nn \\
& \times \Tr \left[       \hat{\s}^\a \hat{G}_{13}^A(\ep) \, \hat{s} \, \hat{\rho}_3 \, \hat{s}  \,  \hat{G}_{32}^A(\ep) \, \hat{\s}^\a \,\hat{G}_{24}^R(\ep)  \hat{s} \hat{\rho}_4 \, \hat{s} \, \hat{G}_{41}^R(\ep) \right. \nn \\
&\hspace{0.8cm} + \hat{\s}^\a \hat{G}_{13}^R(\ep) \, \hat{s} \, \hat{\rho}_3 \, \hat{s}  \,  \hat{G}_{32}^R(\ep) \, \hat{\s}^\a \,\hat{G}_{24}^A(\ep)  \hat{s} \hat{\rho}_4 \, \hat{s} \, \hat{G}_{41}^A(\ep) \nn \\              
&\hspace{.8cm} - \hat{\s}^\a \hat{G}_{13}^R(\ep) \, \hat{s} \, \hat{\rho}_3 \, \hat{s}  \,  \hat{G}_{32}^R(\ep) \, \hat{\s}^\a \,\hat{G}_{24}^R(\ep)  \hat{s} \hat{\rho}_4 \, \hat{s} \, \hat{G}_{41}^R(\ep)  \nn \\
&\hspace{.8cm} - \left. \hat{\s}^\a \hat{G}_{13}^A(\ep) \, \hat{s} \, \hat{\rho}_3 \, \hat{s}  \,  \hat{G}_{32}^A(\ep) \, \hat{\s}^\a \,\hat{G}_{24}^A(\ep)  \hat{s} \hat{\rho}_4 \, \hat{s} \, \hat{G}_{41}^A(\ep) \right]. \label{b-ii}
\end{align}
\end{subequations}
At zero temperature Eqs.~(\ref{b-i}) and (\ref{b-ii}) can be expressed entirely in terms of retarded Green's functions,
\begin{subequations}
\begin{align} \label{b-i-c}
& \delta\s_\mathrm{s}^{\a\a}[\mathrm{b(i)}] =  -\frac{\Gamma_{\rms,\rmc}^2}{2\pi\mathcal{V}} \sum_{\b=1}^{2} \int_{\bfr_1,\bfr_2,\bfr_3,\bfr_4} \nn \\ 
& \quad \times \Tr \left[ \hat{G}_{41}^R \, \hat{\s}^\b \, \hat{G}_{12}^R \, \hat{\s}^\b \,  \hat{G}_{23}^R \,\hat{s} \, \hat{\rho}_3 \, \hat{s} \,\hat{G}_{34}^R \, \hat{s} \, \hat{\rho}_4 \, \hat{s} \right], \\
& \delta\s_\mathrm{s}^{\a\a}[\mathrm{b(ii)}]= -\frac{\Gamma_{\rms,\rmc}^2}{2\pi\mathcal{V}}\sum_{\b=1}^{2} \int_{\bfr_1,\bfr_2,\bfr_3,\bfr_4} \nn \\
& \quad \times \Tr \left[\hat{G}_{41}^R  \,  \hat{\s}^\b \, \hat{G}_{13}^R \, \hat{s} \, \hat{\rho}_3 \, \hat{s}  \,  \hat{G}_{32}^R \, \hat{\s}^\b \,\hat{G}_{24}^R  \hat{s} \hat{\rho}_4 \, \hat{s}\right],
\end{align}
\end{subequations}
where we have introduced the abbreviation $\hat{G}_{12}^{R/A} \equiv \hat{G}_{12}^{R/A}(\ep=0)$. Annihilating the current vertices by the Ward identity (\ref{ward-1}), we obtain
\be
\begin{split}
& \, 2 \times \delta\s_\mathrm{s}^{\a\a}[\mathrm{b(i)}] = - \delta\s_\mathrm{s}^{\a\a}[\mathrm{b(ii)}] \\
& \,= \frac{\Gamma_\mathrm{s,c}^2}{2\pi\mathcal{V}} \sum_{\b=1}^2 \int_{\bfr_3,\bfr_4} \, \left[(\bfr_4-\bfr_3)^\b\right]^2 \,\Tr 
\left[ \, \hat{G}_{43}^R  \,  \hat{s} \, \hat{\rho}_3 \, \hat{s} \,\hat{G}_{34}^R \hat{s} \hat{\rho}_4 \, \hat{s} \right].
\end{split}
\ee

\subsubsection{Category c}
The current-current correlation function given by c(i) reads
\begin{subequations}\label{ci_and_p43}
\begin{align}
& \delta\Pi_{\a\a}^{\mathrm{[c(i)]}}(i\W_n) =  - \Gamma_\mathrm{s,c}^2 \, T \sum_{i\w_p} \int_{\bfr_1,\bfr_2,\bfr_3,\bfr_4} \nn \\
& \times \Tr \left[\hat{G}_{31}(i \w_p + i \W_n)\, \hat{\s}^\a \hat{G}_{12}(i \w_p)  \right. \nn \\ 
& \hspace{0.8cm} \times \left. \hat{\s}^\a \,  \hat{G}_{24}(i \w_p + i \W_n) \hat{s} \, \hat{\mathcal{P}}_{43}(i\w_p+i\W_n) \,\hat{s}  \right], \label{f1-1}
\end{align}
where the self-energy operator $\hat{\mathcal{P}}_{43}(i\w_p)$ takes the form
\be   \label{def-p43}
\begin{split}
\hat{\mathcal{P}}_{43}(i\w_p) =  T^2 \! \! \sum_{i\w_q, i\w_r} &\, \hat{G}_{43}(i \w_q) \, \hat{s} \,\hat{G}_{34}(i \w_r) \\ 
&\, \times \hat{s} \,\hat{G}_{43}(i \w_r- i \w_q + i\w_p).
\end{split}
\ee
\end{subequations}
The diagrams c(ii) and c(iii) read
\begin{subequations}
\begin{align}
& \delta\Pi_{\a\a}^{\mathrm{[c(ii)]}}(i\W_n) = - \Gamma_\mathrm{s,c}^2 \, T^3  \! \! \!  \sum_{i\w_p, i\w_q, i\w_r}\int_{\bfr_1,\bfr_2,\bfr_3,\bfr_4} \nn \\
& \times \Tr \left[\hat{G}_{41}(i \w_p + i \W_n) \hat{\s}^\a \hat{G}_{13}(i \w_p) \, \hat{s} \,\hat{G}_{32}(i \w_q)  \right. \nn \\ 
& \hspace{0.8cm} \times \left. \hat{\s}^\a \,\hat{G}_{24}(i \w_q + i \W_n)  \, \hat{s} \, \hat{G}_{43}(i \w_r) \right. \nn \\ 
& \hspace{0.8cm} \times \left. \hat{s} \, \hat{G}_{34}(i \w_r + i \w_p - i \w_q) \, \hat{s}  \right], \label{h1-1} \\
& \delta\Pi_{\a\a}^{\mathrm{[c(iii)]}}(i\W_n) = - \Gamma_\mathrm{s,c}^2 \, T^3 \!\!\!  \sum_{i\w_p, i\w_q, i\w_r}\int_{\bfr_1,\bfr_2,\bfr_3,\bfr_4} \!\!\!\! \Tr \left[ \hat{G}_{31}(i \w_p)  \right. \nn \\ 
& \hspace{0.3cm} \times         \hat{\s}^\a \, \hat{G}_{14}(i \w_p- i \W_n) \, \hat{s} \,\hat{G}_{43}(i \w_q+ i \w_p - i \w_r )  \nn \\ 
& \hspace{0.3cm} \times \left.  \hat{s} \, \hat{G}_{32}(i \w_q) \, \hat{\s}^\a \, \hat{G}_{24}(i \w_q+i\W_n) \, \hat{s} \, \hat{G}_{43}(i \w_r ) \, \hat{s}  \right] \label{s-1} \\
& = - \Gamma_\mathrm{s,c}^2 \,  T^3 \!\!\!  \sum_{i\w_p, i\w_q, i\w_r}\int_{\bfr_1,\bfr_2,\bfr_3,\bfr_4} \Tr \left[ \hat{G}_{31}(i \w_p+i \W_n) \right. \nn \\ 
& \hspace{0.3cm} \times \left. \hat{\s}^\a \hat{G}_{14}(i \w_p) \, \hat{s} \,\hat{G}_{43}(i \w_q+ i \w_p - i \w_r ) \right. \nn \\ & \hspace{0.3cm} \times \left. \hat{s} \, \hat{G}_{32}(i \w_q-i\W_n) \, \hat{\s}^\a \, \hat{G}_{24}(i \w_q) \, \hat{s} \, \hat{G}_{43}(i \w_r ) \, \hat{s} \right], \label{s-2}
\end{align}
\end{subequations}
where from Eq.~(\ref{s-2}) to Eq.~(\ref{s-1}) we have used the chiral condition (\ref{sym-mgf}). Applying the Ward identity (\ref{ward-2}), up to order of $\W$, we split Eq.~(\ref{h1-1}) into two terms
\begin{subequations}
\begin{align}  \label{h1-1-c}
& \delta\Pi_{\a\a}^{\mathrm{[c(ii)]}}(i\W_n) = -  \Gamma_\mathrm{s,c}^2 \, T \,  \sum_{i\w_p}\int_{\bfr_2,\bfr_3,\bfr_4} \Tr \left\{ \hat{s} \, \hat{\mathcal{Q}}_{43,\a}(i\w_p) \, \hat{s} \right. \nn \\ 
& \times \left. \left[ \hat{G}_{32}(i \w_p) \, \hat{\s}^\a \,\hat{G}_{24}(i \w_p + i \W_n) \right. \right. \nn \\
& \hspace{0.5cm} \left. \left. + \hat{G}_{32}(i \w_p+i\W_n) \, \hat{\s}^\a \,\hat{G}_{24}(i \w_q)\right] \right\} + \mathcal{O}(\W^2), 
\end{align}
and write Eqs.~(\ref{s-1}) and (\ref{s-2}) as
\begin{align}
& \delta\Pi_{\a\a}^{\mathrm{[c(iii)]}}(i\W_n) = 2 \, \Gamma_\mathrm{s,c}^2 \, T \,  \sum_{i\w_p}\int_{\bfr_2,\bfr_3,\bfr_4} \Tr \left\{ \hat{s} \, \hat{\mathcal{Q}}_{43,\a}(i\w_p) \, \hat{s} \right. \nn \\ 
& \hspace{0.5cm} \left. \times \hat{G}_{32}(i \w_p) \, \hat{\s}^\a \,\hat{G}_{24}(i \w_p+i \W_n)   \right\} + \mathcal{O}(\W^2) \label{s-3} \\
& = 2 \, \Gamma_\mathrm{s,c}^2 \, T \,  \sum_{i\w_p}\int_{\bfr_2,\bfr_3,\bfr_4} \Tr \left\{ \hat{s} \, \hat{\mathcal{Q}}_{43,\a}(i\w_p) \, \hat{s} \, \hat{G}_{32}(i \w_p+i\W_n) \, \right. \nn \\ 
& \hspace{0.5cm} \left.  \times \hat{\s}^\a \, \hat{G}_{24}(i \w_p) \right\} + \mathcal{O}(\W^2), \label{s-4}
\end{align}
where 
\be   \label{def-Q43}
\hat{\mathcal{Q}}_{43,\a}(i\w_p) = -i \, (\bfr_4 - \bfr_3)^\a \, \hat{\mathcal{P}}_{43}(i\w_p).
\ee
\end{subequations}
Adding Eqs.~(\ref{s-3}) and (\ref{s-4}) and comparing the result to Eq.~(\ref{h1-1-c}) one has
\be \label{cii-ciii}
\delta\Pi_{\a\a}^{\mathrm{[c(ii)]}}(i\W_n) = - \delta\Pi_{\a\a}^{\mathrm{[c(iii)]}}(i\W_n) + \mathcal{O}(\W^2). 
\ee 

By the spectral representation, the analytically continued self-energy matrix $\hat{\mathcal{P}}_{43}(z)$ [Eq.~(\ref{def-p43})] takes the form  
\be     \label{P-43}
\begin{split}
& [\hat{\mathcal{P}}_{43}(z)]_{ab} \\
&  = \sum_{i,j,k}  A_{ab,43}^{ijk} \frac{[f_F(\ep_j)-f_F(\ep_i)] \, [f_F(\ep_k) + f_B(\ep_j- \ep_i)]}{z- \ep_i + \ep_j-\ep_k},
\end{split}
\ee    
where $f_B(\ep_i)$ is the Bose-Einstein distribution function and the spectral weight $A_{ab,43}^{ijk}$ depends on single-particle wave functions. Important properties of $\hat{\mathcal{P}}_{43}(z)$ are manifest via Eqs.~(\ref{def-p43}) and (\ref{P-43}). The branch cut of  $\hat{\mathcal{P}}_{43}(z)$ is on the real axis and the retarded/advanced sector can be defined as $\hat{\mathcal{P}}_{43}^{R/A}(\ep) = \hat{\mathcal{P}}_{43}(\ep \pm i \delta)$. The Hermitian conjugation and the chiral condition are represented $\hat{\mathcal{P}}_{43}^\dagger(z) = \hat{\mathcal{P}}_{34}(z^\ast)$ and $-\hat{\s}^3  \, \hat{\mathcal{P}}_{43}(z) \, \hat{\s}^3 = \hat{\mathcal{P}}_{43}(-z)$, respectively. The matrix $\hat{\mathcal{Q}}_{43,\a}(z)$ [Eq.~(\ref{def-Q43})] follows similar properties.

After analytic continuation Eqs.~(\ref{f1-1}) and (\ref{h1-1-c}) lead to
\begin{subequations}
\begin{align}
& \delta\Pi_{\a\a}^{\mathrm{[c(i)]}}(\w) = \Gamma_\mathrm{s,c}^2 \, \int_{-\infty}^{+\infty}\frac{\rmd\ep}{2 \pi i} \, \int_{\bfr_1,\bfr_2,\bfr_3,\bfr_4} \nn \\
& \times \Tr \left\{ \left[ f(\ep) - f(\ep-\w) \right] \hat{\s}^\a \hat{G}_{12}^A(\ep-\w) \, \hat{\s}^\a \,  \hat{G}_{24}^R(\ep) \,  \hat{s} \, \hat{\mathcal{P}}_{43}^R(\ep) \right. \nn \\
& \hspace{0.8cm} \times \hat{s} \, \hat{G}_{31}^R(\ep)  +    f(\ep-\w) \, \hat{\s}^\a \hat{G}_{12}^R(\ep-\w) \, \hat{\s}^\a \,  \hat{G}_{24}^R(\ep) \nn \\
& \hspace{0.8cm} \times \hat{s} \, \hat{\mathcal{P}}_{43}^R(\ep) \, \hat{s} \, \hat{G}_{31}^R(\ep) - f(\ep) \, \hat{\s}^\a \hat{G}_{12}^A(\ep-\w) \, \hat{\s}^\a \,  \hat{G}_{24}^A(\ep) \nn \\
& \hspace{0.8cm} \times \left. \! \hat{s} \, \hat{\mathcal{P}}_{43}^A(\ep) \, \hat{s} \, \hat{G}_{31}^A(\ep)   \right\},
\end{align}
\begin{align}
& \delta\Pi_{\a\a}^{\mathrm{[c(ii)]}}(\w) = \Gamma_\mathrm{s,c}^2 \, \int_{-\infty}^{+\infty}\frac{\rmd\ep}{2 \pi i} \, \int_{\bfr_2,\bfr_3,\bfr_4}  \nn \\
& \times \Tr \left\{ \left[ f(\ep) - f(\ep-\w) \right] \hat{s} \, \hat{\mathcal{Q}}_{43,\a}^A(\ep-\w) \hat{s} \left[ \hat{G}_{32}^A(\ep-w) \right. \right. \nn \\ 
& \hspace{0.8cm} \times \left.\left. \hat{\s}^\a \, \hat{G}_{24}^R(\ep) + \hat{G}_{32}^R(\ep)  \hat{\s}^\a \, \hat{G}_{24}^A(\ep-\w ) \right] \right. \nn \\
& \hspace{0.8cm} +   f(\ep-\w) \, \hat{s} \, \hat{\mathcal{Q}}_{43,\a}^R(\ep-\w) \hat{s} \left[ \hat{G}_{32}^R(\ep-w)  \hat{\s}^\a \, \hat{G}_{24}^R(\ep) \right. \nn \\ 
& \hspace{0.8cm} \left. \left. + \hat{G}_{32}^R(\ep)  \hat{\s}^\a \, \hat{G}_{24}^R(\ep-\w )\right]- f(\ep) \, \hat{s} \, \hat{\mathcal{Q}}_{43,\a}^A(\ep-\w)  \right. \nn \\
& \hspace{0.8cm} \left. \times \hat{s} \left[ \hat{G}_{32}^A(\ep-\w)  \hat{\s}^\a \, \hat{G}_{24}^A(\ep) + \hat{G}_{32}^A(\ep)  \hat{\s}^\a \, \hat{G}_{24}^A(\ep-\w )\right]  \right\}.
\end{align}
\end{subequations}
Therefore, the conductivity corrections read
\begin{subequations}
\begin{align}   \label{d-c-i}
&\delta\s_\mathrm{s}^{\a\a}(\mathrm{[c(i)]}) = -\frac{\Gamma_\mathrm{s,c}^2}{4\pi\mathcal{V}} \int_{-\infty}^{+\infty}\!\rmd \ep \frac{\rmd f(\ep)}{\rmd \ep} \int_{\bfr_1,\bfr_2,\bfr_3,\bfr_4} \nn \\
& \times \Tr \left[ \hat{\s}^\a \hat{G}_{12}^A(\ep) \, \hat{\s}^\a \,  \hat{G}_{24}^R(\ep) \,  \hat{s} \, \hat{\mathcal{P}}_{43}^R(\ep) \, \hat{s} \, \hat{G}_{31}^R(\ep) \right. \nn \\
&\hspace{0.8cm} + \hat{\s}^\a \hat{G}_{12}^R(\ep) \, \hat{\s}^\a \,  \hat{G}_{24}^A(\ep) \,  \hat{s} \, \hat{\mathcal{P}}_{43}^A(\ep) \, \hat{s} \, \hat{G}_{31}^A(\ep) \nn \\
&\hspace{0.8cm} - \hat{\s}^\a \hat{G}_{12}^R(\ep) \, \hat{\s}^\a \,  \hat{G}_{24}^R(\ep) \,  \hat{s} \, \hat{\mathcal{P}}_{43}^R(\ep) \, \hat{s} \, \hat{G}_{31}^R(\ep) \nn \\
&\hspace{0.8cm} - \left. \hat{\s}^\a \hat{G}_{12}^A(\ep) \, \hat{\s}^\a \,  \hat{G}_{24}^A(\ep) \,  \hat{s} \, \hat{\mathcal{P}}_{43}^A(\ep) \, \hat{s} \, \hat{G}_{31}^A(\ep) \right],
\end{align}
\begin{align}  \label{d-c-ii}
&\delta\s_\mathrm{s}^{\a\a}[\mathrm{c(ii)}] = -\frac{\Gamma_\mathrm{s,c}^2}{4\pi\mathcal{V}} \int_{-\infty}^{+\infty}\!\rmd \ep \frac{\rmd f(\ep)}{\rmd \ep} \int_{\bfr_2,\bfr_3,\bfr_4} \nn \\
& \times \Tr \left\{ \left[  \hat{s} \, \hat{\mathcal{Q}}_{43,\a}^A(\ep) \hat{s} +  \hat{s} \, \hat{\mathcal{Q}}_{43,\a}^R(\ep) \hat{s}\right] \left[ \hat{G}_{32}^A(\ep)  \hat{\s}^\a \, \hat{G}_{24}^R(\ep) \right. \right. \nn \\ 
& \hspace{0.8cm} \left.\left. + \hat{G}_{32}^R(\ep)  \hat{\s}^\a \, \hat{G}_{24}^A(\ep)\right]  -   2 \, \hat{s} \, \hat{\mathcal{Q}}_{43,\a}^R(\ep) \hat{s} \, \hat{G}_{32}^R(\ep)  \hat{\s}^\a \, \hat{G}_{24}^R(\ep) \right. \nn \\
& \hspace{0.8cm} \left. - 2 \, \hat{s} \, \hat{\mathcal{Q}}_{43,\a}^A(\ep) \hat{s} \, \hat{G}_{32}^A(\ep)  \hat{\s}^\a \, \hat{G}_{24}^A(\ep) \right\}.
\end{align}
\end{subequations}

At zero temperature Eq.~(\ref{d-c-i}) leads to
\be \label{delta-d1}
\begin{split}
& 2 \times \delta\s_\mathrm{s}^{\a\a}[\mathrm{c(i)}] \\
& =
\frac{\Gamma_\mathrm{s,c}^2}{2\pi \mathcal{V}} \sum_{\b=1}^2 \int_{\bfr_3,\bfr_4} [(\bfr_4-\bfr_3)^\b]^2 \Tr 
\left[ \hat{s} \, \hat{\mathcal{P}}_{43}^R \, \hat{s} \, \hat{G}_{34}^R \, \right],
\end{split}
\ee
where we have applied the Ward identity (\ref{ward-1}). On the other hand, Eq.~(\ref{d-c-ii}) gives
\be  \label{delta-v0}
\delta\s_\mathrm{s}^{\a\a}[\mathrm{c(ii)}] = \frac{1}{2}\sum_{\b=1}^{2} \delta\s_\mathrm{s}^{\b\b}(\mathrm{[c(ii)]}) = - 2 \times \delta\s_\mathrm{s}^{\a\a}(\mathrm{[c(i)]}),
\ee
where we have used the relation
\begin{align}
& \, \sum_{\a=1}^{2} \int_{\bfr_2} \Tr \left[ \hat{s} \, \hat{\mathcal{Q}}_{43,\a}^{R/A} \, \hat{s} \, \hat{G}_{32}^A \,  \hat{\s}^\a \, \hat{G}_{24}^R \right]  \nn \\
 = & \, i \,  \Tr \left[  \hat{s} \, \hat{\mathcal{P}}_{43}^{R/A} \, \hat{s} \, \hat{G}_{34}^A \, \hat{\s}^3 \, \right] \times \sum_{\a,\b=1}^{2} \varepsilon_{\a\b} (\bfr_4 - \bfr_3)^\a \, (\bfr_4 - \bfr_3)^\b \nn \\
 =& \,0.
\end{align}
Combining Eqs.~(\ref{cii-ciii}) and (\ref{delta-v0}) we finally prove that $\delta\s_\mathrm{s}^{\a\a}[2\times \mathrm{c(i)}+2\times \mathrm{c(ii)} + \mathrm{c(iii)}] = 0$.

\subsubsection{Category d}
The current-current correlation function represented by d(i) reads
\begin{subequations}
\begin{align}  \label{d-i}
& \delta\Pi_{\a\a}^{\mathrm{[d(i)]}}(i\W_n) =  \Gamma_\mathrm{s,c}^2 \, T \sum_{i\w_p} \int_{\bfr_1,\bfr_2,\bfr_3,\bfr_4} \nn \\
& \times \Tr \left[\hat{G}_{31}(i \w_p + i \W_n)\, \hat{\s}^\a \hat{G}_{12}(i \w_p)  \right. \nn \\ 
& \hspace{0.5cm} \times \left. \hat{\s}^\a \,  \hat{G}_{24}(i \w_p + i \W_n) \hat{s} \, \hat{\mathcal{R}}_{43}(i\w_p+i\W_n) \,\hat{s}  \right],
\end{align}
where the self-energy operator $\hat{\mathcal{R}}_{43}(i\w_p)$ takes the form 
\begin{align} \label{def-r43}
\hat{\mathcal{R}}_{43}(i\w_p) =  T^2 \! \! \sum_{i\w_q, i\w_r} &\, \Tr\left[ \hat{s} \, \hat{G}_{43}(i \w_r- i \w_q + i\w_p) \right. \nn \\ 
                                                               &\, \times \left. \hat{s} \, \hat{G}_{34}(i \w_r)\right] \,\hat{G}_{43}(i \w_q).
\end{align}
\end{subequations}
Clearly, after analytic continuation $\hat{\mathcal{R}}_{43}(z)$ has similar properties to those of $\hat{\mathcal{P}}_{43}(z)$ [Eqs.~(\ref{def-p43}) and (\ref{P-43})]. The diagrams d(ii)-d(iv) read
\begin{subequations}
\begin{align}   \label{d-ii}
& \delta\Pi_{\a\a}^{[\mathrm{d(ii)}]}(i\W_n) = \Gamma_\mathrm{s,c}^2 \, T^3 \! \! \sum_{i\w_p, i\w_q, i\w_r} \int_{\bfr_1,\bfr_2,\bfr_3,\bfr_4} \nn \\
&  \times \Tr \left[ \hat{s} \, \hat{G}_{41}(i \w_p + i \W_n) \, \hat{\s}^\a \hat{G}_{13}(i \w_p) \, \hat{s} \,  \hat{G}_{32}(i \w_q) \right. \nn \\  
& \hspace{0.8cm} \times \left. \hat{\s}^\a \, \hat{G}_{24}(i \w_q + i \W_n) \right] \nn \\ 
&  \times \Tr\left[ \hat{s} \, \hat{G}_{43}(i \w_r+ i \w_p - i \w_q) \, \hat{s} \, \hat{G}_{34}(i \w_r)\right],
\end{align}
\begin{align}   \label{d-iii}
& \delta\Pi_{\a\a}^{[\mathrm{d(iii)}]}(i\W_n) = \Gamma_\mathrm{s,c}^2 \, T^3 \!\!\!\!  \sum_{i\w_p, i\w_q, i\w_r} \int_{\bfr_1,\bfr_2,\bfr_3,\bfr_4} \nn \\
& \times \Tr \left[ \hat{s} \, \hat{G}_{31}(i \w_p + i \W_n) \, \hat{\s}^\a \, \hat{G}_{14}(i \w_p) \, \hat{s} \,\hat{G}_{43}(i \w_r)\right]  \nn \\
& \times \Tr \left[ \hat{s} \, \hat{G}_{42}(i \w_q)\hat{\s}^\a \, \hat{G}_{23}(i\w_q+i\W_n) \right. \nn \\ 
&  \hspace{0.8cm} \times \left. \hat{s} \, \hat{G}_{34}(i \w_r + i \w_q - i \w_p) \right], 
\end{align}
\begin{align}
& \delta\Pi_{\a\a}^{[\mathrm{d(iv)}]}(i\W_n) = \Gamma_\mathrm{s,c}^2 \, T^3 \!\!\!\!  \sum_{i\w_p, i\w_q, i\w_r} \int_{\bfr_1,\bfr_2,\bfr_3,\bfr_4} \nn \\
& \times \Tr \left[ \hat{G}_{31}(i \w_p + i \W_n) \, \hat{\s}^\a \hat{G}_{14}(i \w_p) \, \hat{s} \,\hat{G}_{43}(i \w_r) \, \hat{s} \right] \nn \\
& \times \Tr \left[ \hat{G}_{32}(i \w_q+i\W_n) \, \hat{\s}^\a \, \hat{G}_{24}(i\w_q) \right. \nn \\
& \hspace{0.8cm} \times \left. \hat{s} \, \hat{G}_{43}(i \w_r + i \w_q - i \w_p) \, \hat{s} \right]   \label{d-iv1} \\
& = \Gamma_\mathrm{s,c}^2 \, T^3 \!\!\!\!  \sum_{i\w_p, i\w_q, i\w_r} \int_{\bfr_1,\bfr_2,\bfr_3,\bfr_4} \nn \\
& \times \Tr \left[ \hat{G}_{31}(i \w_p ) \, \hat{\s}^\a \hat{G}_{14}(i \w_p+ i \W_n) \, \hat{s} \,\hat{G}_{43}(i \w_r) \, \hat{s} \right] \nn \\
& \times \Tr \left[ \hat{G}_{32}(i \w_q) \, \hat{\s}^\a \, \hat{G}_{24}(i\w_q+i\W_n) \right. \nn \\
& \hspace{0.8cm} \times \left. \hat{s} \, \hat{G}_{43}(i \w_r + i \w_q - i \w_p) \, \hat{s} \right].  \label{d-iv2} 
\end{align}
\end{subequations}
From Eq.~(\ref{d-iv1}) to Eq.~(\ref{d-iv2}) we have used the chirality (\ref{sym-mgf}). Applying the Ward identity (\ref{ward-1}), up to order of $\W$, we can write Eq.~(\ref{d-ii}) as
\begin{align}   \label{d-ii-c}
& \delta\Pi_{\a\a}^{[\mathrm{d(ii)}]}(i\W_n) = \Gamma_\mathrm{s,c}^2 \, T \,  \sum_{i\w_p}\int_{\bfr_2,\bfr_3,\bfr_4} \!\!\!\Tr \left\{ \hat{s} \, \hat{\mathcal{S}}_{43,\a}(i\w_p) \, \hat{s} \right. \nn \\ 
& \times \left. \left[ \hat{G}_{32}(i \w_p) \, \hat{\s}^\a \,\hat{G}_{24}(i \w_p + i \W_n) \right. \right. \nn \\
& \hspace{0.5cm} \left. \left. + \hat{G}_{32}(i \w_p+i\W_n) \, \hat{\s}^\a \,\hat{G}_{24}(i \w_q)\right] \right\} + \mathcal{O}(\W^2),
\end{align}
where
\be
\hat{\mathcal{S}}_{43,\a}(z) = -i \, (\bfr_4 - \bfr_3)^\a \, \hat{\mathcal{R}}_{43}(z).
\ee
Moreover, we have the relations
\begin{subequations} \label{rel-d}
\begin{align}
& \delta\Pi_{\a\a}^{[\mathrm{d(iii)}]}(i\W_n) = \delta\Pi_{\a\a}^{[\mathrm{d(ii)}]}(i\W_n) + \mathcal{O}(\W^2), \\ 
& \delta\Pi_{\a\a}^{[\mathrm{d(iv)}]}(i\W_n) = -\delta\Pi_{\a\a}^{[\mathrm{d(ii)}]}(i\W_n) + \mathcal{O}(\W^2).
\end{align}
\end{subequations}
Repeating the procedure for evaluating c(i) and c(ii), at zero temperature we have
\begin{align}  \label{delta-di}
& 2 \times \delta\s_\mathrm{s}^{\a\a}[\mathrm{d(i)}] = - \delta\s_\mathrm{s}^{\a\a}[\mathrm{d(ii)}] \nn \\ 
& = \frac{\Gamma_\mathrm{s,c}^2}{2\pi\mathcal{V}} \sum_{\b=1}^2 \int_{\bfr_3,\bfr_4} [(\bfr_4-\bfr_3)^\b]^2 \Tr 
\left[ \hat{s} \, \hat{\mathcal{R}}_{43}^R \, \hat{s} \, \hat{G}_{34}^R \, \right].
\end{align}
Combining Eqs.~(\ref{rel-d}) and (\ref{delta-di}) we prove that $\delta\s_\mathrm{s}^{\a\a}[2 \times \mathrm{d(i)} + \mathrm{d(ii)} + \mathrm{d(iii)}+ \mathrm{d(iv)}] =0$.


\subsubsection{Category e}

The diagrams e(i) and e(ii) read
\begin{subequations}  \label{delta-ln-e}
\begin{align}
& \delta\Pi_{\a\a}^{[\mathrm{e(i)}]}(i\W_n) =  \Gamma_\mathrm{s,c}^2 \, T^2 \sum_{i\w_p,i\w_q} \int_{\bfr_1,\bfr_2,\bfr_3,\bfr_4} \nn \\
& \times  \Tr \left[ \hat{G}_{41}(i \w_p) \, \hat{\s}^\a \, \hat{G}_{13}(i \w_p-i \W_n)  \, \hat{s} \, \hat{G}_{34}(i \w_p) \, \hat{s} \,\hat{\rho}_4 \, \hat{s} \right] \nn \\ &\, \times \Tr\left[ \hat{\s}^\a \,  \hat{G}_{23}(i \w_q + i \W_n) \, \hat{s} \,  \hat{G}_{32}(i \w_q) \right], \\
& \delta\Pi_{\a\a}^{[\mathrm{e(ii)}]}(i\W_n) =  \Gamma_\mathrm{s,c}^2 \, T^2 \sum_{i\w_p,i\w_q} \int_{\bfr_1,\bfr_2,\bfr_3,\bfr_4} \nn \\
&\times \Tr \left[  \hat{G}_{31}(i \w_p+ i \W_n) \, \hat{\s}^\a \, \hat{G}_{14}(i \w_p)  \, \hat{s} \,\hat{\rho}_
4 \, \hat{s} \, \hat{G}_{43}(i \w_p) \, \hat{s}  \right] \nn \\ &\, \times \Tr\left[ \hat{\s}^\a \,  \hat{G}_{23}(i \w_q + i \W_n) \, \hat{s} \, \hat{G}_{32}(i \w_q)   \right].
%
\end{align}
\end{subequations}
Up to order of $\W$, Eq.~(\ref{delta-ln-e}) leads to
\begin{align}
	\delta\Pi_{\a\a}^{[\mathrm{e(ii)}]}(i\W_n) = & \, - \delta\Pi_{\a\a}^{[\mathrm{e(i)}]}(i\W_n) + \mathcal{O}(\W^2). \label{delta-l2} 
\end{align}
Therefore, we prove that $\delta\s_\mathrm{s}^{\a\a}[ \mathrm{e(i)}+ \mathrm{e(ii)}] =0$.

\subsubsection{Category f}

The diagrams f(i) and f(ii) read
\begin{subequations}  \label{delta-ln-f}
\begin{align}
& \delta\Pi_{\a\a}^{[\mathrm{f(i)}]}(i\W_n) = - \Gamma_\mathrm{s,c}^2 \, T^2 \sum_{i\w_p,i\w_q} \int_{\bfr_1,\bfr_2,\bfr_3,\bfr_4} \nn \\
& \times \Tr \left[ \hat{G}_{31}(i \w_p+ i \W_n)  \, \hat{\s}^\a \, \hat{G}_{13}(i \w_p) \, \hat{s} \,  \hat{G}_{32}(i \w_q- i \W_n)  \right. \nn \\ 
& \hspace{0.8cm} \left.  \times   \hat{\s}^\a \,  \hat{G}_{24}(i \w_q) \, \hat{s} \,  \hat{\rho}_4 \, \hat{s} \,  \hat{G}_{43}(i \w_q) \hat{s} \,  \right], \\
& \delta\Pi_{\a\a}^{[\mathrm{f(ii)}]}(i\W_n) = - \Gamma_\mathrm{s,c}^2 \, T^2 \sum_{i\w_p,i\w_q} \int_{\bfr_1,\bfr_2,\bfr_3,\bfr_4} \nn \\
& \times \Tr \left[ \hat{G}_{31}(i \w_p+i \W_n)  \, \hat{\s}^\a \, \hat{G}_{13}(i \w_p) \, \hat{s} \,  \hat{G}_{34}(i \w_q) \right. \nn \\ 
& \hspace{0.8cm} \left.  \times \hat{s}\, \hat{\rho}_4 \, \hat{s} \, \hat{G}_{42}(i \w_q) \, \hat{\s}^\a\,    \hat{G}_{23}(i \w_q+ i \W_n)  \hat{s} \, \right].
\end{align}
\end{subequations}
Up to order of $\W$, Eq.~(\ref{delta-ln-f}) leads to
\begin{align}
	\delta\Pi_{\a\a}^{[\mathrm{f(ii)}]}(i\W_n) = & \,- \delta\Pi_{\a\a}^{[\mathrm{f(i)}]}(i\W_n) + \mathcal{O}(\W^2),  \label{delta-n3}
\end{align}
Therefore, we prove that $\delta\s_\mathrm{s}^{\a\a}[\mathrm{f(i)} + \mathrm{f(ii)}] =0$.

\subsection{Higher-order corrections \label{high-sec}}
\begin{figure}
\centering
\includegraphics[width=0.47\textwidth]{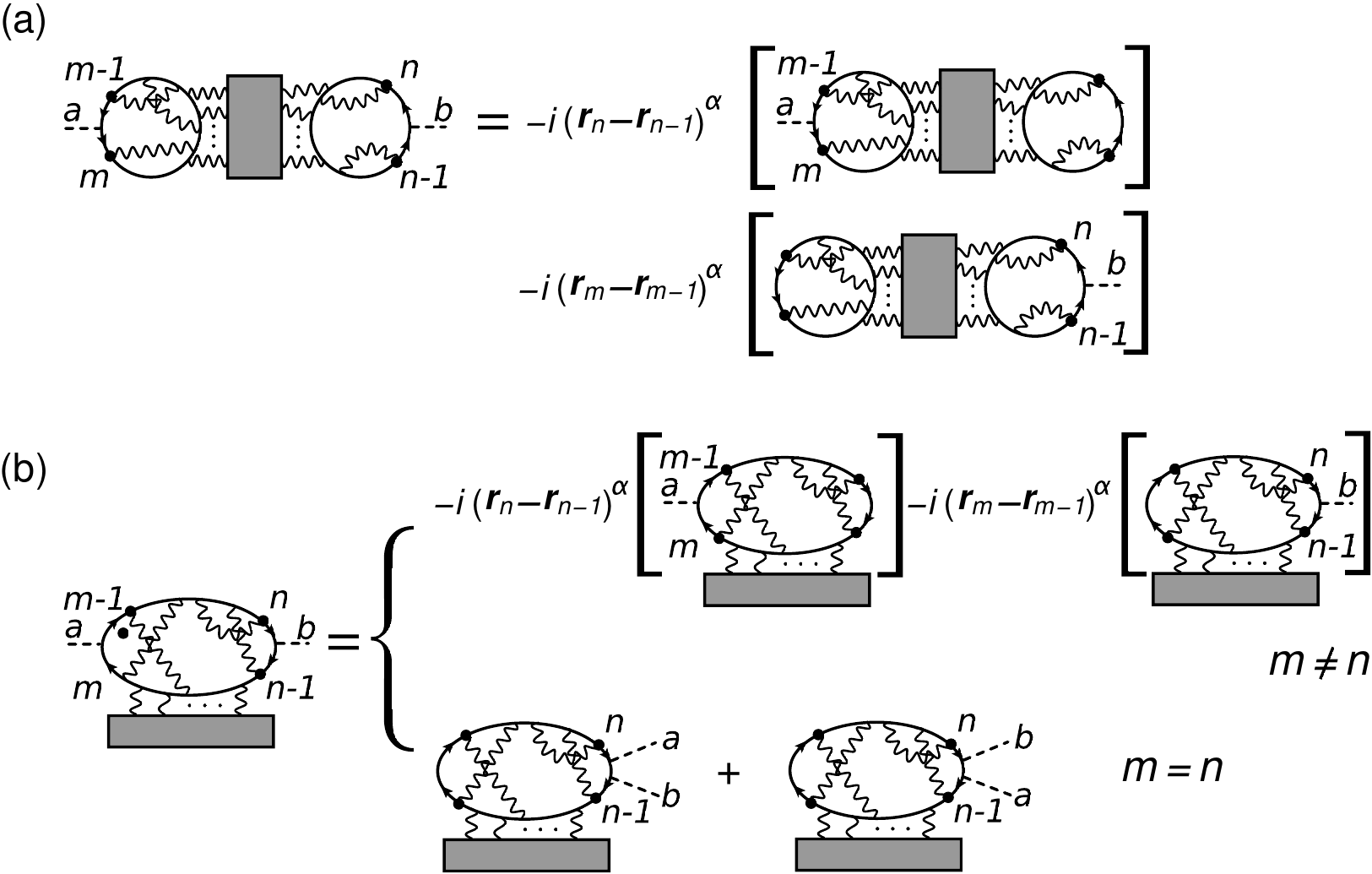} \\
\caption{High-order interaction corrections to the spin current-current correlation function in 
class AIII. 
The equations shown are valid up to order of $\W$. (a) The current-current correlation function 
$\delta\Pi_{\a\a,mn}^{(\mathrm{a})}(i\W)$, where the two current density operators $j_\a(\bfr_a)$ 
and $j_\a(\bfr_b)$ locate on different fermion loops. (b) The current-current correlation 
function $\delta\Pi_{\a\a,mn}^{(\mathrm{b})}(i\W)$, where the current density operators are 
on one fermion loop.}   \label{high-order}
\end{figure}

Regardless of details, at a fixed perturbative order in interactions, a current-current correlation diagram should take 
the form of 
one of the two expressions as shown in Fig.~\ref{high-order}. (a) The two current densities $j_\a(\bfr_a)$ and $j_\a(\bfr_b)$ are allocated on different fermion loops. For second order this corresponds to, for example, Fig.~\ref{second-order}(d)(iii). (b) The two current densities are on the same fermion loop. For second order this corresponds to, for example, Fig.~\ref{second-order}(c)(ii). Along a circular direction on the loops that support the current densities, we label the coordinates of interaction vertices as $\bfr_1,\bfr_2,\bfr_3,...$, subsequently.   
Although we treat the $\{\bfr_i\}$ as independent labels, some of these are in fact constrained in pairs by the interaction. 
This does not affect the following argument.

As shown in Fig.~\ref{high-order}(a), up to the order of $\W$, the correlation function $\delta\Pi_{\a\a,mn}^{(\mathrm{a})}(i\W)$ splits into two terms due to the Ward identity (\ref{ward-2}). For second order this decomposition corresponds to, for example, Eq.~(\ref{rel-d}). Summing over the coordinate indices $m$ and $n$, one readily has
\be  \label{all-d-a} 
\sum_{n,m} \delta\Pi_{\a\a,mn}^{(\mathrm{a})}(i\W) = \mathcal{O}(\W^2),
\ee  
since 
\be \label{loopsum} \sum_{n \in \mathrm{loop}}(\bfr_n-\bfr_{n-1})^\a = 0. \ee
Therefore, the dc conductivity corrections arising from the diagrams in Fig.~\ref{high-order}(a) vanish for all temperatures \cite{ZNA2001}.

As shown in Fig.~\ref{high-order}(b), up to order of $\W$, the correlation function $\delta\Pi_{\a\a,m n}^{(\mathrm{b})}(i\W)$ has to be analyzed in two cases,  
\be
\delta\Pi_{\a\a,m n}^{(\mathrm{b})}(i\W) = \begin{cases} \delta\Pi_{\a\a,m n}^{(\mathrm{b}1)}(i\W), \quad m \neq n, \vspace{0.3cm} \\  \delta \Pi_{\a\a,n}^{(\mathrm{b}2)}(i\W), \quad m=n, \end{cases}
\ee
where the expressions of $ \delta\Pi_{\a\a,m n}^{(\mathrm{b}1)}(i\W)$ and $\delta\Pi_{\a\a,n}^{(\mathrm{b}2)}(i\W)$ are shown in the figure respectively. Summing over the indices $m$ and $n$ we obtain
\begin{align} \label{delta-b}
& \sum_{n, m} \delta\Pi_{\a\a,m n}^{(\mathrm{b})}(i\W) = \sum_{n \neq m} \delta\Pi_{\a\a,m n}^{(\mathrm{b}1)}(i\W) + \sum_{n}\delta\Pi_{\a\a, n}^{(\mathrm{b}2)}(i\W) \nn \\
& =  \sum_{n}\left[\delta\Pi_{\a\a, n}^{(\mathrm{b}2)}(i\W) -\delta{\Pi}_{\a\a, nn}^{(\mathrm{b}1)}(i\W)\right] + \mathcal{O}(\W^2),
\end{align}
where we have used $\sum_{n, m} \delta\Pi_{\a\a,m n}^{(\mathrm{b}1)}(i\W) = 0$ due to the circular property (\ref{loopsum}). 
Defining a 
dressed propagator matrix analogous to $\hat{\mathcal{P}}_{43}(i\w_p)$ in Eq.~(\ref{ci_and_p43})
and repeating the procedure for evaluating the diagrams in Fig.~\ref{second-order}(c)(i) and~\ref{second-order}(c)(ii), one can show that at zero temperature the conductivity corrections arising from $\delta\Pi_{\a\a, n}^{(\mathrm{b}2)}(i\W)$ and $\delta{\Pi}_{\a\a, nn}^{(\mathrm{b}1)}(i\W)$ are identical for any $n$. 
The branch cut of the dressed propagator
is on the real axis for all finite orders.
Therefore, the zero-temperature conductivity corrections $\delta\s_{\mathrm{s},mn}^{\a\a}(\mathrm{b})$ cancel each other,
\be \label{delta-c}
\sum_{n,m} \delta\s_{\mathrm{s},mn}^{\a\a}(\mathrm{b}) = \mathcal{O}(\W^2).
\ee 

We point out possible caveats of our analysis above. The precondition for applying the Ward 
identity is that the diagrams as shown in the brackets in Fig.~\ref{high-order} should be finite. 
Short-ranged interactions satisfy this precondition as long as the spin- and mass-squared operators 
are irrelevant at low energies. However, for Coulomb interactions these diagrams should exhibit 
an
infrared divergence at any order (marginal) and eventually lead to a minimal dc conductivity \cite{SUP-MinCond-1,SUP-MinCond-2,SUP-MinCond-3}. 
For 
the
disorder-free and noninteracting case, the Ward identity 
must also be used with care, since it leads to an
ultraviolate divergence \cite{mirlin2006}. Moreover, the Taylor expansion based on the Ward 
identity (\ref{ward-1}) applies in the dc limit ($\W \to 0$ before $T \to 0$) but fails in 
the optical limit ($\W \to 0$ after $T \to 0$).

\section{Large winding number expansion \label{Fnlsm-sec}}

\subsection{Symmetry structure: replicated path integral \label{target-mani}}

We write an imaginary time fermion path integral for Eq.~(\ref{LET-ham-nd}) 
to encode disorder-averaged Green's functions,
\bsub\label{Schiral}
\begin{align}
	S
	\equiv&\,
	S_h 
	+
	S_{\epsilon},
	\\
	S_h
	=&\,
	\int \! \frac{d \omega_n \rmd^2 \bfr}{2 \pi}  \, 
	\left[
	\begin{aligned}
	&\,\bar{L}\left( - i \bar{\partial} + \bar{A}_j \tmfr^j + \bar{\mathcal{A}} \right) L
	\\
	&\,+
	\bar{R}\left( - i \partial + A_j \tmfr^j + \mathcal{A} \right) R
	\end{aligned}
	\right],
	\label{ShDef}
	\\
	S_{\epsilon}
	=&\,
	\int \! \frac{d \omega_n \rmd^2 \bfr}{2 \pi}  \,
	\left[
	-i \omega_n
	\left(
	\bar{L} R + \bar{R} L
	\right)
	\right],
	\label{SeDef}
\end{align}
\esub
where $L \rightarrow L_{v,a}(\omega_n,\vex{r})$ carries color $v \in \{1,2,\ldots,|\nu|\}$ and
replica $a \in \{1,2,\ldots,n\}$ indices. (Replicas are introduced to facilitate disorder averaging
\cite{SUP-ASbook}.) These are related to the fermion field in Eq.~(\ref{LET-ham-nd}) via 
\begin{align}
\begin{aligned}
	\eta_{v,a}(\omega_n,\vex{r}) 
	=&\, 
	\begin{bmatrix} 
	L_{v,a}(\omega_n,\vex{r}) \\
	R_{v,a}(\omega_n,\vex{r})
	\end{bmatrix},
	\\
	\bar{\eta}_{v,a}(\omega_n,\vex{r})
	=&\,
	\begin{bmatrix}
	\bar{R}_{v,a}(\omega_n,\vex{r}) & \bar{L}_{v,a}(\omega_n,\vex{r})
	\end{bmatrix}.
\end{aligned}
\end{align}
In Eq.~(\ref{Schiral}), we have introduced the chiral notations
\[
\begin{gathered}
	\{\partial,\bar{\partial}\} \equiv \partial_x \mp i \partial_y,
	\\
	\{A_j,\bar{A}_j\} \equiv A_{j,x} \mp i A_{j,y},
	\;\;
	\{\mathcal{A},\bar{\mathcal{A}}\} \equiv \mathcal{A}_{x} \mp i \mathcal{A}_{y}.
\end{gathered}
\]

The structure of the low-energy effective theory for the critically delocalized \cite{SRFL2008,Foster2012,FXCup,CFup}
surface Majorana fermions follows largely from symmetry analysis. As in standard localization
physics, this will be a nonlinear sigma model with a target manifold determined by
the set of transformations that preserves the action, in every fixed realization of disorder
\cite{SUP-ASbook,wm2008,foster0608,dellanna2006}. The manifold is the quotient of the 
symmetry of $S_h$ (the ``Hamiltonian piece'') relative to the symmetry of $S$ \cite{foster0608}.

We consider class AIII. Equation~(\ref{ShDef}) is invariant under independent left and right
unitary transformations
\begin{align}
	L \mapsto \hat{U}_L L,\;\; 
	\bar{L} \mapsto \bar{L} \hat{U}_L^\dagger,\;\;
	R \mapsto \hat{U}_R R,\;\; 
	\bar{R} \mapsto \bar{R} \hat{U}_R^\dagger.
\end{align}
Here $\hat{U}_{L,R}$ are $\text{U}(n N)$ transformations that act on the product of 
replica $\otimes$ Matsubara frequency labels. ($N$ is the number of Matsubara frequencies. 
Formally we must take $N \rightarrow \infty$ and $n\rightarrow 0$, such that 
$N n \rightarrow 0$.) The ``energy piece'' of the action $S_{\epsilon}$ further 
restricts $\hat{U}_R = \hat{U}_L$. Thus the target manifold is \cite{foster0608}
\begin{align}\label{AIIISym}
	\frac{\text{U}(n N) \times \text{U}(n N)}{\text{U}(n N)} 
	\simeq
	\text{U}(n N).
\end{align}

Equation~(\ref{AIIISym}) is almost sufficient to determine the form of the nonlinear sigma model.
The exact solution \cite{SUP-NTW1994,MCW96,CKT1996}
to the noninteracting, disorder-only problem in $2+0$ dimensions via non-Abelian 
bosonization shows that the standard sigma model action must be 
supplemented by a Wess-Zumino-Novikov-Witten (WZNW) term at level $K$, 
where $K = |\nu|$ for class AIII \cite{FXCup}. 
The analysis for classes CI and DIII is similar, leading to 
\begin{align}\label{Embed}
\begin{array}{lcc}
	\textrm{Class } & \textrm{Target manifold } G & \textrm{WZNW level } K\\
\hline
	\textrm{CI}   & \textrm{Sp}(2 n N) 	& |\nu|/2 \\
	\textrm{AIII} & \textrm{U}(n N)  	& |\nu| \\
	\textrm{DIII} & \textrm{O}(n N) 	& |\nu| \\
\hline
\end{array}
\end{align}

\subsection{Wess-Zumino-Novikov-Witten Finkel'stein nonlinear $\boldsymbol{\sigma}$ model} \label{Fnlsm-WZW}

The Finkel'stein nonlinear 
sigma
model with a WZNW term 
(WZNW-FNLsM)
incorporates three sectors:
\be  \label{act-sept}
	S[\htQ] = S_0 [\htQ] + S_\rmI[\htQ] + \Gamma_K[\htQ],  
\ee     
where $S_0 [\htQ]$ represents the 
standard dynamical sigma
model on 
the 
target manifold 
$G$ 
defined in Eq.~(\ref{Embed}), 
$S_\rmI[\htQ]$ 
encodes 
the 
four-fermion interactions shown in Eq.~(\ref{LET-action-int}), 
and $\Gamma_K[\htQ]$ is the WZNW term at level $K$. 

The 
WZNW term takes the unique form \cite{SUP-ASbook}
\be  \label{act-wzw}
\begin{split}
	\Gamma_K[\htQ] =& 
	-i \, K \int \! \frac{\rmd^3 \bfr}{12 \pi l_\phi} \, \epsilon^{\alpha \beta \gamma} \, 
	\\
	&\;\;\,\times\!
	\Tr \left[ (\hat{Q}^{-1} \partial_\alpha \hat{Q}) (\hat{Q}^{-1} \partial_\beta \hat{Q}) 
	(\hat{Q}^{-1} \partial_\gamma \hat{Q}) \right]\!,
\end{split}
\ee
where the spatial integral is over the 3D bulk of a superconductor that is surrounded 
by the surface we are considering, and $l_\phi$ is the Dynkin index of the corresponding 
symmetry group, 
\be  \label{dynkin}
	l_\phi = 1 \;\; (\text{classes CI, AIII}), \quad 
	l_\phi = 2 \;\; (\text{class DIII}).
\ee

The nontopological action and 
matrix field $\htQ(\bfr)$ 
target space 
distinguish the three universality classes.  \\   
(i) \emph{Class CI.} 
$S_0 [\htQ]$ and $S_\rmI [\htQ]$ take the forms \cite{dellanna2006}
\begin{subequations}  \label{action-normal}
\begin{align}
	S_{0} [\htQ] 
	= & \frac{1}{2\lambda}\int_{\mathbf{r}} \text{Tr}\left[ \nabla \hat{Q}^\dagger(\mathbf{r}) \cdot \nabla \hat{Q}(\mathbf{r}) \right] 
	\nn \\ 
          & - h \int_{\mathbf{r}} 
	\text{Tr}\left( |\hat{\omega}| \ots \hat{\Sigma}^3 
	\left[ 
	i \hat{Q}(\mathbf{r}) - i \hat{Q}^\dagger(\mathbf{r})
	\right] \right),  \label{ci-nonint}
	\\
	S_\rmI [\htQ] 
	= 
	& \sum_a\!\int_{\bfr, \tau} 
	\left[ 
		\frac{\Gamma_\rms}{2} 
                \left( \Tr_\mu 
		\left\{ \hat{\boldsymbol{\mu}} 
		\left[ \htQ_{\tau \, \tau}^{a \, a}(\bfr) 
		- 
		(\htdQ)_{\tau \, \tau}^{a \, a}(\bfr) \right] 
		\right\} \right)^2
	\right. 
	\nn \\ 
           & 
	\left. 
		+  
		\frac{\Gamma_\rmc}{2}
                \left\{ \Tr_\mu 
		\left[ \htQ_{\tau \, \tau}^{a \, a}(\bfr) 
		+
		(\htdQ)_{\tau \, \tau}^{a \, a}(\bfr)\right] 
		\right\}^2
	\right]. 
	\label{ci-int}
\end{align}
\end{subequations}
In Eq.~(\ref{action-normal}), 
we introduce two sets of Pauli matrices: 
$\hat{\boldsymbol{\mu}}=(\hmu^1,\hmu^2,\hmu^3)$ 
act on physical spin space, while
$(\hat{\Sigma}^1,\hat{\Sigma}^2,\hat{\Sigma}^3)$ act on the sign of the Matsubara frequency;   
i.e., $\bra{\omega_n} \hat{\Sigma}^3 \ket{\omega_m} = \delta_{m,n} \sgn(\omega_m)$. 
$\hat{Q}(\mathbf{r})$ denotes a
square matrix taking indices in replica space 
with $a,b \in \{1,2,\cdots, n\}$, physical spin space with $\mu,\mu^\prime \in \{\uaw, \daw \}$, 
and the imaginary time $\tau, \tau^\prime$ or modulus Matsubara frequencies $|\omega|$, $|\omega^\prime|$ and their sign 
$\Sigma$, $\Sigma^\prime$ spaces:
\be
	\hat{Q}(\bfr) \to 
	\begin{cases} 
		Q^{\mu a, \mu^\prime b}_{\tau, \tau^\prime }(\bfr) , & \mbox{Temporal basis}, \\ 
		Q^{\mu a, \mu^\prime b}_{\Sigma |\omega|, \Sigma^\prime |\omega^\prime| }(\bfr) , & \mbox{Frequency basis}. 
	\end{cases}
\ee
The matrix field $\hat{Q}(\bfr)$ belongs to Sp($2nN$) group, which
satisfies the unitary condition 
\begin{subequations}
\be  \label{u-const}
	\hat{Q}^\dagger(\bfr)  \hat{Q}(\bfr) =\hat{1},
\ee
and 
in the temporal basis
the symplectic condition
\be  \label{sp-const}
	\hmu^2 \, \htQ^\trasp(\bfr)  \, \hmu^2 = \htQ^\dagger(\bfr).
\ee
\end{subequations}
In a perturbative expansion, the physical saddle point is set by the frequency term in Eq.~(\ref{ci-nonint}) and 
is given by \cite{dellanna2006,foster0608}
\be  \label{sp-ci}
	\htQ_{\text{sp}} (\bfr) 
	= 
	 - i \hat{\Sigma}^3.
\ee
(ii) \emph{Class AIII.} 
$S_0 [\htQ]$ and $S_\rmI [\htQ]$ take the forms \cite{foster0608}
\begin{subequations}  \label{fnlsm-aiii-c}
\begin{align}
	S_{0} [\htQ] = 
	& \,\frac{1}{2\lambda}\int_{\mathbf{r}} 
	\text{Tr}\left[ \nabla \hat{Q}^\dagger(\mathbf{r}) \cdot \nabla \hat{Q}(\mathbf{r}) \right] 
	\nn \\
        &\, 
	-
	\frac{\lambda_\rmA}{ 2 \lambda^2} \int_{\mathbf{r}} 
	\text{Tr}\left[ \hat{Q}^\dagger(\mathbf{r}) \nabla \hat{Q}(\mathbf{r})\right] \cdot \text{Tr}\left[ \hat{Q}^\dagger(\mathbf{r}) \nabla \hat{Q}(\mathbf{r})\right] 
	\nn \\
        & \, 
	- h \int_{\mathbf{r}} \text{Tr}\left( |\hat{\omega}| \ots \hat{\Sigma}^3 
	\left[i \hat{Q}(\mathbf{r}) - i \hat{Q}^\dagger(\mathbf{r}) \right] 
	\right), 
	\label{aiii-nonint}
	\\
	S_\rmI [\htQ] = 
	& \, \sum_a\!\int_{\bfr, \tau} \left\{ 
	\Gamma_\rms 
	\left[ \htQ_{\tau \, \tau}^{a \, a}(\bfr) 
	- 
	(\htdQ)_{\tau \, \tau}^{a \, a}(\bfr) \right]^2 \right. 
	\nn \\ 
        & \, \left. 
	+ 
	\Gamma_\rmc 
	\left[ \htQ_{\tau \, \tau}^{a \, a}(\bfr) 
	+ 
	(\htdQ)_{\tau \, \tau}^{a \, a}(\bfr) \right]^2 \right\}. 
	\label{aiii-int}
\end{align}
\end{subequations}
In Eq.~(\ref{fnlsm-aiii-c}), 
$\hat{Q}(\mathbf{r})$ 
takes indices in replica space, 
and imaginary time 
or 
Matsubara frequency 
space:
\be  \label{aiii-Q}
	\hat{Q}(\bfr) \to 
	\begin{cases}  
	Q^{ a, b}_{\tau, \tau^\prime }(\bfr), & \mbox{Temporal basis}, \\ 
	Q^{a,  b}_{\Sigma |\omega|, \Sigma^\prime |\omega^\prime| }(\bfr), & \mbox{Frequency basis}. 
	\end{cases}
\ee
The matrix field $\hat{Q}(\bfr) \in \text{U}(nN)$ satisfies only the unitary constraint (\ref{u-const}). 
The saddle point of the sigma model still takes the form of Eq.~(\ref{sp-ci}). 

The 
``Gade '' 
term~\cite{SUP-Gade9193} in the second line of Eq.~(\ref{aiii-nonint}) 
is special to class AIII. It is proportional to the disorder variance $\lambda_A$ of the 
Abelian vector potential in Eq.~(\ref{LET-ham-nd}), taken to be Gaussian white noise correlated:
\begin{align}
	\overline{\mathcal{A}_\a(\bfr) \mathcal{A}_\b(\bfr^\prime)} & =  \lambda_A  \del_{\a\b} \, \del^{(2)}(\bfr - \bfr^\prime). 
	\label{ab-dis}
\end{align}	
(iii) \emph{Class DIII.} In this class the form of $S_0 [\htQ]$ is the same as that of 
Eq.~(\ref{ci-nonint}), 
where the matrix field $\hat{Q}(\mathbf{r})$ possesses the same indices as those for class AIII [see Eq.~(\ref{aiii-Q})]. 
Because physical spin is not a conserved quantity any longer, $S_\rmI [\htQ]$ only incorporates the Cooper interaction channel:
\be
	S_\rmI [\htQ] = 
	\sum_a\!\int_{\bfr, \tau} \Gamma_\rmc 
	\left[ \htQ_{\tau \, \tau}^{a \, a}(\bfr) 
	+ 
	(\htdQ)_{\tau \, \tau}^{a \, a}(\bfr) \right]^2. \label{diii-int}
\ee
In the temporal basis,
the matrix field $\hat{Q}(\bfr) \in \text{O}(nN)$ satisfies the orthogonal condition 
\be  \label{o-const}
	\htQ^\ast(\bfr)=\htQ(\bfr), \quad \htQ^\trasp(\bfr)\htQ(\bfr)=\hat{1},
\ee
and the saddle point 
is given by 
Eq.~(\ref{sp-ci}).

\subsection{One-loop renormalization group analysis}

In the limit of 
large topological winding numbers $K \gg 1$, the 
WZNW-FNLsMs are amenable 
to a perturbative RG analysis with $1/K$ as the small parameter. In this section, we perform a 
one-loop RG calculation via the background field method, 
as employed in 
Ref.~\cite{foster0608}. 

We shift the saddle point 
in Eq.~(\ref{sp-ci}) to 
the
identity:
\be  \label{sp-shift}
	\hat{Q}(\bfr) \to -i \hat{\Sigma}^3 \, \hat{Q} (\bfr), 
	\quad 
	\hat{Q}^\dagger(\bfr) \to \hat{Q}^\dagger(\bfr) \, i \hat{\Sigma}^3,
	\quad
	\hat{Q}_{\text{sp}}(\bfr) \to \hat{1}.
\ee
With respect to the transformation (\ref{sp-shift}), the action of the 
sigma model does not change except for the frequency and interaction sectors: The frequency sector transforms as
\be 
	\int_{\mathbf{r}} \text{Tr}
	\left[ |\hat{\omega}| \ots \hat{\Sigma}^3 
	\left(i \hat{Q} - i \hat{Q}^\dagger \right) 
	\right]
	\to  \int_{\mathbf{r}} \text{Tr}\left[ |\hat{\omega}| \left( \hat{Q}^\dagger + \hat{Q} \right) \right].
\ee
In frequency space, the interaction sector transforms as
\be\label{act-int-c1-f}
\begin{aligned} 
	S_\rmI [\htQ] 
	\to & 
	- \sum_a\! 
	\intl{\substack{\omega_1,\omega_2 \\ \omega_3,\omega_4}}
	\int_{\mathbf{r}} \delta_{1+3,2+4} 
	\\
	&
	\times
	\Big\{ 
	\frac{\Gamma_\rms}{2} \, 
		\begin{aligned}[t]	
		&\Tr_\mu \left[ \hat{\boldsymbol{\mu}} \left( s_1 \, \htQ_{1,2}^{a,a} + s_2 \left.\htdQ \right._{1,2}^{a,a}\right) \right] 
		\\
		&
		\cdot 
		\Tr_\mu \left[ \hat{\boldsymbol{\mu}} \left( s_3 \, \htQ_{3,4}^{a,a} + s_4 \left.\htdQ \right._{3,4}^{a,a}\right) \right]   
		\end{aligned}
	\\ 
	& 
	\phantom{\Big\{}
	+
	\frac{\Gamma_\rmc}{2} \, 
		\begin{aligned}[t]
		&\Tr_\mu \left[ \left( s_1 \, \htQ_{1,2}^{a,a} - s_2 \left.\htdQ \right._{1,2}^{a,a}\right) \right]
		\\
		&
		\times
		\Tr_\mu \left[ \left( s_3 \,\htQ_{3,4}^{a,a} - s_4 \left.\htdQ \right._{3,4}^{a,a}\right) \right] 
		\Big\},
		\end{aligned}\!\!
\end{aligned}
\ee 
in class CI, and similarly in classes AIII and DIII. The abbreviated symbols in Eq.~(\ref{act-int-c1-f}) are defined as 
\begin{align}
\begin{aligned}
	& \hat{Q}_{1,2}^{a,b} \equiv \hat{Q}_{\omega_1,\omega_2}^{a,b}, \\
	& \delta_{1+3,2+4}  \equiv 
	2 \pi \delta(\omega_1 + \omega_3 - \omega_2 - \omega_4),
	\\
	& 
	\intl{\omega_1,\omega_2,\ldots} \equiv \int \frac{d \omega_1}{2 \pi} \int \frac{d \omega_2}{2 \pi} \times \cdots, 
	\\
	& s_\a \equiv \text{sgn}(\omega_\a),\quad \a \in \{1, 2,3,4\}.
\end{aligned}
\end{align}

We 
split 
$\htQ$ into ``fast'' $\htQ_\sfF$ and ``slow'' $\htQ_\sfS$ modes,
\be \label{fssept} 
	\hat{Q}(\mathbf{r}) = \hat{Q}_{\sfF}(\mathbf{r}) \, \hat{Q}_{\sfS}(\mathbf{r}) 
	= 
	\hat{Q}_{\sfF}(\mathbf{r}) \left[\hat{1} + \delta{\hat{Q}_{\sfS}(\mathbf{r})}\right], 
\ee   
where $\hat{Q}_{\sfF,\sfS}(\mathbf{r})$ 
belong to the same symmetry group 
as 
$\hat{Q}(\mathbf{r})$. 
We further decompose the slow field into the homogeneous saddle point 
``$\hat{1}$'' plus a slow variation $\del\hat{Q}_{\sfS}(\mathbf{r})$. 
The fast field $\hat{Q}_\sfF(\mathbf{r})$ 
will
be parameterized by unconstrained coordinates 
$\hat{Y}(\mathbf{r})$. 
The slow mode fluctuation 
$\del\hat{Q}_{\sfS}(\bk) \to \del\hat{Q}_{\sfS \w,\w^\prime}^{a,a^\prime}(\bk)$ 
possesses support within a cube of linear size $\td{\Lambda}$ in the 
space $\left( |\w|, \, |\w^\prime|, \, D\bk^2 \right)$ \cite{foster0608},
where 
\begin{align}\label{def-difconst}
	D = 1/(\lambda h) 
\end{align}
is the heat diffusion constant.
The fast mode coordinates $\hat{Y}(\bk) \to \htY_{\w,\w^\prime}^{a,a^\prime}(\bk)$ 
lie within
a surrounding shell of thickness $\Lambda - \td{\Lambda}$. Here	
$\Lambda/\td{\Lambda} \approx 1 + 2 \rmd \ell$ is a ratio of energy cutoffs 
with 
$0 < \rmd \ell \ll 1$.

Via Eq.~(\ref{fssept}) the action 
is 
\be  \label{fs-sep}
	S[\htQ] = S_\sfS[\hat{Q}_\sfS] + S_\sfF[\htQ_\sfF] + S_{\sfS/\sfF}[ \del\hat{Q}_\sfS, \htQ_\sfF],
\ee
where
\be
	S_\sfS[\hat{Q}_{\sfS}] = S[\hat{Q} \to \hat{Q}_\sfS].
\ee

The 
topological number $K$ enters into the renormalization 
equations only for the disorder parameters $\l$ and $\l_\rmA$ 
(the latter only for class AIII), 
because the WZNW term modifies only the 
``stiffness''
vertex.
The interaction parameters $\Gamma_\rms$ (classes CI and AIII) and $\Gamma_\rmc$ obey the same RG 
equations as those for the 
FNLsM
lacking 
the
WZNW term in the corresponding symmetry class \cite{foster0608,dellanna2006}. 
In the remainder, we only present the components of $S_{\sfS/\sfF}[ \del\hat{Q}_\sfS, \htQ_\sfF]$ 
that renormalize the disorder parameters $\l$ and $\l_\rmA$.

\subsubsection{Spin U(1) symmetry: Class AIII}  \label{fnlsm-aiii}

(i) \emph{Parametrization and Feynman rules.} We parametrize the fast field $\hat{Q}_\sfF(\bfr)$ by 
\be \label{param-unitary}
\begin{split}
\hat{Q}_\sfF(\mathbf{r}) = & \, \exp{[i \hat{Y}(\mathbf{r})]}  \\ 
                         = & \, \hat{1} + i \, \hat{Y}(\mathbf{r}) - \frac{1}{2}\hat{Y}^2(\mathbf{r}) + \mathcal{O}[\| \hat{Y} \|^3], 
\end{split}
\ee 
where $\hat{Y}$ is a Hermitian matrix belonging to the unitary Lie algebra $\mathfrak{u}(nN)$:
\be
\htY^\dagger = \htY.
\ee 

Substituting Eq.~(\ref{fssept}) together with Eq.~(\ref{param-unitary}) into the action described 
by Eqs.~(\ref{act-sept}), (\ref{act-wzw}), and (\ref{fnlsm-aiii-c}), and retaining up to 
quadratic terms in the fast mode coordinates $\hat{Y}(\mathbf{r})$, we obtain the fast mode action: 
\be
S_\sfF [\hat{Y}] = S_\sfF^{(0)}[\hat{Y}] + S_\sfF^{(\rmI)}[\hat{Y}],
\ee
where 
\begin{subequations}  \label{act-fast-a3}
\begin{align}
	S_\sfF^{(0)}[\hat{Y}] 
	=\,& 
	\frac{1}{2}\int_{\mathbf{r}} \text{Tr}\left[ \frac{1}{\lambda} (\nabla \hat{Y})^2 + h \, |\hat{\omega}| \, \hat{Y}^2 + h \, \hat{Y} \, |\hat{\omega}| \, \hat{Y} \right] 
	\nn \\
        & + \frac{\lambda_A}{2\lambda^2} \int_{\mathbf{r}} \text{Tr} (\nabla \hat{Y}) \cdot \text{Tr} (\nabla \hat{Y}), 
	\label{sfint} \\
	S_\sfF^{(\rmI)}[\hat{Y}] = \, & \sum_a 
	\intl{\substack{\omega_1,\omega_2 \\ \omega_3,\omega_4}}		
	\int_{\mathbf{r}} \delta_{1+3,2+4} \left[ \Gamma_\rms \,(s_1-s_2) \, (s_3-s_4) \right. 
	\nn \\ 
	& \left. + \Gamma_\rmc \, (s_1+s_2)\,(s_3+s_4) \right] \, Y_{1,2}^{a,a} \, Y_{3,4}^{a,a}. \label{sfint-I}
\end{align}
\end{subequations}
with $S_\sfF^{(0)}$ and $S_\sfF^{(\rmI)}$ arising from Eqs.~(\ref{aiii-nonint}) and (\ref{aiii-int}), respectively. 

\begin{figure}
\centering
\includegraphics[width=0.42\textwidth]{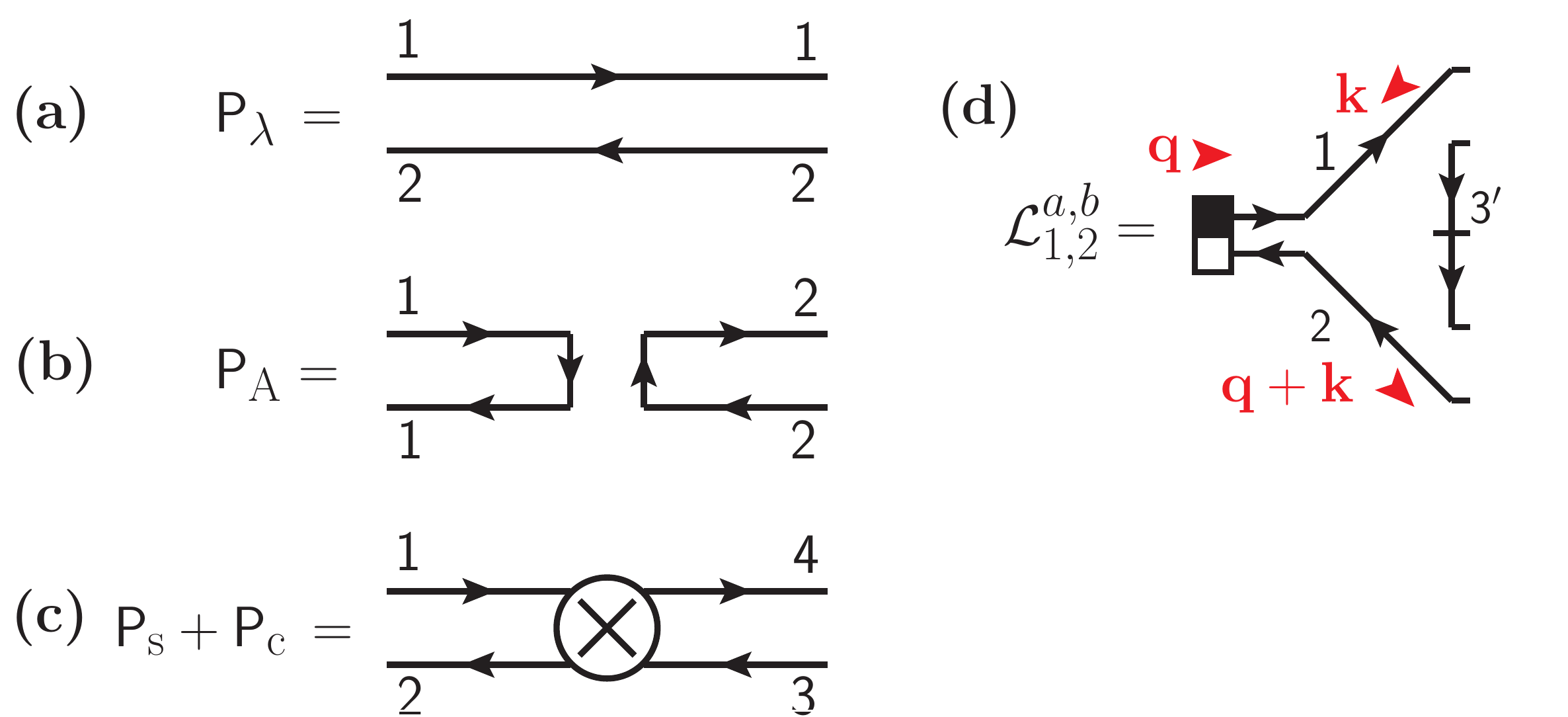}
\caption{Feynman rules for 
class AIII.
(a)--(c) Fast field propagators taking the corresponding expressions in Eq.~(\ref{fast-props}). 
(d) Stiffness vertex arising from the coupling between fast and slow fields represented by Eq.~(\ref{stifnes-a3}). 
The numeric labels appearing at the 
terminals of fermion lines encode replica and frequency indices of fast field $\htY$. For the 
vertex in panel (d) the index ``$3^\prime$'' 
can carry a
fast frequency, while ``$1$'' and ``$2$'' 
carry
only the slow ones.}\label{fr-a3}
\end{figure}

The fast field propagator decomposes into \cite{foster0608},
\begin{subequations}  \label{fast-props}
\begin{align}
	\Big\langle Y_{1,2}^{a,b}&(-\mathbf{k}) \, Y_{3,4}^{c,d} (\mathbf{k}) \Big\rangle 
	= 
	\mathsf{P}_\lambda + \mathsf{P}_\rmA + \mathsf{P}_\rms + \mathsf{P}_\rmc, \label{fast-prop}
	\\
	\mathsf{P}_\lambda = & \, \Delta_\rmO (|\omega_1|, |\omega_2|, \mathbf{k}) \delta_{1,4} \delta_{2,3} \delta_{ad} \delta_{bc}, \label{a3-plam} \\
	\mathsf{P}_\rmA  = & \, -\frac{\lambda_\rmA}{\lambda^2} \mathbf{k}^2 
				\Delta_\rmO (|\omega_1|, |\omega_1|, \mathbf{k}) \, 
				\Delta_\rmO (|\omega_3|, |\omega_3|, \mathbf{k}) \nn \\ 
			& \, \phantom{-} \times 
				\delta_{1,2} \delta_{3,4} \delta_{ab} \delta_{cd}, \label{a3-lama} \\
	\mathsf{P}_\rms = & \, - 2 \, \Gamma_\rms (s_1-s_2)(s_3-s_4)  \delta_{1+3,2+4} \delta_{ab} \delta_{bc} \delta_{cd} \nn \\ 
			& \, \phantom{-} \times 
				\Delta_\rmO (|\omega_1-\omega_2|, 0, \mathbf{k}) \, 
				\Delta_\rmS (|\omega_1-\omega_2|, \mathbf{k}), \label{ps-a3}\\
	\mathsf{P}_\rmc = & \, - 2 \, \Gamma_\rmc (s_1+s_2)(s_3+s_4) \delta_{1+3,2+4} \delta_{ab} \delta_{bc} \delta_{cd} \nn \\ 
			& \, \phantom{-} \times 
				\frac{\Delta_\rmO (|\omega_1|,|\omega_2|, \mathbf{k}) \, 
				\Delta_\rmO (|\omega_3|,|\omega_4|, \mathbf{k})}{1+\gamma_\rmc f(|\w_1-\w_2|,\bk,\Lambda)}, \label{pc-a3}
\end{align}
\end{subequations}
where
\bsub
\begin{align}	
	\Delta_\rmO (|\omega_1|, |\omega_2|, \mathbf{k}) 
	=&\, 
	\frac{1}{h} \frac{1}{D \mathbf{k}^2 + (|\omega_1| + |\omega_2|)}, 
	\label{delta-o}
	\\
	\Delta_\rmS (|\omega|, \mathbf{k}) 
	=&\, 
	\frac{1}{h}\frac{1}{D \mathbf{k}^2 + (1-\gamma_\rms)|\omega|}. 
	\label{delta-s}
\end{align}
\esub
In Eqs.~(\ref{delta-s}) and (\ref{pc-a3}) and the following, $\gamma_{\rms,\rmc}$ are defined by 
\begin{align}\label{gammaReDefs}
	\gamma_{\rms,\rmc} \equiv \frac{4 \Gamma_{\rms,\rmc}}{\pi h}.
\end{align}
The function $f(|\w|,\bk,\Lambda)$ appearing in $\sfP_\rmc$ reads 
\be  \label{def-f}
f(|\w|,\bk,\Lambda) = \ln \left( \frac{2 \Lambda}{D \bk^2 + |\w|} \right),
\ee 
with $\Lambda$ being the hard cutoff in frequency-momentum space \cite{foster0608}. 
We note that in Eq.~(\ref{pc-a3}) and in the following, the Cooper channel propagator 
$\sfP_\rmc$ (and $\td{\sfP}_\rmc$ for class CI) is evaluated up to 
logarithmic
accuracy in the cutoff $\Lambda$.  
The propagators are depicted in Fig.~\ref{fr-a3}.

The fast and slow modes are coupled by the term $S_{\sfS/\sfF}[ \del\hat{Q}_\sfS, \htQ_\sfF]$ in 
Eq.~(\ref{fs-sep}). Equations~(\ref{act-wzw}) and (\ref{aiii-nonint}) give the stiffness vertex shown in Fig.~\ref{fr-a3}(d),
\begin{subequations}  \label{stifnes-a3}  
\be  \begin{split}
	S_{\sfS/\sfF}^{(0)}
	= & 
	\frac{1}{2\lambda}\int_\mathbf{r} 
	\left( \delta^{\alpha\beta} 
	+  
	\frac{i \l K}{4 \pi l_\phi} \ep^{\alpha \beta} \right) 
	\text{Tr} \left\{ \hat{Q}_\sfS \left(\partial_\alpha \hat{Q}_\sfS^{\dagger}\right) \left[\hat{Y}, \partial_\beta \hat{Y} \right]  \right\} \\
	= & 
	\int \frac{d^2\mathbf{k}}{(2\pi)^2} \int \frac{d^2\mathbf{q}}{(2\pi)^2} \text{Tr} \left[ \hat{Y}(-\mathbf{k}-\mathbf{q}) \, \hat{\mathcal{L}} (\mathbf{k}, \mathbf{q}) \, \hat{Y}(\mathbf{k}) \right],
  \end{split}\ee
where
\be \label{stif-vert-mat}
    \hat{\mathcal{L}} (\mathbf{k}, \mathbf{q}) 
     = 
	\frac{i}{2 \lambda} \left[  (2 \mathbf{k}+\mathbf{q}) 
	- 
	\frac{ i \lambda K}{4 \pi l_\phi} (2 \mathbf{k}+\mathbf{q}) \times \mathbf{e}_z \right]   \cdot \hat{\mathbf{L}}(\mathbf{q}), 
\ee
with $\hat{Y}(\mathbf{k})$ and $\hat{\mathbf{L}}(\mathbf{q})$ being the Fourier transforms of the fast field $\hat{Y}(\bfr)$ and the vector operator 
 \be \hat{\mathbf{L}}(\mathbf{r}) = \hat{Q}_\sfS(\mathbf{r}) \nabla \hat{Q}_\sfS^\dagger(\mathbf{r}), \ee
\end{subequations}
respectively. 
The stiffness vertex in classes DIII or CI takes a similar form as in Eq.~(\ref{stifnes-a3}). 
We only need
this
vertex to derive the one-loop renormalization equations 
for the spin resistance $\lambda$ and the Gade parameter $\l_\rmA$. A full 
list of vertices coupling fast and slow modes 
(in 
the
Keldysh formalism) can 
be found in Ref.~\cite{foster0608}.

\begin{figure}[b]
\centering
\includegraphics[width=0.42\textwidth]{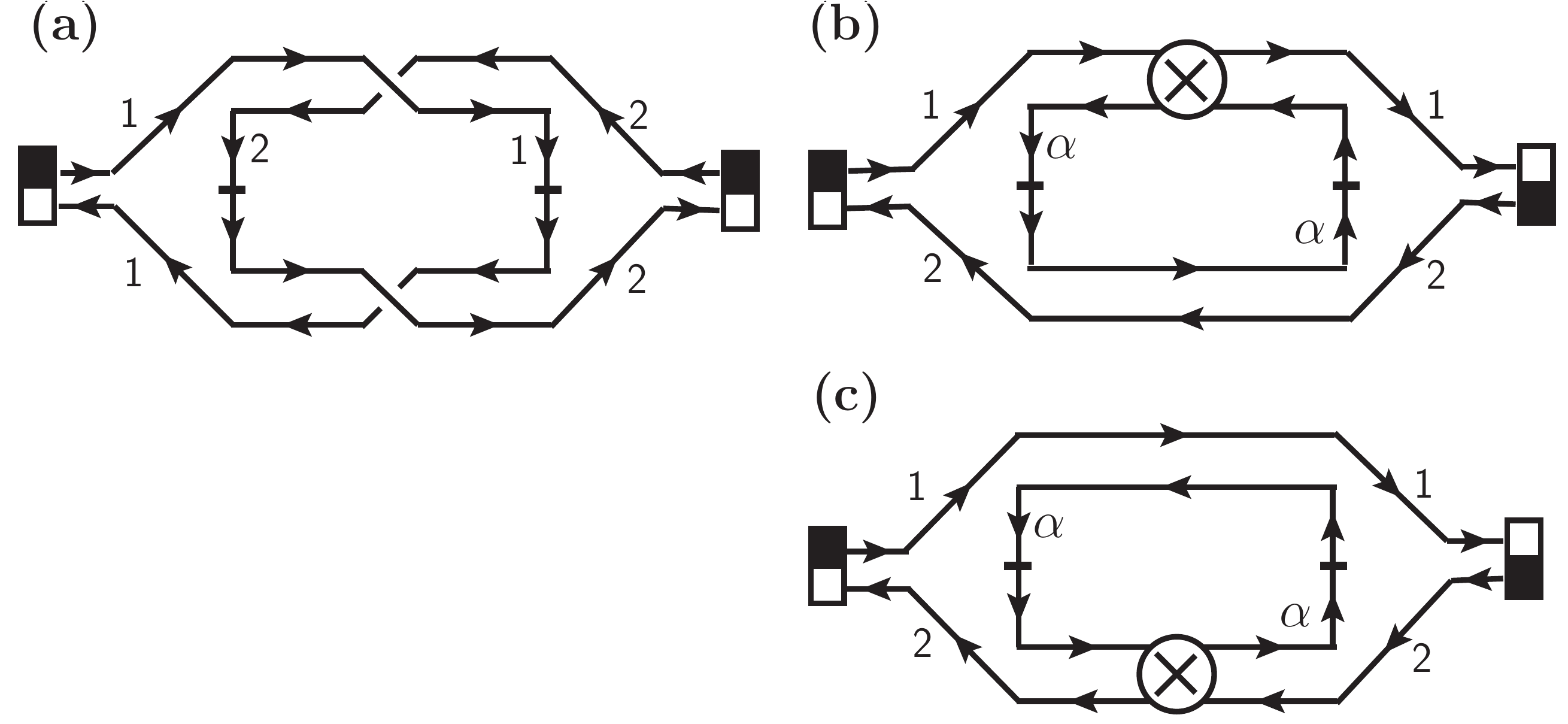}
\caption{Category $\mfrD_\text{AIII}$: Diagrams renormalizing $\l$ and $\l_\rmA$ for class AIII in 
the
replica limit $n \to 0$. 
$\mfrD_\text{AIII}$(a) renormalizes $\l_\rmA$, and $\mfrD_\text{AIII}$(b) and $\mfrD_\text{AIII}$(c) renormalize $\l$. }\label{AA-a3}
\end{figure}

(ii) \emph{Renormalization of $\lambda$ and $\l_\rmA$.} 
Diagram $\mfrD_\text{AIII}$(a) appearing in Fig.~\ref{AA-a3} renormalizes 
$\l_\rmA$,
\be  \label{rg-lamA-a3}
	\mfrD_\text{AIII}(\text{a})  
	=
	F
	\int_{\mathbf{r}} [ \text{Tr}( \hat{Q}_\sfS \nabla \hat{Q}_\sfS^\dagger) ]^2,
\ee
where 
\be\label{WZNWF}
	F 
	\equiv 
	\left( \frac{\rmd \ell}{8 \pi} \right) 
	\bigg[1- \left( \frac{\lambda K}{4 \pi l_\phi} \right)^2 \bigg].
\ee
Diagrams $\mfrD_\text{AIII}$(b) and $\mfrD_\text{AIII}$(c) in Fig.~\ref{AA-a3} renormalize the spin resistance $\lambda$,
\be  \label{rg-lam-a3}
\begin{split}
	& \mfrD_\text{AIII} \text{(b)} + \mfrD_\text{AIII}\text{(c)} \\ 
	& = 
	F 
	\left\{2\left[1+ \frac{1-\gamma_\rms}{\gamma_\rms} \ln(1-\gamma_\rms)\right] -\frac{1}{2}\mathcal{K}(\gamma_\rmc)  \right\} \\ 
	& \quad \times \int_{\mathbf{r}} \text{Tr}( \nabla \hat{Q}_\sfS^\dagger \cdot \nabla \hat{Q}_\sfS),
\end{split}
\ee 
where 
\be\label{KCooperDef}
	\mathcal{K}(\gamma_\rmc) = 2 e^{-1 / \gmc}\left[E_i\left(\frac{1}{\gmc}+ \ln 2\right) - E_i\left(\frac{1}{\gmc}\right) \right]
\ee
with $E_i(z)$ being the exponential integral. 
 
(iii) \emph{Full one-loop RG equations.} 
Substituting 
$l_\phi =1$ 
[Eq.~(\ref{dynkin})] into Eqs.~(\ref{rg-lamA-a3}) and (\ref{rg-lam-a3}) 
and performing a trivial rescaling, 
\be \label{rescale-a3} 
	\l \to 4\pi\l, \quad \la \to 4 \pi \la, 
\ee
we obtain the one-loop RG equations for 
class AIII: 
\begin{subequations}  \label{RG-eqs-AIII}
\begin{align} 
\frac{\rmd \l}{\rmd \ell} = & \, \l^2 \left[1-(K\l)^2\right] \mti(\gms,\gmc), \label{eq-lambda} \\
\frac{\rmd \la}{\rmd \ell} = & \, \l^2 \left[ 1 - (K\l)^2 \right] \left[ 1 + \frac{2 \la}{\l} \mti(\gms,\gmc)\right], \label{eq-la} \\
\frac{\rmd \gms}{\rmd \ell} = & \, (1-\gms)  \left[ \la (\gms + 2\gmc - 2 \gms\gmc)  \right. \nn \\ 
                              & \left. - \l (\gms + \gmc -2 \gms\gmc) \right], \label{eq-gms} \\
\frac{\rmd \gmc}{\rmd \ell} = & \, \la (2 \gms+\gmc) -\l (\gms+\gmc) \nn\\ 
   & + \l \left[ 2 \gmc \ln{(1-\gms)} + \gms \gmc  \right] - 2 \gmc^2,  \label{eq-gmc}
\end{align}
\end{subequations}
where
\be
\mti(\gms,\gmc) = 2 \left[ 1 + \left( \frac{1-\gms}{\gms} \right) \ln{(1-\gms)}  \right]-\frac{1}{2}\mathcal{K}(\gamma_\rmc) ,
\ee
is  the interaction correction to the dc conductivity. 
The RG equations for $\gamma_\rms$ and $\gamma_\rmc$ in Eqs.~(\ref{eq-gms}) and (\ref{eq-gmc}) have been obtained in Ref.~\cite{foster0608}
and 
are
computed only to the lowest nontrivial order in $\lambda$ and $\gamma_\rmc$.
The implications of Eq.~(\ref{RG-eqs-AIII}) for the stability of class AIII surface states
were discussed in Ref.~\cite{FXCup}.

\subsubsection{No spin symmetry: Class DIII}  \label{fnlsm-diii}

\begin{figure}[b]
\centering
\includegraphics[width=0.42\textwidth]{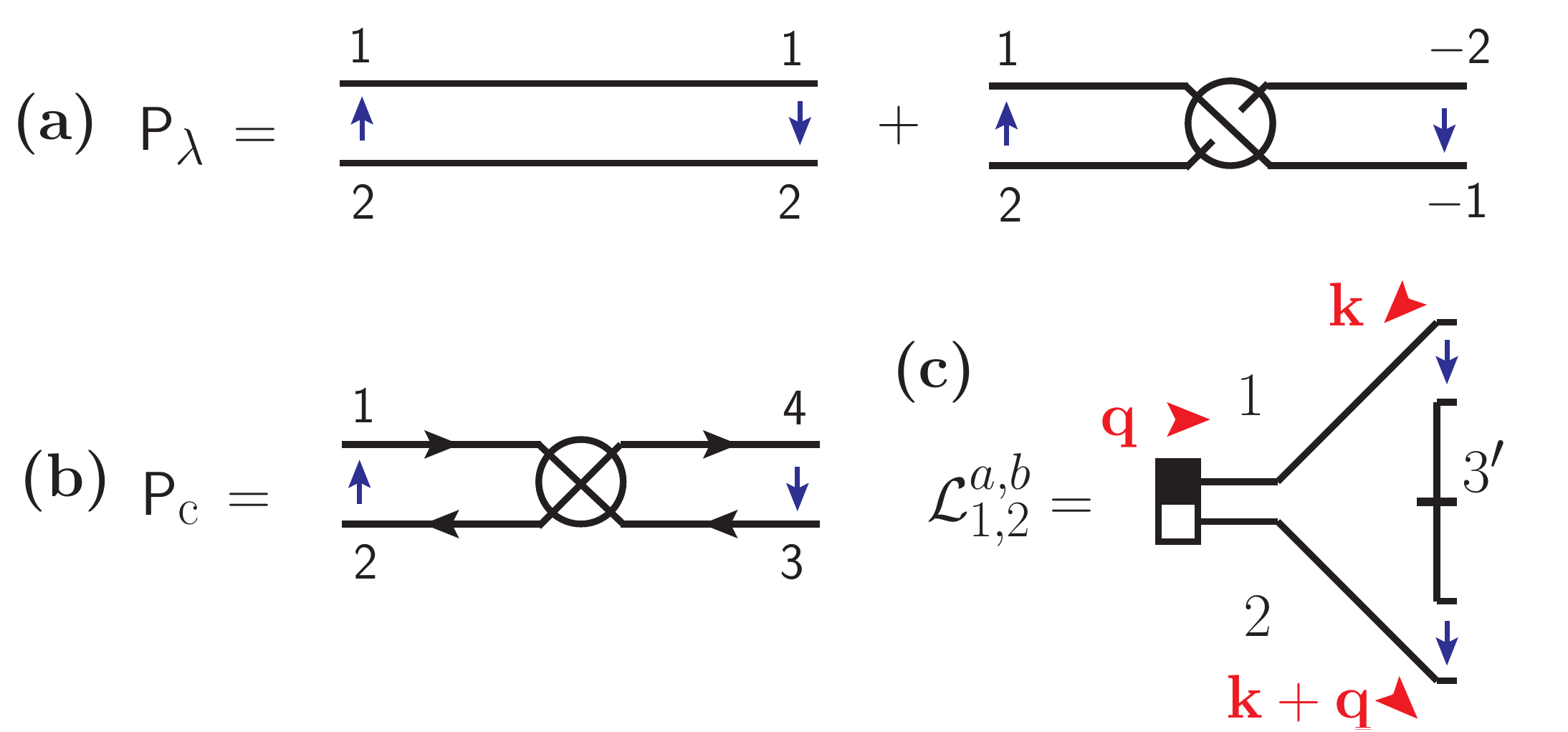}
\caption{Feynman rules for 
class DIII.
(a) The propagator $\sfP_\l$ in Eq.~(\ref{d3-plam}). 
The first 
(second)
diagram is associated with the 
contraction rule $\delta_{1,4} \delta_{2,3} \delta_{ad} \delta_{bc}$ 
($\delta_{1,-3} \delta_{2,-4} \delta_{ac} \delta_{bd}$). 
We use 
the ``blob'' 
to indicate sign flips of frequencies.
(b) The propagator $\sfP_\rmc$ in Eq.~(\ref{d3-pc}). 
The diagram represents the term 
with the frequency conservation law $\delta_{1+3,2+4}$.  
(c) The stiffness vertex represented by Eq.~(\ref{stifnes-a3}) supplemented with a minus sign.
}
\label{fr-d3}
\end{figure}

(i) \emph{Parametrization and Feynman rules.} For class DIII the fast field $\htQ_{\sfF} \in \text{O}(nN)$ 
can be parametrized by the so-called ``$\s$-$\pi$'' coordinates:
\be  \label{d3-fpara}
\begin{split}
\htQ_{\sfF}(\bfr) = & \, \htY(\bfr) + [ \hat{1} + \htY^2(\bfr) ]^{1/2} \\
            = & \, \hat{1} + \htY(\bfr) + \frac{1}{2} \htY^2(\bfr) + \mathcal{O}(\| \htY \|^3), 
\end{split} 
\ee
where 
$\htY$ satisfies the antisymmetric condition in the frequency basis [Eq.~(\ref{aiii-Q})]
\be  \label{d3-coords}
\htY^\trasp = -\hat{\Sigma}^{1} \htY \hat{\Sigma}^{1}.
\ee

The fast mode action is
\be
S_\sfF [\hat{Y}] = S_\sfF^{(0)}[\hat{Y}] + S_\sfF^{(\rmI)}[\hat{Y}],
\ee
where 
\begin{subequations}  \label{act-fast-d3}
\begin{align}
	S_\sfF^{(0)}[\hat{Y}] 
	= \,& 
	-\frac{1}{2}\int_{\mathbf{r}} \text{Tr}\left[ \frac{1}{\lambda} (\nabla \hat{Y})^2 + h \, |\hat{\omega}| \, \hat{Y}^2 + h \, 
	\hat{Y} \, |\hat{\omega}| \, \hat{Y} \right], 
\end{align}
\begin{align}
	S_\sfF^{(\rmI)}[\hat{Y}] 
	= \, & 
	-\sum_a 
	\intl{\substack{\omega_1,\omega_2 \\ \omega_3,\omega_4}}		
	\int_{\mathbf{r}} \delta_{1+3,2+4}  \nn \\ & \times \Gamma_\rmc \, (s_1+s_2)\,(s_3+s_4) \, Y_{1,2}^{a,a} \, Y_{3,4}^{a,a}, \label{sfint-d3}
\end{align}
\end{subequations}
with $S_\sfF^{(0)}$ and $S_\sfF^{(\rmI)}$ arising from Eqs.~(\ref{ci-nonint}) and (\ref{diii-int}), 
respectively. We note that Eq.~(\ref{sfint-d3}) only incorporates the Cooper interaction channel.
The propagators are
\begin{subequations}  \label{fast-props-d3}

\begin{align}
	\Big\langle Y_{1,2}^{a,b}&(-\mathbf{k}) \, Y_{3,4}^{c,d} (\mathbf{k}) \Big\rangle 
	= 
	\mathsf{P}_\lambda + \mathsf{P}_\rmc, \label{fast-prop-d3} \\
	\mathsf{P}_\lambda = 
	& \, -\frac{1}{2}\left(1-\del_{1,-2}\del_{ab}\right) 
        \nn \\ 
	& \phantom{-}\times 
	\left(\delta_{1,4} \delta_{2,3} \delta_{ad} \delta_{bc}  - \delta_{1,-3} \delta_{2,-4} \delta_{ac} \delta_{bd} \right) 
	\nn \\ 
	& \phantom{-}\times 
	\Delta_\rmO (|\omega_1|, |\omega_2|, \mathbf{k}), \label{d3-plam} \\
	\mathsf{P}_\rmc 
	= & \, 2\Gamma_\rmc \, (s_1+s_2)(s_3+s_4) 
	\delta_{1+3,2+4}
	\delta_{ab} \delta_{bc} \delta_{cd} 
	\nn \\
	& \, \times \frac{\Delta_\rmO (|\omega_1|,|\omega_2|, \mathbf{k}) 
	\,
	\Delta_\rmO (|\omega_3|,|\omega_4|, \mathbf{k})}{1+\gamma_\rmc f(|\w_1-\w_2|,\bk,\Lambda)},  \label{d3-pc}
        \end{align}
\end{subequations} \\
where $\Delta_\rmO(|\omega_1|, |\omega_2|, \mathbf{k})$ and $f(|\w|,\bk,\Lambda)$
are defined in
Eqs.~(\ref{delta-o}) and (\ref{def-f}), respectively. 
Note that $\sfP_\lambda$ and $\sfP_\rmc$ are \emph{antisymmetric} [Eq.~(\ref{d3-coords})] in frequency and replica spaces 
and hence 
possess
vanishing diagonal terms.
The 
spin diffusion kernel $\Delta_\rmS (|\omega|, \mathbf{k})$ [Eq.~(\ref{delta-s})] 
is absent compared to class AIII since spin is not conserved 
in 
DIII. 

In Fig.~\ref{fr-d3}(a) we represent the two components of $\sfP_\l$ as the two diagrams that 
exhibit the corresponding frequency and replica structures: 
the first diagram 
corresponds
to 
$\delta_{1,4} \delta_{2,3} \delta_{ad} \delta_{bc}$ 
and the second to 
$\delta_{1,-3} \delta_{2,-4} \delta_{ac} \delta_{bd}$ [Eq.~(\ref{d3-plam})].  
Notice that the first diagram should carry a minus sign. 
The propagator 
$\sfP_\rmc$ is represented by the diagram in Fig.~\ref{fr-d3}(b). 
The arrows along the fermion lines indicate the frequency combination $\delta_{1+3,2+4}$ [Eq.~(\ref{d3-pc})]. 
The stiffness vertex, as pictured in Fig.~\ref{fr-d3}(c), takes the form of Eq.~(\ref{stif-vert-mat}) 
with a sign flip $\hat{\mathcal{L}} (\mathbf{k}, \mathbf{q}) \to -\hat{\mathcal{L}} (\mathbf{k}, \mathbf{q})$ and satisfies the antisymmetric constraint $\hat{\mathcal{L}}^\trasp = - \hat{\Sigma}^1\hat{\mathcal{L}}\hat{\Sigma}^1$ in frequency and replica space.  

\begin{figure}
\centering
\includegraphics[width=0.42\textwidth]{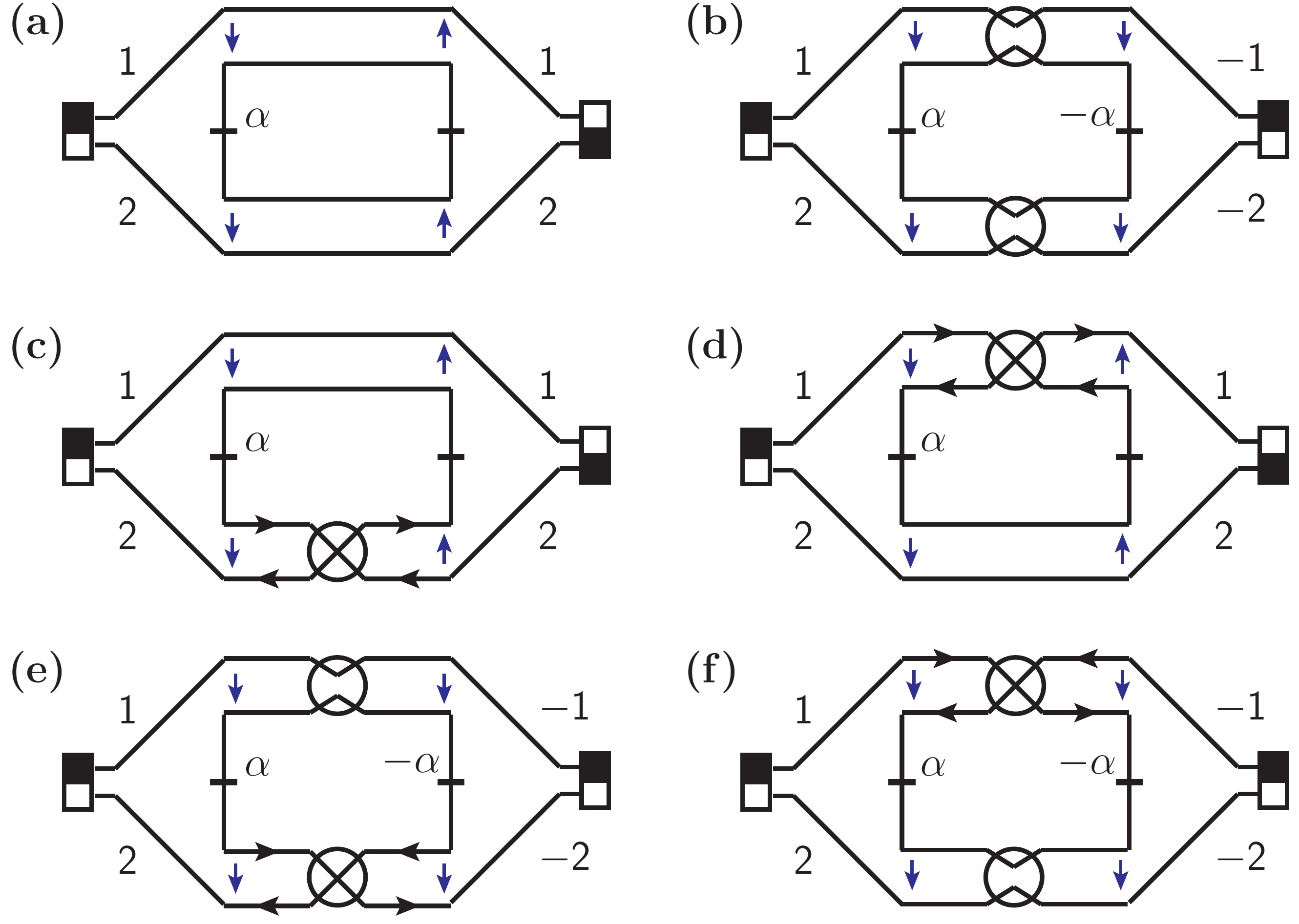}
\caption{Category $\mfrD_{\text{DIII}}$: Diagrams renormalizing $\l$ 
in class DIII.
(a) and (b) Weak 
antilocalization 
contribution [Eq.~(\ref{donly-d3})]. 
(c)--(f) Altshuler-Aronov correction [Eq.~(\ref{AA-lam-d3})].}\label{AA-d3}
\end{figure}

(ii) \emph{Renormalization of $\l$.} The one-loop diagrams that renormalize $\l$ are depicted 
in Fig.~\ref{AA-d3}. $\mfrD_{\text{DIII}}$(a) and $\mfrD_{\text{DIII}}$(b) involve only 
$\sfP_\l$ and give identical contributions. At finite replica ($n \neq 0$) we obtain
\be  \label{donly-d3}
\begin{split}
	\mfrD_{\text{DIII}} & \text{(a)} + \mfrD_{\text{DIII}}\text{(b)} \\ 
               =& 
		\frac{F (n N -2)}{2} 
		\int_{\mathbf{r}} \text{Tr}( \nabla \hat{Q}_\sfS^\dagger \cdot \nabla \hat{Q}_\sfS),
\end{split} 
\ee
where $N$ is the number of Matsubara frequencies, and $F$ is defined by Eq.~(\ref{WZNWF}).
In Eq.~(\ref{donly-d3}) the prefactor 
``$n N-2$''
arises from the 
replica-frequency 
loop, where the diagonal 
terms of the two propagators are removed due to the factor 
``$1 - \del_{1,-2}\del_{ab}$''
in Eq.~(\ref{d3-plam})~\cite{SUP-rep-d3}.

Diagrams $\mfrD_{\text{DIII}}$(c)--$\mfrD_{\text{DIII}}$(f) in Fig.~\ref{AA-d3} incorporate 
one basic diffusion operator $\sfP_\l$ and one interaction-dressed operator $\sfP_\rmc$, and 
give identical contributions:
\be  \label{AA-lam-d3}
\begin{split}
	\sum_{\mathfrak{d} = \text{c}}^{\text{f}}\mfrD_{\text{DIII}}(\mathfrak{d}) 
		= 
		& 
		- 
                \frac{F}{2} 
		\mathcal{K}\left( \gmc\right)
		\int_{\mathbf{r}} \text{Tr}( \nabla \hat{Q}_S^\dagger \cdot \nabla \hat{Q}_S),
\end{split}
\ee
which can be understood as the Altshuler-Aronov correction by Cooper interactions.
The function $\mathcal{K}\left(\gamma\right)$ is defined in Eq.~(\ref{KCooperDef}).

(iii) \emph{One-loop RG equation for $\l$.} 
Substituting 
$l_\phi =2$ [Eq.~(\ref{dynkin})] into 
Eqs.~(\ref{donly-d3}) and (\ref{AA-lam-d3}), taking the replica limit $n \to 0$, 
and performing a trivial rescaling 
\be 
	\l \to 8\pi\l, 
\ee
we obtain the RG equation for $\l$:
\be \label{eq-lambda-d3}
	\frac{\rmd \l}{\rmd \ell} = \, -\l^2 \left[1-(K\l)^2\right] \left[ 2+\mathcal{K}\left(\gmc\right) \right].
\ee 
In class DIII the Cooper interactions 
are always
irrelevant \cite{FXCup}, 
so we do not compute the full beta function for $\gmc$.

\subsubsection{Spin SU(2) symmetry: Class CI} \label{fnlsm-ci}

(i) \emph{Parametrization and Feynman rules.} We parametrize the fast field $\hat{Q}_\sfF \in \text{Sp}(2nN)$ 
via Eq.~(\ref{param-unitary}). In the frequency basis, the unitary and symplectic conditions (\ref{u-const}) and (\ref{sp-const}) imply   
\be \label{c1-fpara}
 \htY^\dagger = \htY, \quad \hat{Y}^\trasp = - ( \hmu^2 \otimes \hat{\Sigma}^1 ) \, \hat{Y} \, ( \hmu^2 \otimes \hat{\Sigma}^1 ).  
\ee
As a consequence, $\hat{Y}$ takes the following block form in spin  space:
\be \label{decomp-c1}
\hat{Y}= \frac{1}{\sqrt{2}}\left( i \hat{W}_0 \ots \hat{1} + \sum_{\a=1}^{3} \hat{W}_\a \ots \hmu^\a \right), 
\ee
where the $nN \times nN$ matrices $\hat{W}_{\rmj=0,1,2,3}$ satisfy
\be
\begin{split}
\hat{W}_0^\dagger  = -\hat{W}_0,  & \,\quad  \hat{W}_0^\trasp  = -\hat{\Sigma}^1 \, \hat{W}_0 \, \hat{\Sigma}^1, \\
\hat{W}_\a^\dagger  = \hat{W}_\a, & \,\quad  \hat{W}_\a^\trasp =  \hat{\Sigma}^1 \, \hat{W}_\a \, \hat{\Sigma}^1, 
\end{split}
\ee 
with $\a \in \{1,2,3\}$. 
The parametrization (\ref{decomp-c1}) naturally separates the interactions to one 
Cooper channel and three equivalent spin channels as shown in Eqs.~(\ref{act-ff-c1-I0}) and (\ref{act-ff-c1-Ij}).
  
The fast mode action consists of four decoupled modes described by the matrix fields $\hat{W}_{\rmj=0,1,2,3}$: 
\be  \label{act-ff-c1}
S_\sfF [\{\hat{W}_\rmj\}] = \sum_{\rmj =0}^{3} \left( S_{\sfF,\rmj}^{(0)}[\hat{W}_\rmj] +  S_{\sfF,\rmj}^{(\text{I})}[\hat{W}_\rmj] \right),
\ee
where the antisymmetric sector $\htW_0$ reads
\begin{subequations}  \label{act-fast-c1} 
\begin{align}
	S_{\sfF,0}^{(0)}[\hat{W}_0] 
	=& \, 
	- \frac{1}{2}\int_{\mathbf{r}} \wtd{\text{Tr}}\left[ \frac{1}{\lambda} (\nabla \hat{W}_0)^2 + h \, |\hat{\omega}| \, \hat{W}_0^2 \right. \nn \\ 
	& \left. + h \, \hat{W}_0 \, |\hat{\omega}| \, \hat{W}_0 \right], \label{act-ff-c1-00}  \\
	S_{\sfF,0}^{(\text{I})}[\hat{W}_0] 
	=& \, - \, \sum_a 
	\intl{\substack{\omega_1,\omega_2 \\ \omega_3,\omega_4}}	 	
	\int_{\mathbf{r}} \delta_{1+3,2+4} \nn \\
        & \times \Gamma_\rmc  (s_1+s_2)\,(s_3+s_4) \, [W_0]_{1,2}^{a,a} \, [W_0]_{3,4}^{a,a} , \label{act-ff-c1-I0}
\end{align}
and the symmetric
sectors
$\htW_{\a=1,2,3}$ read 
\begin{align}
	S_{\sfF,\a}^{(0)}[\hat{W}_\a] 
	=& \, 
	\frac{1}{2} \int_{\mathbf{r}} \wtd{\text{Tr}}\left[ \frac{1}{\lambda} (\nabla \hat{W}_\a)^2 + h \, |\hat{\omega}| \, \hat{W}_\a^2 \right. \nn \\ 
	& \left. +h \, \hat{W}_\a \, |\hat{\omega}| \, \hat{W}_\a \right], \label{act-ff-c1-0j} \\
	S_{\sfF,\a}^{(\text{I})}[\hat{W}_\a] 
	= & \,  
	\, \,\sum_a 
	\intl{\substack{\omega_1,\omega_2 \\ \omega_3,\omega_4}}			
	\int_{\mathbf{r}} \delta_{1+3,2+4} \nn \\ 
	& \times  \Gamma_\rms  (s_1-s_2) \, (s_3-s_4)\, [W_\a]_{1,2}^{a,a} \, [W_\a]_{3,4}^{a,a}. 
	\label{act-ff-c1-Ij}
\end{align}
\end{subequations}
In Eq.~(\ref{act-fast-c1}), $S_{\sfF,\rmj}^{(0)}$ and $S_{\sfF,\rmj}^{(\rmI)}$ arise from Eqs.~(\ref{ci-nonint}) and (\ref{act-int-c1-f}), 
respectively, and $\wtd{\Tr}$ denotes matrix trace operation over replica and frequency spaces.  
Equations~(\ref{act-ff-c1}), (\ref{act-ff-c1-I0}), and (\ref{act-ff-c1-Ij}) exhibit the orthogonality 
among the Cooper channel and the three equivalent spin channels. 
\begin{figure}[b]
\centering
\includegraphics[width=0.42\textwidth]{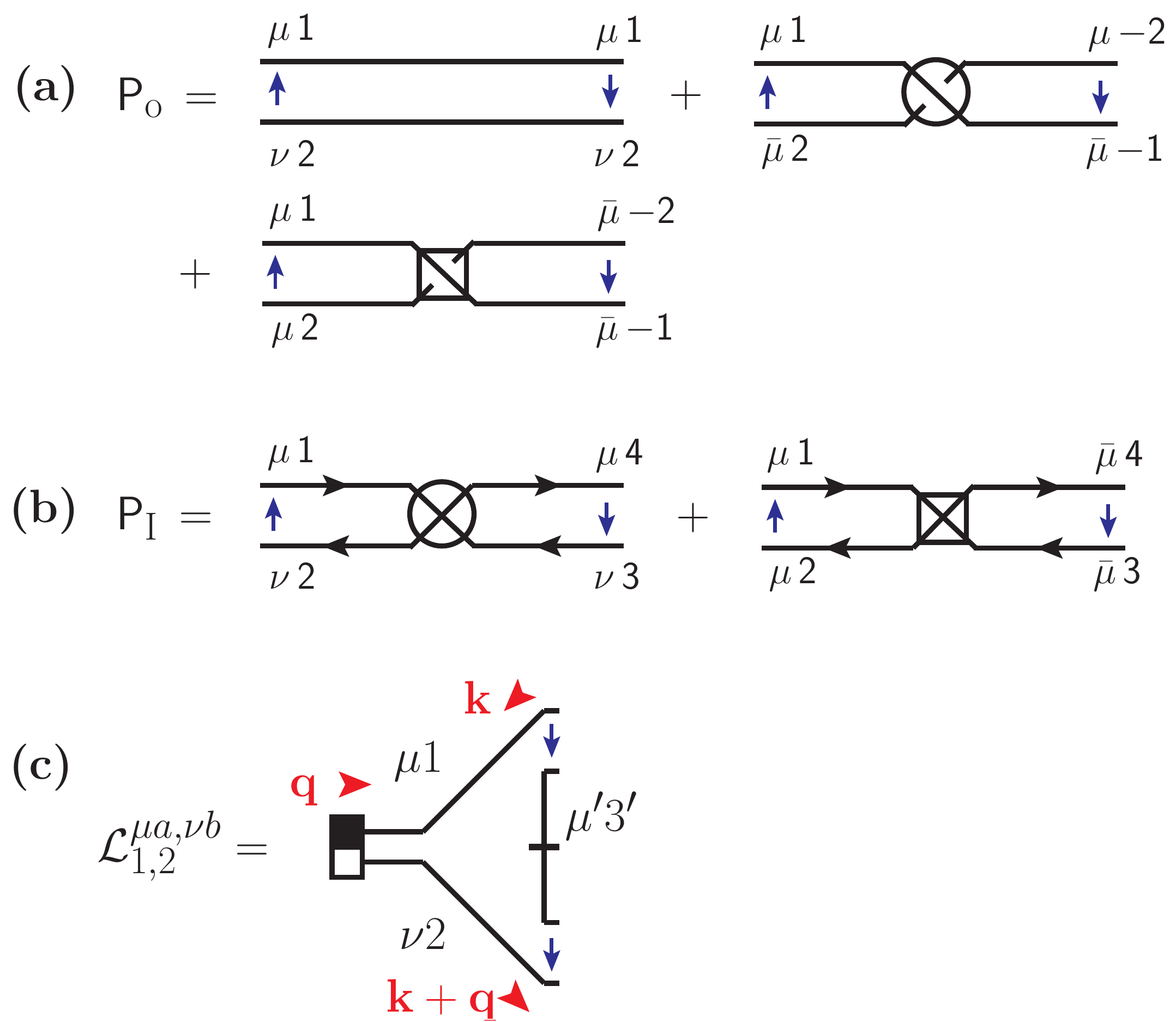}
\caption{Feynman rules for 
class CI.
(a) Disorder-only propagator $\sfP_\rmO$ in Eq.~(\ref{Y-sept-prop}). 
The three diagrams represent the three terms (\ref{c1-dsonly-1})--(\ref{c1-dsonly-3}) in respective order. 
We use blob and square nodes to distinguish spin exchange channels. 
(b) Interaction-dressed propagator $\sfP_\rmI$ in Eq.~(\ref{Y-sept-prop}). 
The two diagrams represent the two terms (\ref{c1-int-1}) and (\ref{c1-int-2}) in respective order.
(c) The stiffness vertex that arises from the coupling between fast and slow 
fields represented by Eq.~(\ref{stifnes-a3}).} 
\label{fr-c1}
\end{figure}

Via the procedure leading to Eq.~(\ref{fast-prop}), the propagators of the $\hat{W}_{\rmj=0,1,2,3}$ fields are readily obtained:
\begin{subequations}  \label{w-prop-c1}
\begin{align}
\left\langle [W_0]_{1,2}^{a,b} (-\mathbf{k}) \, [W_0]_{3,4}^{c,d} (\bk) \right\rangle = & \, \sfP_{0,\l}+ \td{\sfP}_{0,\l}+ \sfP_{\rmc},\\
\left\langle [W_\a]_{1,2}^{a,b} (-\mathbf{k}) \, [W_\a]_{3,4}^{c,d} (\bk) \right\rangle = & \, \sfP_{\l} + \td{\sfP}_{\l} + \sfP_{\rms},
\end{align} 
\end{subequations}
where the disorder-only components are  
\begin{subequations}  \label{prop-f-c1-0}

\begin{align}  
\sfP_{0,\l} = & \, -\frac{1}{2} (1-\del_{ab}\del_{1,-2}) \delta_{1,4} \delta_{2,3} \delta_{ad} \delta_{bc} \, \Delta_\mathrm{O}(|\omega_1|, |\omega_2|, \mathbf{k}), \\
\td{\sfP}_{0,\l} = & \, \frac{1}{2} (1-\del_{ab}\del_{1,-2})  \del_{1,-3}\del_{2,-4} \, \del_{ac}\del_{bd}\, \Delta_\mathrm{O} (|\omega_1|, |\omega_2|, \mathbf{k}),\\
\sfP_{\l} = & \, \frac{1}{2} \delta_{1,4} \delta_{2,3} \delta_{ad} \delta_{bc} \, \Delta_\mathrm{O} (|\omega_1|, |\omega_2|, \mathbf{k}), \\
\td{\sfP}_{\l} = & \, \frac{1}{2} \delta_{1,-3} \delta_{2,-4} \delta_{ac} \delta_{bd} \, \Delta_\mathrm{O} (|\omega_1|, |\omega_2|, \mathbf{k}),
\end{align} 
and the interaction-dressed components are
\begin{align}             
	\sfP_{\rmc} = & \,2 \Gamma_\rmc (s_1+s_2)(s_3+s_4) \delta_{1+3,2+4}  \nn \\ 
	& 
	\times  \delta_{ab} \delta_{bc} \delta_{cd} 
	\frac{\Delta_\rmO (|\omega_1|,|\omega_2|, \mathbf{k}) 
	\, 
	\Delta_\rmO (|\omega_3|,|\omega_4|, \mathbf{k})}{1+\gamma_\rmc f(|\w_1-\w_2|,\bk,\Lambda)},\\
	\sfP_{\rms} = & \, -2 \Gamma_\rms (s_1-s_2)(s_3-s_4)   \delta_{1+3,2+4} \nn \\ 
	& 
	\times \delta_{ab} \delta_{bc} \delta_{cd} 
	\Delta_\rmS (|\omega_1-\omega_2|, \mathbf{k}) 
	\, 
	\Delta_\rmO (|\omega_1-\omega_2|, 0, \mathbf{k}),
\end{align} 
\end{subequations}
with the diffusion kernels $\Delta_\rmO (|\omega_1|,|\omega_2|, \mathbf{k})$ and $\Delta_\rmS (|\w|, \mathbf{k})$ and the logarithmic function $f(|\w|,\bk,\Lambda)$ 
defined by Eqs.~(\ref{delta-o}), (\ref{delta-s}), and (\ref{def-f}), respectively. Similarly to the situation in class DIII, 
the propagators of the antisymmetric field $\htW_0$ possess vanishing diagonal terms.

Via Eqs.~(\ref{decomp-c1}), (\ref{w-prop-c1}), and (\ref{prop-f-c1-0}) we obtain the propagators of the $\hat{Y}$ fields:
\begin{subequations}
\be
\left\langle Y_{1,2}^{\mu a, \mu^\prime b} (-\mathbf{k}) \, Y_{3,4}^{\nu c,\nu^\prime d} (\bk) \right\rangle = \sfP_\mathrm{O} + \sfP_\mathrm{I}.
\ee
\end{subequations}
$\sfP_\mathrm{O}$ and $\sfP_\mathrm{I}$ can be organized to the following forms that exhibit distinguished spin exchange channels:  
\begin{subequations}  \label{Y-sept-prop}
\begin{align}
\sfP_\mathrm{O}  = & \, \frac{1}{2} \left[ \del_{\mu^\prime \mu} \del_{\nu^\prime \nu} \del_{\mu \nu} \left( \sfP_\l - \sfP_{0,\l} \right) + 2 \del_{\mu^\prime \bar{\mu}} \del_{\nu^\prime \bar{\nu}} \del_{\nu \mu^\prime} \sfP_\l \right]  \label{c1-dsonly-1} \\
 &+ \del_{\mu^\prime \bar{\mu}} \del_{\nu^\prime \bar{\nu}} \del_{\nu \mu^\prime} \td{\sfP}_\l    \label{c1-dsonly-2} \\ 
 &- \frac{1}{2}\del_{\mu^\prime \mu} \del_{\nu^\prime \nu} \del_{\mu \bar{\nu}} \left( \td{\sfP}_\l + \td{\sfP}_{0,\l} \right),      \label{c1-dsonly-3} \\
 \sfP_\mathrm{I} = & \,  \frac{1}{2} \left[ \del_{\mu^\prime \mu} \del_{\nu^\prime \nu} \del_{\mu \nu} \left( \sfP_\rms - \sfP_\rmc \right) + 2 \del_{\mu^\prime \bar{\mu}} \del_{\nu^\prime \bar{\nu}} \del_{\nu \mu^\prime} \sfP_\rms \right] \label{c1-int-1} \\ 
                   & -\frac{1}{2} \del_{\mu^\prime \mu} \del_{\nu^\prime \nu} \del_{\mu \bar{\nu}} \left( \sfP_\rms + \sfP_\rmc \right) \label{c1-int-2}
\end{align}
\end{subequations}
where the barred spin index $\bar{\mu}$ is the ``spin flip'' of index $\mu$; e.g., $\{\mu,\bar{\mu}\} = \{\uparrow,\downarrow\}$.

\begin{figure}
\centering
\includegraphics[width=0.42\textwidth]{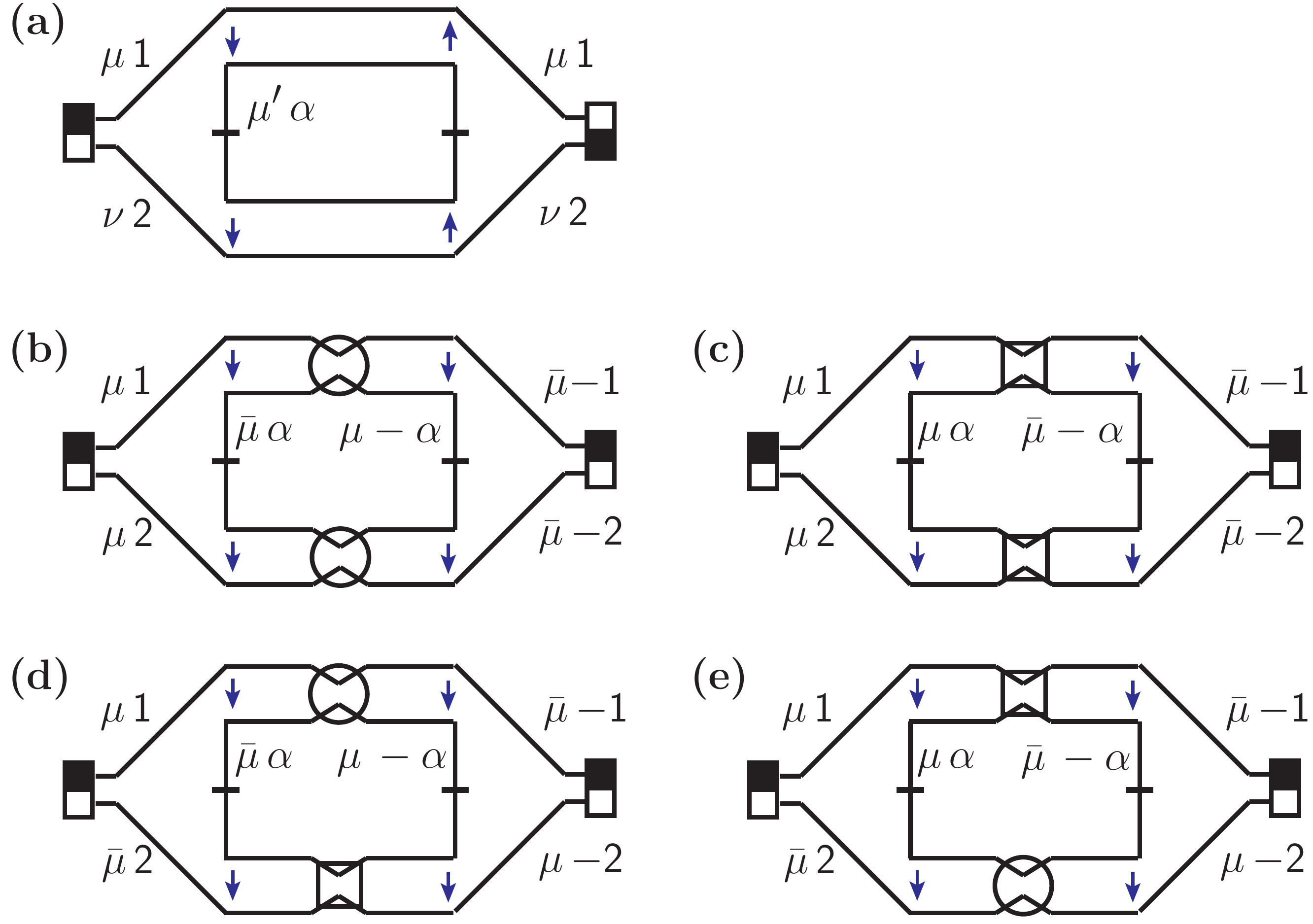}
\caption{Category $\mfrD_\text{CI}1$: Disorder-only (weak localization) diagrams renormalizing 
$\l$ 
in class CI. The
sum of the diagrams (b)--(e) gives the same contribution as (a). Summation is implied over repeated spin indices.
The amplitude is evaluated in Eq.~(\ref{c1-lam-donly}).}
\label{AA-c1-d}
\end{figure}

\begin{figure}
\centering
\includegraphics[width=0.42\textwidth]{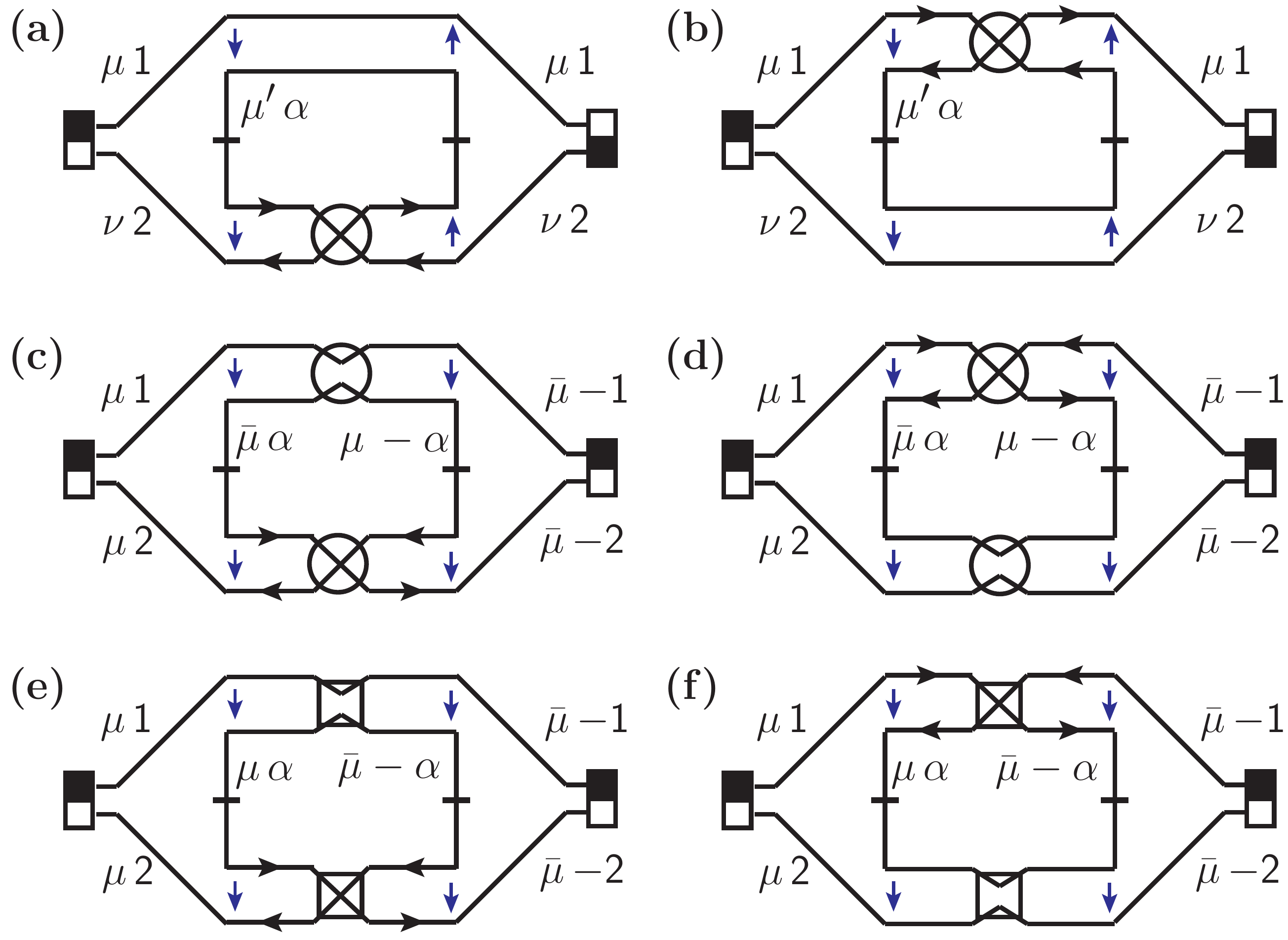}\\
\vspace{0.3cm}
\includegraphics[width=0.42\textwidth]{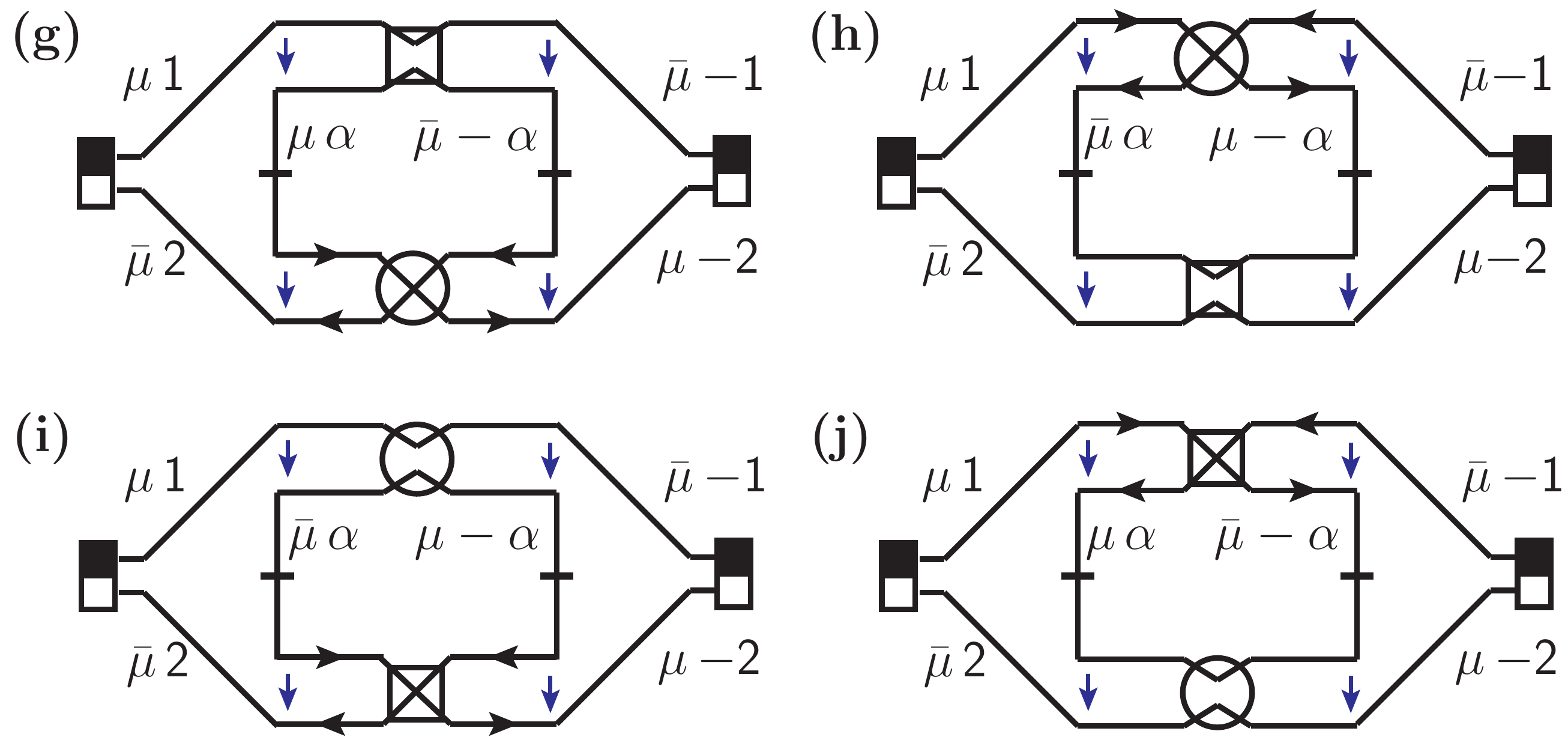}
\caption{Category $\mfrD_\text{CI}2$: Interaction-dressed (Altshuler-Aronov) diagrams renormalizing 
$\l$ 
in class CI.
The diagrams in the left and right columns give identical contributions. 
Moreover, the sum of the diagrams (c), (e), (g), and (i)
gives the same contribution as (a). Summation is implied over repeated
spin indices. 
The amplitude is evaluated in Eq.~(\ref{c1-lam-int}).}\label{AA-c1-I}
\end{figure}

We represent the three terms (\ref{c1-dsonly-1})-(\ref{c1-dsonly-3}) 
by the three diagrams in Fig.~\ref{fr-c1}(a) in respective order. 
We note that the term~(\ref{c1-dsonly-1}) preserves spin indices along the fermion lines, and (\ref{c1-dsonly-2}) and (\ref{c1-dsonly-3}) possess spin flips where the blob and square nodes distinguish spin exchange channels ``$\mu\bar{\mu} \to \mu\bar{\mu}$" and ``$\mu\mu \to \bar{\mu}\bar{\mu}$," respectively. Similarly, the two terms (\ref{c1-int-1}) and (\ref{c1-int-2}) of the interaction-dressed propagator $\sfP_\rmI$ are represented by the two diagrams in Fig.~\ref{fr-c1}(b). We use the blob and square nodes to distinguish spin exchange channels.

The stiffness vertex takes the form in Eq.~(\ref{stifnes-a3}) albeit involves extra spin indices, as depicted in Fig.~\ref{fr-c1}(c). Via the symplectic condition (\ref{sp-const}) one can show that the stiffness vertex matrix $\hat{\mathcal{L}}$ defined by Eq.~(\ref{stif-vert-mat}) satisfies constraints in spin space
$\hat{\mathcal{L}}_{\mu\mu} =- \hat{\Sigma}^1 \hat{\mathcal{L}}_{\bar{\mu}\bar{\mu}}^\trasp \hat{\Sigma}^1$ and $\hat{\mathcal{L}}_{\mu\bar{\mu}} = \hat{\Sigma}^1 \hat{\mathcal{L}}_{\bar{\mu}\mu}^\trasp \hat{\Sigma}^1$,
which is useful for evaluating 
the amplitudes of Feynman diagrams as discussed below.

(ii) \emph{Renormalization of $\lambda$.} Diagrams $\mfrD_\text{CI}1$(a)--(e) appearing in Fig.~\ref{AA-c1-d} 
represent the weak-localization corrections to spin resistance $\l$ with amplitude 
\be  \label{c1-lam-donly}
\begin{split}
	\sum_{\mathfrak{d} =\text{a}}^{\text{e}}\mfrD_\text{CI}1(\mathfrak{d})= 
	& \, 
	(n N+1) F 
	\int_{\mathbf{r}} \text{Tr}( \nabla \hat{Q}_\sfS^\dagger \cdot \nabla \hat{Q}_\sfS). 
\end{split}
\ee

Diagrams $\mfrD_\text{CI}2$(a)-(j) appearing in Fig.~\ref{AA-c1-I} represent the Altshuler-Aronov correction to 
the spin resistance $\l$ with amplitude
\be  \label{c1-lam-int}
\begin{split}
	& \sum_{\mathfrak{d} =\text{a}}^{\text{j}}\mfrD_\text{CI}2(\mathfrak{d}) \\ 
	& = \, 
	F
	\left\{ 3 \left[1+\frac{1-\gamma_s}{\gamma_s} \ln(1-\gamma_s)\right] -\frac{1}{4}\mathcal{K}(\gamma_\rmc)  \right\} 
	\\
	& \quad \times 
	\int_{\mathbf{r}} \text{Tr}( \nabla \hat{Q}_\rmS^\dagger \cdot \nabla \hat{Q}_\rmS).
\end{split}
\ee

(iii) \emph{Full one-loop RG equations.} 
Substituting 
$l_\phi =1$ [Eq.~(\ref{dynkin})] into 
Eqs.~(\ref{c1-lam-donly}) and (\ref{c1-lam-int}) 
and 
rescaling as in Eq.~(\ref{rescale-a3}),
we obtain the full one-loop RG equations:
\begin{subequations}\label{RG-eqs-CI}
\begin{align}
\frac{\rmd \l}{\rmd \ell} = & \, \l^2 \left[1-(K\l)^2\right] [1+ \mathcal{J}(\gms,\gmc)], \label{eq-lambda-c1} \\
\frac{\rmd \gms}{\rmd \ell} = & \, - \frac{\l}{2}\gmc(1-\gms)(1-2\gms), \label{eq-gms-c1}\\
\frac{\rmd \gmc}{\rmd \ell} = & \, \frac{\l}{2}\left\{ -3\gms-2\gmc+3 \gmc \left[ \ln{(1-\gms)} +\gms \right] \right\}-\gmc^2, \label{eq-gmc-c1}
\end{align}
\end{subequations}
where 
\be  \label{AA-correction-c1}
\mathcal{J}(\gms,\gmc) = 3 \left[ 1 + \frac{1-\gms}{\gms} \ln{(1-\gms)}  \right]-\frac{1}{4}\mathcal{K}(\gamma_\rmc),
\ee 
is
the Altshuler-Aronov correction to the spin resistance $\l$. Equations~(\ref{eq-gms-c1}) and (\ref{eq-gmc-c1}) 
have been obtained in Ref.~\cite{dellanna2006} and
are valid
to the lowest nontrivial order in $\lambda$ and $\gamma_\rmc$. 
The implications of Eq.~(\ref{RG-eqs-CI}) for the stability of class CI surface states
were discussed in Refs.~\cite{FXCup, Foster2012}.

\begin{acknowledgments}

We are grateful to I.~Gornyi, I.~Gruzberg, V.~E.~Kravtsov, A.~Mirlin, M.~M\"uller and A.~Scardicchio 
for helpful discussions. This research was supported by the Welch Foundation under Grant No.~C-1809 and 
by an Alfred P. Sloan Research Fellowship (No.~BR2014-035).

\end{acknowledgments}



\begin{thebibliography}{99}
\bibitem{footnote--charge density}
	In the microscopic model presented in 
	Sec.~\ref{micro-sec},
	it can be shown that the long-wavelength 
	electric charge density appears as a Cartan subalgebra component of the non-abelian 
	color current $\vex{J}^j = \eta^\dagger \tmfr^j \htbs \eta$ at the TSC surface.
\bibitem{SUP-rep-d3} 
	For class DIII we exploit the formula $\sum_{c=1}^{n}(1-\delta_{ac})(1-\delta_{cb}) = n-2$ 
	for $a \neq b$.
\end{thebibliography}
\end{document}